\pdfoutput=1 
\documentclass[11pt]{article}
\usepackage{amsmath}
\usepackage{amssymb}
\usepackage{slashed}
\usepackage{graphicx,epstopdf}
\usepackage{cite}
\usepackage{pbox}
\usepackage{geometry}
\usepackage{float}
\usepackage{array}
\usepackage{amsfonts}
\usepackage{graphicx}
\usepackage[font={small}]{caption}
\usepackage[table]{xcolor}
\usepackage{xcolor}
\usepackage{hyperref}
\hypersetup{
    colorlinks=true,
    linkcolor=red,
    filecolor=magenta,      
    urlcolor=blue,
   citecolor=blue,
} 

\providecommand{\xlink}[1]
  {\href{http://arxiv.org/abs/#1}{arXiv:#1}}
\definecolor{RawSienna}{cmyk}{0,0.72,1,0.45}
\definecolor{dgreen}{rgb}{0.0,0.42,0.13}
\definecolor{darkblue}{rgb}{0.0, 0.0, 0.55}
\definecolor{cornellred}{rgb}{0.7, 0.11, 0.11}
\definecolor{calpolypomonagreen}{rgb}{0.08, 0.5, 0.5}

\def\beq{\begin{equation}}
\def\eeq{\end{equation}}
\def\bea{\begin{eqnarray}}
\def\eea{\end{eqnarray}}
\makeatletter
\@addtoreset{equation}{section}

\begin{document}
\title{\LARGE \bf A study on a minimally broken residual TBM-Klein symmetry with its implications on flavoured leptogenesis and ultra high energy neutrino flux ratios }
\author{{Rome Samanta$^{\rm a,b}$\footnote{R.Samanta@soton.ac.uk}, Mainak Chakraborty$^{\rm c}$\footnote{mainak.chakraborty2@gmail.com}
}\\
a) Physics and Astronomy, University of Southampton, Southampton, SO17 1BJ, U.K. \\
  b) Saha Institute of Nuclear Physics, 1/AF Bidhannagar,
  Kolkata 700064, India\\
c) Centre of Excellence in Theoretical and Mathematical Sciences\\
Siksha \textquoteleft O\textquoteright Anusandhan (Deemed to be University) \\Khandagiri Square, Bhubaneswar 751030, India\\
  }
\maketitle
\begin{abstract}
We present a systematic study on   minimally perturbed neutrino mass matrices which at the leading order give rise to Tri-BiMaximal (TBM) mixing due to a 
residual $\mathbb{Z}_2\times \mathbb{Z}_2^{\mu\tau}$ Klein symmetry  in the neutrino mass term of the  low energy effective seesaw Lagrangian. Considering 
only the breaking of $\mathbb{Z}_2^{\mu\tau}$ with two relevant breaking parameters ($\epsilon_{4,6}^\prime$), after a comprehensive numerical analysis, we
show that  the phenomenologically viable case  in this scenario is a special case of TM1 mixing.  For this class of models, from the phenomenological 
perspective, one always needs large breaking (more than $ 45\%$) in one of the breaking parameters. However, to be consistent the maximal mixing of $\theta_{23}$,
while more than $ 35\%$ breaking is needed in the other, a range $49.4^\circ-53^\circ$ and $38^\circ-40^\circ$ could be probed allowing breaking up to  $ 25\%$ 
in the same parameter. Thus though this model cannot distinguish the octant of $\theta_{23}$, non-maximal mixing is preferred from the viewpoint of small breaking.
The model is also interesting from leptogenesis perspective. Unlike the standard $N_1$-leptogenesis scenario, here  all the  RH neutrinos contribute to  lepton 
asymmetry due to  the small mass splitting controlled by the $\mathbb{Z}_2^{\mu\tau}$ breaking parameters. Inclusion of flavour coupling effects (In general, which
have been  partially included in all the leptogenesis studies in perturbed TBM framework) makes our analysis and results pertaining to a successful leptogenesis  
more accurate than any other studies in existing literature. Finally, in the context of recent discovery of the ultra high energy (UHE) neutrino events at IceCube, 
assuming UHE neutrinos originate from purely astrophysical sources, we obtain prediction on the neutrino flux ratios at neutrino telescopes.
\end{abstract}
\section{Introduction}
The structure of the leptonic mixing matrix $U_{\rm PMNS}$  has always been  the center of attraction in the flavour model 
building landscape. Until the experimental discovery of a nonvanishing value of the reactor mixing angle 
$\theta_{13}$\cite{An:2012eh,Ahn:2012nd}, it was the paradigm of Tri-BiMaximal (TBM) Ansatz of 
$U_{\rm PMNS}$\cite{Harrison:2002er}, that dominated almost for a decade\cite{Xing:2002sw} with the 
prediction $\theta_{13}=0$, particularly in the approaches of model building with the discrete non-Abelian 
symmetries such as $A_4$, $S_4$ etc.\cite{Ishimori:2010au}. Now the neutrino mixing parameters, particularly 
two mass squared differences and three mixing angles have entered in the precision era. Thus, as far as the TBM mixing 
is concerned, it has been rendered outdated at least by the nonvanishing value of $\theta_{13}$. However, due to its other 
two predictions,  $\tan\theta_{23}=1$ and $\sin\theta_{12}=1/\sqrt{3}$ which are still close to their respective experimental 
best-fit values, the TBM mixing cannot be just overthrown  from the way in  search for an viable flavour model of
neutrino masses and mixing. Several theories for the modification of TBM mixing in terms of perturbations to the TBM 
mass matrix  have been proposed \cite{He:2006qd,models}. A brief recall of few of the  existing theories which deal with schemes of the modified TBM mixing would be worthwhile in our context. Ref.\cite{He:2006qd} has discussed the 
consequences of perturbation to the effective $M_\nu^{TBM}$ 
but with less emphasis on the high energy symmetries. 
On the other hand, in some of the works in Ref.\cite{models}, a high energy symmetry group  $A_4$ is perturbed
softly such that the effective residual symmetries are unable to generate the exact TBM mixing. Alternative moderated versions, such as 
TM1\cite{Albright:2008rp,Xing:2006ms}, TM2\cite{Albright:2008rp,Grimus:2008tt} mixing with an additional prediction 
on the Dirac CP phase $\delta$, have also been considered to comply with the existing neutrino oscillation data. Elaborate descriptions of direct models mainly focusing on TM1 and TM2 mixings have been given in references \cite{tm1} and \cite{tm2}.\\


Besides all these, there exists a residual symmetry approach\cite{Lam:2007qc,Lam:2008rs,Lam:2008sh,Ge:2011qn,He:2011kn}. In \cite{Lam:2007qc,Lam:2008rs,Lam:2008sh} it has been shown 
that a $3\times 3$ neutrino Majorana  mass matrix with nondegenerate eigenvalues always enjoys 
a $\mathbb{Z}_2\times \mathbb{Z}_2$ residual Klein symmetry accompanied with a diagonal charged lepton 
mass matrix which is further protected by a $\mathbb{Z}_n$ symmetry for $n>2$. Thus, given a neutrino mixing 
matrix, one can always construct the corresponding residual Klein symmetry generators for each of 
the $\mathbb{Z}_2$ symmetries.  For a  mixing of TBM kind, these generators are found to be identical to the generators 
of the $S_4$ group in a three dimensional irreducible representation\cite{Lam:2008sh}. 
The diagonal $\mathbb{Z}_3$ type symmetry acts as the residual symmetry in the charged lepton sector 
while the other two generate a TBM-Klein symmetry with one of them being the $\mu\tau$ interchange 
symmetry\cite{mutaus}.  In\cite{CP1}, the consequences of a Scaling-Klein symmetry  have also been  worked out.\\

It is clear that the vanishing value of $\theta_{13}$ in TBM mixing is caused due to the existence of the $\mu\tau$ 
interchange symmetry as one of the TBM-Klein symmetry generators. Thus to be consistent with the oscillation data,  
one has to relax the constraints arising from the exact $\mu\tau$ symmetry. One way is to consider a $\mu\tau$ flavoured 
nonstandard CP symmetry (${\rm CP}^{\mu\tau}$)\cite{mutau} instead of an exact $\mu\tau$ symmetry with the other 
TBM-Klein generator being completely broken. Introduction of such a symmetry leads to a 
co-bimaximal ($\theta_{23}=\pi/4$, $\cos\delta=0$) mixing\cite{Ma:2015fpa}.  This approach has 
drawn a lot of attention\cite{CPt} after the recent  hint from T2K about a maximal Dirac CP 
violation\cite{Abe:2017uxa}. Following this approach, alternatives to TBM mixing have been 
proposed recently in\cite{joshi} and \cite{rode}. In both the papers, the $\mu\tau$ interchange 
symmetry has been replaced by a $\mu\tau$ CP symmetry keeping the remaining generator of the TBM-Klein
symmetry intact. This further makes the ${\rm CP}^{\mu\tau}$  more predictive with the added predictions 
of the unbroken TBM generator. 
In our present work we follow this direction, i.e., we keep a generator 
of the residual TBM-Klein symmetry unbroken and study modifications of the $\mu\tau$ interchange symmetry. 
However, instead of replacing the $\mu\tau$ interchange symmetry by ${\rm CP}^{\mu\tau}$, we have shown 
how minimal the breaking of the former could be to be consistent with the recent global fit neutrino oscillation data\cite{Capozzi:2017ipn} or in other words, we have performed a study on the goodness of the $\mu\tau$ symmetry while keeping the other $\mathbb{Z}_2$ of the TBM-Klein symmetry unbroken.
Unlike in \cite{joshi,rode}, here a nonmaximal value of $\theta_{23}$ is also allowed. We show that among the two relevant breaking parameters, one should always be large (more than 0.45). However,  to be consistent with the other global fit data, whilst  a maximality or a near maximality in $\theta_{23}$ requires  large breaking (more than 0.35) in the other,  a range $49.4^\circ-53^\circ$  or $38^\circ-40^\circ$ could be probed if we allow the same breaking parameter up to 0.25  (to be consistent with neutrino oscillation global fit data, at least $16\%$ breaking of the $\mu\tau$ symmetry is required in our scenario). Thus our model could be tested shortly in the experiments such as  NO$\nu$A\cite{Adamson:2017qqn}. We have studied
the breaking of the $\mu\tau$ interchange symmetry from the  Lagrangian level of the Type-I seesaw. This in turn 
has allowed us to explore a scenario of leptogenesis with quasi-degenerate heavy RH neutrinos and to work out 
the consequences pertaining  to a successful leptogenesis in this scheme. 

%

Without loss of any generality we choose to work in the diagonal basis of charged leptons where the right
handed neutrino mass matrix is also  diagonal unless it is perturbed. The Lagrangian for the neutrino mass 
terms (Dirac+Majorana) is denoted as
\bea
-\mathcal{L}_{mass}^{\nu,N}= \overline{\slashed{L}}_{L\alpha}(m_D)_{l\alpha}N_{Rl} +\frac{1}{2}\overline{N_{Rl}^C}(M_R)_l \delta _{lm}N_{Rm} + {\rm h.c}., \label{seesawlag}
\eea
where $\slashed{L}_{L\alpha}=\begin{pmatrix}\nu_{L\alpha} & e_{L\alpha}\end{pmatrix}^T$ is the SM lepton doublet of 
flavour $\alpha$. The effective light neutrino mass matrix is then given by the 
well known Type-I seesaw formula 
\bea
M_\nu = -m_DM_R^{-1}m_D^T. \label{seesaweq}
\eea  
A unitary matrix  $U$  diagonalizes  $M_\nu$ in (\ref{seesaweq}) as
\begin{equation}
U^T M_\nu U=M_\nu^d \equiv diag\hspace{1mm}(m_1,m_2,m_3),\label{e0}
\end{equation} 
where $m_i\hspace{1mm}(i=1,2,3)$ are real positive mass eigenvalues of light neutrinos. Since $M_\ell$  is diagonal, $U$ is simply 
equivalent to the leptonic ${\rm PMNS}$ mixing matrix $U_{\rm PMNS}$:  
\bea
U=P_\phi U_{\rm PMNS}\equiv 
P_\phi \begin{pmatrix}
c_{1 2}c_{1 3} & e^{i\frac{\alpha}{2}} s_{1 2}c_{1 3} & s_{1 3}e^{-i(\delta - \frac{\beta}{2})}\\
-s_{1 2}c_{2 3}-c_{1 2}s_{2 3}s_{1 3} e^{i\delta }& e^{i\frac{\alpha}{2}} (c_{1 2}c_{2 3}-s_{1 2}s_{1 3} s_{2 3} e^{i\delta}) & c_{1 3}s_{2 3}e^{i\frac{\beta}{2}} \\
s_{1 2}s_{2 3}-c_{1 2}s_{1 3}c_{2 3}e^{i\delta} & e^{i\frac{\alpha}{2}} (-c_{1 2}s_{2 3}-s_{1 2}s_{1 3}c_{2 3}e^{i\delta}) & c_{1 3}c_{2 3}e^{i\frac{\beta}{2}}
\end{pmatrix},\label{eu}
\eea
where $P_\phi={\rm diag}~(e^{i\phi_1},~e^{i\phi_2},~e^{i\phi_3})$ is an unphysical diagonal  phase matrix and 
$c_{ij}\equiv\cos\theta_{ij}$, $s_{ij}\equiv\sin\theta_{ij}$ with the mixing angles $\theta_{ij}=[0,\pi/2]$. 
Here we have followed the PDG convention\cite{Agashe:2014kda}  but denote our Majorana phases by $\alpha$ and $\beta$. CP-violation enters in the leptonic sector through  nontrivial values of the Dirac phase $\delta$ and Majorana 
phases $\alpha,\beta$  with $\delta,\alpha,\beta=[0,2\pi]$. 
\paragraph{}
We first derive the constraint equations emerging from a 
residual Klein symmetry in the case of a general $\mu\tau$ interchange symmetry and then discuss the implications of 
those equations to  the TBM mixing scheme plus the related modifications. The unbroken TBM generator in this scenario
leads to a TM1 and a TM2 type mixing\cite{Albright:2008rp}. Since the predicted solar mixing angle ($35.8^0$) for a
TM2 type mixing is disfavored at $\sim 3\sigma$\cite{Capozzi:2017ipn}, we devote our entire numerical section only to the 
TM1 type scenario arising in our analysis.  Notice that,  though the unbroken $\mathbb{Z}_2$ leads to a TM1 type mixing,  here  the existence of another partially broken $\mu\tau$ symmetry makes this scenario more predictive than the pure TM1.
\paragraph{}
Predominance matter over antimatter has become a proven fact by several experimental observations. All our known structures of 
universe (like stars, galaxies and clusters) are made up of matter, where as existence of antimatter hasn't been 
confirmed yet. The dynamical process of generation of baryon asymmetry from baryon symmetric era of early Universe is known as 
baryogenesis. Among the various possible mechanisms of baryogenesis, the most interesting and also relevant to our 
present neutrino mass model is baryogenesis through leptogenesis. For successful generation of baryon asymmetry Sakharov
conditions\cite{Sakharov:1967dj} must be satisfied. The necessary CP violation is provided by the complex Yukawa coupling between heavy singlet 
right handed neutrinos and left handed doublet neutrinos. Existence of Majorana mass term of the right handed neutrinos 
ensures lepton number violation. Departure from thermal equilibrium is achieved whenever the interaction rate of 
Yukawa coupling is smaller than the Hubble expansion  rate. Thus the model under consideration possesses 
all the necessary ingredients to satisfy Sakharov conditions and is able to produce lepton asymmetry at a very high
scale which is further converted into baryon asymmetry through Sphaleronic transitions. 
\paragraph{}
In this work we examine qualitatively as well as quantitatively, how efficiently our model can address 
the low energy neutrino phenomenology and cosmological baryon asymmetry within the same frame work. Therefore 
Lagrangian parameters once constrained by the oscillation data are used thereafter in the computations of
leptogenesis in a bottom up approach. Since we plan to study leptogenesis over a wide range of right handed neutrino mass, we explore the
possibilities of both flavour dependent and flavour independent leptogenesis. To track the evolution of the baryon asymmetry 
from very high temperature down to very low temperature (present epoch) we use  network of most general flavour 
dependent coupled Boltzmann Equations (BEs) where contributions from all three generations of RH neutrinos 
are taken into account. Implications of nondiagonal right handed neutrino mass matrix have also been dealt with great care. \\

Recently IceCube detector\cite{Aartsen:2013bka,Aartsen:2016oji} at the south pole has detected ultra high energy (UHE ) neutrino events which in turn has opened a new era in neutrino astronomy. Though the present data points those neutrinos to be of extraterrestrial origin, the sources of those neutrinos are still unknown. Assuming the sources to be pure astrophysical (we consider the conventional $pp$ and $p\gamma$ sources), we calculate the flavour ratios at the neutrino telescope. Due to the broken $\mu\tau$ symmetry in this model,  commonly predicted democratic flavour distribution 1:1:1 changes. Thus the prediction of the flavour ratios in this model will be tested hopefully with enhanced statistics in neutrino telescopes such as IceCube.\\

So, the  main and new features of this work could be summarised as follows: \\

i) Unlike the previous literatures, the model under consideration deals with neither arbitrarily broken TBM \cite{He:2006qd} nor an exact 
TM1 mixing\cite{Albright:2008rp,Xing:2006ms,tm1,tm2}. To be precise, for an arbitrarily broken TBM, both the symmetries in
$\mathbb{Z}_2\times \mathbb{Z}_2^{\mu\tau}$ are broken softly\cite{He:2006qd} whereas, for an exact or pure TM1 mixing,  
$\mathbb{Z}_2^{\mu\tau}$ is broken completely while the other $\mathbb{Z}_2$ remains unbroken. In our scenario, similar 
to pure TM1 mixing, we keep  $\mathbb{Z}_2$ unbroken, however, instead of breaking  $ \mathbb{Z}_2^{\mu\tau}$ completely, 
we restrict to the fact that how minimal the breaking of the $\mathbb{Z}_2^{\mu\tau}$ could be to be consistent with the existing 
neutrino data. Thus the scenario is a special case of a pure TM1 mixing and is more predictive than the said mixing due the 
existence of another partially broken  $\mathbb{Z}_2^{\mu\tau}$. This separates our work from any previous analysis and thus after 
constraining the model parameter space with neutrino oscillation data, whatever predictions emerge, are entirely novel. 
Though Ref.\cite{Mohapatra:2004hta}, and Ref.\cite{Ge} share some common ground with this work, we shall point out the distinction
in the numerical section where we present a comparative study of this work with the works which look similar a priory. \\

ii) For the numerical analysis, we perform an  exact diagonalization of the neutrino mass matrix which in turn
allows us to take into consideration the  
terms which are higher order in the breaking parameters (terms proportional to $\epsilon^2$ and so on are usually neglected in perturbed TBM analysis).
Thus our numerical results are quite robust. If we allow  breaking in one of  the relevant breaking parameters up to $25\%$, our minimal breaking scenario prefers non-maximal mxing, e.g.,  $\theta_{23}$  within the range $\sim (49.4^\circ-53^\circ)$ and hence a particular range of the Dirac CP phase $\delta$, due an analytical correlation predicted by the unbroken $\mathbb{Z}_2$ (though this correlation is also present in case of a pure TM1 mixing).  Thus the goodness of this scenario can easily be tested in the ongoing and forthcoming neutrino experiments. \\

iii) Within the broken TBM scenarios, in general baryogenesis via leptogenesis has been studied assuming $N_1$-dominated scenario where the
heavy neutrino flavour effects are neglected, assuming any asymmetry produced by the heavier neutrinos are significantly washed out 
by $N_1$ and $N_{2,3}$-washout do not affect asymmetry produced by $N_1$ at the production. In addition, to compute the final $Y_B$, approximate
formulae  are used  which include partial flavour coupling effects (an assumption of `$A$' matrix to be diagonal). In our scenario, 
due to the typical structure of the symmetry, the RH neutrino masses are very close to each other which compels us to take into account the
effects of $N_2$ as well as $N_3$. We solve full flavour dependent Boltzmann equations with full flavour coupling effects and show 
how depending upon breaking parameters the next to lightest of the heavy RH neutrinos affects the final asymmetry. With best of our 
knowledge, within the  broken TBM framework, such a diligent computation of leptogenesis has not been done before.\\

iv) Encouraged by the recent discovery of high energy neutrino events at IceCube, we calculate flavour flux ratios at neutrino telescopes
which deviates from the democratic flavour distribution 1:1:1. The predicted flavour flux ratios are either testable with enhanced statistics 
at the neutrino telescopes such as IceCube or could be used as an input to the  astrophysical fits \cite{Sui:2018bbh} to the existing data to test this model.\\

v) This model also predicts a testable range of the neutrino less double beta decay parameter $|(M_\nu)_{ee}|$.

\paragraph{}
The rest of the paper is organized as follows: 
In Sec.\ref{sc2}, we briefly review the basic framework of residual symmetry along with a discussion on the general $\mu\tau$ interchange symmetry which is characterized by a residual $\mathbb{Z}_2^{\lambda}\times\mathbb{Z}_2^{\mu\tau}$ Klein symmetry. Given the general setup in Sec.\ref{sc2}, we further focus on the  TBM mixing ($\lambda=1/\sqrt{3}$) and  phenomenologically consistent minimal breaking pattern of the residual $\mathbb{Z}_2^{\mu\tau}$ in Sec.\ref{pt_tbm}. Sec.\ref{sc4} is entirely devoted to the study of generation of baryon asymmetry through leptogenesis. Its various subsections 
deal with rigorous evaluation of CP asymmetry parameters, setting up the chain of Boltzmann equations applicable in different 
temperature regimes. The extensive numerical study of the viable cases (which includes  : constraining the parameters by $3\sigma$ global fit of oscillation data, computation related to the baryogenesis via leptogenesis and prediction of the flavour flux ratios at the neutrino telescopes) is given in Sec.\ref{sc5}. In Sec.\ref{sc6}, we summarize the entire work and try to highlight the salient features of this study towards addressing neutrino oscillation phenomenology along with major issues such as baryon asymmetry of universe.

\section{Residual symmetry and its implication on $\mu\tau$ variants} \label{sc2}
 A horizontal symmetry $G_i$ of a neutrino  mass matrix is realized through the invariance equation 
\bea
G_i^T M_\nu G_i = M_\nu, \label{horiz}
\eea 
where $G_i$ is an unitary matrix in the neutrino flavour space. Now Eq.(\ref{horiz}) and (\ref{e0}) together imply 
that we can define a new unitary matrix $V_i=G_iU$ such that it also diagonalizes $M_\nu$. 
The matrix $V$ should then be equal to $U d_i$:
\bea
G_iU=Ud_i\equiv U^\dagger G_i U=d_i
 \label{sym1}
 \eea
with $d_i$ being a diagonal rephasing matrix. For neutrinos of Majorana type, $d_i^2=1$. Therefore, $d_i$ can only have 
entries $\pm 1$. Thus there are now eight possible structures of $d_i$ two  of which are a simple unit matrix and its negative. 
The remaining six can be considered as three different pairs, where the two matrices of a pair are identical to 
each other apart from an overall relative negative sign.
Finally, among these three (pairs) 
matrices, only two are independent as each $d_i$ always satisfies the relation $d_i=d_jd_k$, where $i, j$ and $ k$ can take 
value 1, 2 and 3. Now $d_i^2=1$ implies each $d_i$ define a $\mathbb{Z}_2$ symmetry. Therefore, on account of the relation 
in  (\ref{sym1}), each $G_i~(G_i^2=1,~{\rm det}\hspace{1mm}(G_i)=\pm 1)$ also represents $\mathbb{Z}_2$ symmetry and generates the 
residual $\mathbb{Z}_2\times \mathbb{Z}_2$ symmetry (Klein Symmetry) in the neutrino mass term of the Lagrangian. We can now choose the 
two independent $d_i$ matrices as $d_2={\rm diag} \hspace{1mm}(-1,1,-1)$ and $d_3={\rm diag} \hspace{1mm}(-1,-1,1)$ 
for ${\rm det}\hspace{1mm}(G_i)=1$. For ${\rm det}\hspace{1mm} G_i=-1$, $d_2$ and $d_3$ would differ from the previous 
choices only by an overall minus sign.\\ 

Given this basic set up, we first discuss the general $\mu\tau$ interchange symmetry in the framework of 
residual $\mathbb{Z}_2\times \mathbb{Z}_2$. We then  proceed to the discussion of TBM  mixing 
by setting  the solar mixing angle $\theta_{12}=\sin^{-1}(1/\sqrt{3})$ in the $\mu\tau$ interchange scheme. A neutrino 
Majorana mass matrix 
 \bea
M_{\nu}^{\mu\tau}=\begin{pmatrix}
a&b&-b\\
b&c&d\\
-b&d&c
\end{pmatrix}
\eea
invariant under the  $\mu\tau$ interchange symmetry is diagonalized as 
\bea
(U_\nu^{\mu\tau})^T M_\nu^{\mu\tau}U_\nu^{\mu\tau}=M_d^{\mu\tau},
\eea
where
\bea
U_\nu^{\mu\tau}=\begin{pmatrix}
\sqrt{1-\lambda^2}&\lambda & 0\\
-\frac{\lambda}{\sqrt{2}} & \frac{1}{\sqrt{2}}\sqrt{1-\lambda ^2} & \frac{1}{\sqrt{2}}\\
\frac{\lambda}{\sqrt{2}} &-\frac{1}{\sqrt{2}}\sqrt{1-\lambda ^2} & \frac{1}{\sqrt{2}}
\end{pmatrix}.\label{umut}
\eea
The parameter $\lambda$ is related to the solar mixing angle as $\lambda=\sin \theta_{12}$. 
Here we choose the appropriate  minus signs in $M_\nu^{\mu\tau}$ to be in conformity with the PDG 
convention\cite{Agashe:2014kda}. Now from (\ref{sym1}), $G_2$ and $G_3$ corresponding to $d_2$ and $d_3$ can 
be  calculated as
\bea
 G_2^{\lambda}=\begin{pmatrix}
2\lambda^2 -1 &\lambda \sqrt{2(1-\lambda^2)} &-\lambda \sqrt{2(1-\lambda^2)}\\
 \lambda \sqrt{2(1-\lambda^2)}&-\lambda^2 &-(1-\lambda^2)\\
 -\lambda \sqrt{2(1-\lambda^2)}&-(1-\lambda^2) &-\lambda^2
\end{pmatrix},
G_3=\begin{pmatrix}
-1&0&0\\
0&0&1\\
0&1&0
\end{pmatrix}.
\eea 
The relation $d_i=d_jd_k$ makes the construction of $G_1$ simple: 
$G_1^{\lambda}=G_2^{\lambda}G_3$. Thus $G_1^{\lambda}$ will be of form
\bea
G_1^{\lambda}=\begin{pmatrix}
1-2\lambda^2 &-\lambda \sqrt{2(1-\lambda^2)} &\lambda \sqrt{2(1-\lambda^2)}\\
- \lambda \sqrt{2(1-\lambda^2)}&-(1-\lambda^2) &-\lambda^2\\
 \lambda \sqrt{2(1-\lambda^2)}&-\lambda^2 &-(1-\lambda^2)
\end{pmatrix}.
\eea
Since $G_3$ is basically the $\mu\tau$ interchange symmetry in the flavour basis,  we therefore rename the
residual Klein symmetry for this $\mu\tau$ interchange case as $\mathbb{Z}_2^{\lambda}\times \mathbb{Z}_2^{\mu\tau}$. 
We now implement the $\mathbb{Z}_2^{\lambda}\times \mathbb{Z}_2^{\mu\tau}$ on the Dirac mass matrix $m_D$ and the Majorana mass matrix $M_R$ of (\ref{seesawlag}) as 
\bea
(G_{2}^\lambda)^T m_D^0 G_{2}^\lambda = m_D^0,G_{3}^T m_D^0 G_3=m_D^0, \nonumber\\
(G_{2}^\lambda)^T M_R^0 G_{2}^\lambda = M_R^0,G_{3}^T M_R^0 G_3=M_R^0.\label{res1}
\eea
Equations in (\ref{res1}) automatically imply the $G_1^{\lambda}$ invariance of $m_D$ and $M_R$ on account of the 
relation $G_1^{\lambda}=G_2^{\lambda}G_3$. Now one can work out the constraint equations that arise due the invariance 
relations in (\ref{res1}). A most general $3\times3$ mass matrix 
\bea
M^{G}=\begin{pmatrix}
A&B&C\\
D&E&F\\
G&H&I
\end{pmatrix}\label{gen}
\eea
that is invariant under  $G^\lambda_{1,2,3}$ (cf. Eq. \ref{res1}) would lead to the following constraint equations:
for $G_1^\lambda$ invariance,
\bea
B+C&=&\lambda(\sqrt{2(1-\lambda^2)})^{-1}(-H+F+E-I),\nonumber\\
D+G&=&\lambda(\sqrt{2(1-\lambda^2)})^{-1}(H-F+E-I),\nonumber\\
(D-G+B-C)&=&\Big \{\sqrt{2(1-\lambda^2)}(2\lambda^2 -1)\Big \}^{-1} [4\lambda (1-\lambda^2) A +2\lambda (1-\lambda^2)(H+F-E-I)],\nonumber\\
B-C&=&D-G \label{G1f} 
\eea 
for $G_2^\lambda$ invariance,
\bea
B+C&=&-(2\lambda^{-1})\sqrt{2(1-\lambda^2)}(-H+F+E-I),\nonumber\\
D+G&=&(2\lambda^{-1})\sqrt{2(1-\lambda^2)}(-H+F-E+I),\nonumber\\
(D-G+B-C)&=&\Big \{\sqrt{2(1-\lambda^2)}(2\lambda^2 -1)\Big \}^{-1} [4\lambda (1-\lambda^2) A +2\lambda (1-\lambda^2)(H+F-E-I)],\nonumber\\
B-C&=&D-G \label{G2f}
\eea 
and for $G_3^\lambda$ invariance,
\bea
C=-B,G=-D, E=I, F=H. \label{G3f}
\eea
Note that since $m_D^0$ is a general complex $3\times3$ matrix, one can simply consider the constraint equations derived 
in (\ref{G1f}), (\ref{G2f}) and (\ref{G3f}). However, since $M_R^0$ is Majorana type, one has to consider a complex 
symmetric structure for the matrix $M_G$ in (\ref{gen}) which would  require the  replacements $B=D$, $C=G$ and 
$F=H$ in the equations (\ref{G1f}), (\ref{G2f}) and (\ref{G3f}). The invariance equations 
of (\ref{res1}) are the consequences of an assumed $\mathbb{Z}_2^\lambda\times \mathbb{Z}_2^{\mu\tau}$ symmetry 
on both the left ($\nu_L$) and the right chiral ($\nu_R$) fields. It is worthwhile to highlight another interesting 
aspect of this symmetry. The overall invariance of the effective $M_\nu$ that arises from the Type-I seesaw mechanism, can be realized 
by  implementing the residual $\mathbb{Z}_2^\lambda\times \mathbb{Z}_2^{\mu\tau}$ on the left-chiral fields  only.
Since $M_\nu$ arises due to the seesaw relation in (\ref{seesaweq}), the invariance condition on $m_D$ alone
\bea
 G_i^Tm_D^0=-m_D^0\label{par}
 \eea
 implies 
\bea 
 G_i^T M_\nu G_i=M_\nu.
 \eea
For such an invariance, the determinant of $m_D^0$ would be vanishing and therefore one of the neutrinos will become massless\cite{Ge}.
Proceeding in the similar manner, as  in the previous case, we derive the following constraint equations for 
partial $G^\lambda_{123}$ invariance (cf. Eq. \ref{par}). 
For $G_1^\lambda$ invariance we have
\bea
2(1-\lambda^2) A-\lambda\sqrt{2(1-\lambda^2)} (D-G)&=&0,\nonumber\\
2(1-\lambda^2) B-\lambda\sqrt{2(1-\lambda^2)} (E-H)&=&0,\nonumber\\
2(1-\lambda^2) C-\lambda\sqrt{2(1-\lambda^2)} (F-I)&=&0.\label{G1h}
\eea 
Similarly for  $G_2^\lambda$ invariance  
\bea
2\lambda^2 A+\lambda\sqrt{2(1-\lambda^2)} (D-G)&=&0,\nonumber\\
2\lambda^2 B+\lambda\sqrt{2(1-\lambda^2)} (E-H)&=&0,\nonumber\\
2\lambda^2 C+\lambda\sqrt{2(1-\lambda^2)} (F-I)&=&0 \label{G2h}
\eea 
and for $G_3^\lambda$ invariance 
\bea
D=-G,\hspace{1mm}E=-H, \hspace{1mm}F=-I\label{G3h}~
\eea
could be obtained.
Again the minus signs in the invariance equations are used to be consistent with the PDG convention. 
Let us now switch to the analysis on TBM mixing which is a trivial generalization of the above discussion 
with $\lambda=1/\sqrt{3}$. We write the  $\mathbb{Z}_2$ generators for TBM mixing as 
\bea
G_1^{TBM}=\frac{1}{3}\begin{pmatrix}
1&-2&2\\-2&-2&-1\\2&-1&-2
\end{pmatrix},G_2^{TBM}=\frac{1}{3}\begin{pmatrix}
-1&2&-2\\2&-1&-2\\-2&-2&-1
\end{pmatrix},G_3^{\mu\tau}=
\begin{pmatrix}
-1&0&0\\0&0&1\\0&1&0
\end{pmatrix}.
\eea
Similarly, for $\lambda=1/\sqrt{3}$, the well known $U_{TBM}$ mixing simply comes out from (\ref{umut}) as
\bea
U_{TBM}=\begin{pmatrix}
\sqrt{\frac{2}{3}}&\sqrt{\frac{1}{3}}&0\\
-\sqrt{\frac{1}{6}}&\sqrt{\frac{1}{3}}&\sqrt{\frac{1}{2}}\\
\sqrt{\frac{1}{6}}&-\sqrt{\frac{1}{3}}&\sqrt{\frac{1}{2}}
\end{pmatrix}.\label{utbm}
\eea
As mentioned in the introduction, since $G_3^{\mu\tau}$  leads to a vanishing value $\theta_{13}$, the 
full $\mathbb{Z}_2^\lambda\times \mathbb{Z}_2^{\mu\tau}$ can not be a phenomenologically accepted symmetry of the Lagrangian. 
To be more precise,  the nondegenarate eigenvalue of $d_3$, i.e., $(d_3)_{33}=+1$,  fixes the third 
column of $U_\nu^{\mu\tau}$ to  $(0,1/\sqrt{2},1/\sqrt{2})^T$ which implies  a vanishing value of $\theta_{13}$ while
a nonzero value of the latter has been confirmed by the experiments at $5.2\sigma$\cite{An:2015rpe}.  
Thus to generate a nonzero $\theta_{13}$ we break the $\mathbb{Z}_2^{\mu\tau}$ ($G_3^{\mu\tau}$) with small 
breaking parameters keeping the other residual $\mathbb{Z}_2$s (either $G_{1}^{TBM}$ or $G_{2}^{TBM}$) intact. 
In the next section,  depending upon the residual symmetries on the neutrino fields and their breaking pattern, 
we  categorize our discussion into three categories.\\
\section{Breaking of $\mathbb{Z}_2^{\mu\tau}$: perturbation to the TBM mass matrices } \label{pt_tbm}
The residual TBM-Klein symmetry is implemented  in the basis where  $M_R$ is diagonal. This further leads to 
degenerate heavy RH neutrinos. We then consider the most general perturbation matrix that breaks only 
the $\mathbb{Z}_2^{\mu\tau}$ in $M_R$. 
Since these breaking parameters are responsible for generation of nonzero $\theta_{13}$, the
extent of quasidegeneracy between the right handed neutrinos (or in other words the smallness
of the breaking parameters) is now dictated by the $3\sigma$ value of $\theta_{13}$.
A systematic discussion of the breaking scheme is presented in the following subsections.
\subsection{$G_{1,2}^{TBM}$ and $G_3^{\mu\tau}$ on both the fields, $\nu_L$ and $N_R$}\label{brk_pt_tbm}
{ Case 1. {\it $G_1^{TBM}$ invariance of $M_\nu$: breaking of $G_3^{\mu\tau}$ in $M_R$ }}\\

\noindent
At the leading order, i.e., when the effective light neutrino mass matrix $M_\nu$ respects exact TBM-Klein symmetry, 
the most general Dirac mass matrix $m_D^0$ and the Majorana mass matrix $M_R^0$  satisfy the invariance equations
\bea
(G_{1}^{TBM})^T m_D^0 G_{1}^{TBM}=m_D^0, \hspace{1mm}(G_{1}^{TBM})^TM_R^0 G_{1}^{TBM}=M_R^0,\nonumber\\
(G_{3}^{\mu\tau})^T m_D^0 G_{3}^{\mu\tau}=m_D^0, \hspace{1mm}(G_{3}^{\mu\tau})^T M_R^0 G_{3}^{\mu\tau}=M_R^0.
\eea
Now using (\ref{G1f}), (\ref{G2f}) and (\ref{G3f}) one constructs the structures $m_D^0$ and $M_R^0$ as
\bea
m_D^0=\begin{pmatrix}
be^{i\beta}-ce^{i\gamma}-ae^{i\alpha}&ae^{i\alpha}&-ae^{i\alpha}\\
ae^{i\alpha}&be^{i\beta}&ce^{i\gamma}\\
-ae^{i\alpha}&ce^{i\gamma}&be^{i\beta} 
\end{pmatrix}, \hspace{1mm} M_R^0=\begin{pmatrix}
y&0&0\\
0&y&0\\
0&0&y
\end{pmatrix} \label{ttt}
\eea
where $a,b,c,y$ are real positive numbers and $\alpha,\beta,\gamma$ are phase parameters.
To generate a viable neutrino mixing we now consider breaking of $G_3^{\mu\tau}$ in the RH Majorana 
mass matrix only. We modify $M_R^0$ by adding a complex symmetric perturbation matrix $M_R^{G_1\epsilon}$ 
that breaks the $G_3^{\mu\tau}$ but invariant under the  transformation
\bea
(G_{1}^{TBM})^TM_R^{G_1\epsilon} G_{1}^{TBM}=M_R^{G_1\epsilon}
\eea 
to ensure the overall $G_{1}^{TBM}$ invariance of the effective light neutrino $M_\nu$. 
Now with $\lambda=1/\sqrt{3}$, (\ref{G1f}) would imply that a general complex symmetric matrix 
\bea
M^{GCS}_R= \begin{pmatrix}
\epsilon_1 & \epsilon_2 & \epsilon_3\\
\epsilon_2 & \epsilon_4 & \epsilon_5 \\
\epsilon_3& \epsilon_5 & \epsilon_6
\end{pmatrix}\label{gcs}
\eea 
which is invariant under $G_{1}^{TBM}$, follows the constraint equations
\bea
\epsilon_2 &=&\frac{1}{4}(3\epsilon_4+\epsilon_6)-(\epsilon_1+\epsilon_5),\nonumber\\
\epsilon_3 &=&-\frac{1}{4}(3\epsilon_6+\epsilon_4)+(\epsilon_1+\epsilon_5).
\eea 
Note that $\epsilon_1$ and $\epsilon_5$ do not break the $G_3^{\mu\tau}$ symmetry, thus for a simplified 
discussion we take both of them to be of vanishing values.  Thus the perturbation matrix $M_R^{G_1\epsilon}$ can 
be written as 
 \bea
 M_R^{G_1\epsilon} =
\begin{pmatrix}
0 & \frac{1}{4}(3\epsilon_4+\epsilon_6) & -\frac{1}{4}(3\epsilon_6+\epsilon_4)\\
\frac{1}{4}(3\epsilon_4+\epsilon_6) & \epsilon_4 & 0\\
-\frac{1}{4}(3\epsilon_6+\epsilon_4)& 0 & \epsilon_6
\end{pmatrix}.
\eea
Now the effective $M_\nu$ which is invariant under $G_1^{TBM}$ can be written as 
\bea
M_{\nu 1}^{G_{1}^{TBM}}=-m_D^0 M_R^{-1} (m_D^0)^T,\label{G1m}
\eea
where
\bea
M_R=M_R^0+M_R^{G_1\epsilon}.\label{mrg2}
\eea
Since $G_1^{TBM}$ invariance of the effective $M_\nu$ always fixes the first column of the mixing matrix to $(\sqrt{\frac{2}{3}},-\sqrt{\frac{1}{6}},\sqrt{\frac{1}{6}})^T$ up to some phases, a direct comparison of the latter with the $U_{PMNS}$ of (\ref{eu}) leads to  the well known  correlation between $\theta_{12}$ and $\theta_{13}$ for a TM1 mixing as
\bea
\sin^2 \theta_{12}=\frac{1}{3}(1-2\tan^2 \theta_{13}). \label{1strl}
\eea
To introduce CP violation  in a minimal way, it is useful to assume $\beta=\gamma$\cite{Karmakar:2014dva} in the $m_D^0$ matrix (Eq.(\ref{ttt}))
which after the phase rotation $m_D^0\rightarrow e^{-i\gamma}m_D^0$, can be conveniently parametrized as 
\begin{equation}
 m_D^0=\begin{pmatrix}
b-c-ae^{i\beta^\prime}&ae^{i\beta^\prime}&-ae^{i\beta^\prime}\\
ae^{i\beta^\prime}&b&c\\
-ae^{i\beta^\prime}&c&b \label{g1_md}
\end{pmatrix},
\end{equation}
where $\beta^\prime=\alpha-\gamma$. With the parametrization of $M_R$ of (\ref{mrg2}) as
\begin{eqnarray}
 M_R &=& M_R^0+M_R^{G_1\epsilon}=y\begin{pmatrix}
1 & \frac{1}{4}(3\epsilon_4^\prime+\epsilon_6^\prime) & -\frac{1}{4}(3\epsilon_6^\prime+\epsilon_4^\prime)\\
\frac{1}{4}(3\epsilon_4^\prime+\epsilon_6^\prime) & 1+\epsilon_4^\prime & 0\\
-\frac{1}{4}(3\epsilon_6^\prime+\epsilon_4^\prime)& 0 &1+\epsilon_6^\prime \label{tot_mr}
\end{pmatrix},
\end{eqnarray}
where $\epsilon_{4,6}^\prime(=\epsilon_{4,6}/y)$ are dimensionless breaking parameters,
we proceed further to calculate the effective light neutrino mass matrix ($M_{\nu 1}^{G_{1}^{TBM}}$). 
Eq. (\ref{G1m}) can now be simplified as
\begin{eqnarray}
M_{\nu 1}^{G_{1}^{TBM}} &=& m_D^0 M_R^{-1} {m_D^0}^T \nonumber\\
      &=& m_D^0 ~y^{-1}(M_R^\prime)^{-1} {m_D^0}^T\nonumber\\
      &=& {m_D^0}^\prime (M_R^\prime)^{-1} {{m_D^0}^\prime}^T, \label{mnu1}
\end{eqnarray}
where $M_R^\prime=(1/y)M_R$ and ${m_D^0}^\prime=(1/\sqrt{y}){m_D^0}$, i.e a factor of $(1/\sqrt{y})$ is absorbed
in the elements ${m_D^0}^\prime$ matrix however its structure is exactly identical with that of ${m_D^0}$. 
To be precise, the modulus parameters  $a^\prime$, $b^\prime$ and $c^\prime$  are basically $1/\sqrt{y}$ times 
the unprimed parameters $a$, $b$ and $c$ respectively. Thus the elements of the matrix $M_{\nu 1}^{G_{1}^{TBM}}$ now 
become functions of total six parameters, mathematically which can be represented as
\begin{equation}
(M_{\nu 1}^{G_{1}^{TBM}})_{ij}=f_{ij}(a^\prime,b^\prime,c^\prime,\epsilon_4^\prime,\epsilon_6^\prime,\beta^\prime) \label{prm}
\end{equation} 
where $i,j=1,2,3$. Explicit  forms of different $f_{ij}$s are given in the Appendix \ref{appen1}.
\\\\
{ Case 2. {\it  $G_{2}^{TBM}$ invariance of $M_\nu$: breaking of $G_3^{\mu\tau}$ in $M_R$ }}\\

\noindent
In this case we follow the similar prescription as considered in the previous case, i.e., along with the leading order 
invariance equations 
\bea
&&(G_{2}^{TBM})^T m_D^0 G_{2}^{TBM}=m_D^0, \hspace{1mm}(G_{2}^{TBM})^TM_R^0 G_{2}^{TBM}=M_R^0,\nonumber\\
&&(G_{3}^{\mu\tau})^T m_D^0 G_{3}^{\mu\tau}=m_D^0, \hspace{1mm}(G_{3}^{\mu\tau})^T M_R^0 G_{3}^{\mu\tau}=M_R^0,
\eea
we add a perturbation matrix $M_R^{G_2\epsilon}$ to $M_R^0$, where the former satisfies
\bea
(G_{2}^{TBM})^TM_R^{G_2\epsilon} G_{2}^{TBM}=M_R^{G_2\epsilon}.
\eea 
Now  (\ref{G2f}) with $\lambda=1/\sqrt{3}$ and $M_R^{GCS}$ of (\ref{gcs}) together lead to the  constraint equations
\bea
\epsilon_2 &=& \epsilon_6-\epsilon_5-\epsilon_1,\nonumber\\
\epsilon_3 &=& \epsilon_5-\epsilon_4+\epsilon_1. \label{tt2}
\eea
Using the constraint relations in (\ref{tt2})  we get $G_2^{TBM}$ invariant $M_R^{GCS}$ as
 \bea
 M_R^{G_2\epsilon} =
\begin{pmatrix}
0 & \epsilon_6 & -\epsilon_4\\
\epsilon_6 & \epsilon_4 & 0\\
-\epsilon_4& 0 & \epsilon_6
\end{pmatrix},
\eea
where $\epsilon_{1,5}$ are assumed to have vanishing values due to their blindness towards $\mu\tau$ interchange symmetry. 
The effective $ M_\nu $ which is now invariant under $G_2^{TBM}$ comes out as
\bea
M_{\nu 1}^{G_2^{TBM}}=-m_D^0 M_R^{-1} (m_D^0)^T, \label{G2m}
\eea
where 
\bea
M_R=M_R^0+M_R^{G_2\epsilon}.
\eea
Note that $G_2^{TBM}$ invariance of $M_\nu$ always fixes the second column of the mixing matrix to the second column 
of $U^{TBM}$. This leads to the constraint relation between $\theta_{12}$ and $\theta_{13}$ as 
\bea
\sin^2 \theta_{12}=\frac{1}{3}(1+\tan^2 \theta_{13}). \label{2ndrl}
\eea
Given a nonvanishing value of $\theta_{13}$, the solar mixing angle $\theta_{12}$ is always greater 
than $\sin^{-1}(1/\sqrt{3})$ which is disfavored at 3$\sigma$ by the the present oscillation data\cite{Capozzi:2017ipn}.
Therefore, we do not consider this case in our numerical discussion.
\subsection{$G_{1,2}^{TBM}$ on $\nu_L$ and $G_3^{\mu\tau}$ on both the fields, $\nu_L$ and $N_R$}\label{brk_pt_tbm1}
{ Case 1. {\it $G_1^{TBM}$ invariance of $M_\nu$: breaking of $G_3^{\mu\tau}$ in $M_R$ }}\\

\noindent
 In this case, for the effective $M_\nu$ to be invariant under $G_1^{TBM}$ and $G_3^{\mu\tau}$ at the leading order, the 
 constituent mass matrices follow the invariance equations given by
\bea
(G_{1}^{TBM})^T m_D^0 =-m_D^0, \hspace{1mm}(G_{3}^{\mu\tau})^T m_D^0 G_{3}^{\mu\tau}=m_D^0, \hspace{1mm}(G_{3}^{\mu\tau})^T M_R^0 G_{3}^{\mu\tau}=M_R^0.\label{G1m0}
\eea
Now using (\ref{G1h}) and (\ref{G3f}), we find  forms of the most general $m_D^0$ and $M_R^0$ that satisfy (\ref{G1m0}):
\bea
m_D^0=\begin{pmatrix}
ae^{i\alpha} &\frac{1}{2}(be^{i\beta}-ce^{i\gamma})&\frac{1}{2}(ce^{i\gamma}-be^{i\beta})\\
ae^{i\alpha}&be^{i\beta}&ce^{i\gamma}\\
-ae^{i\alpha}&ce^{i\gamma}&be^{i\beta}
\end{pmatrix}, \hspace{2mm} M_R^0=\begin{pmatrix}
x&0&0\\
0&y&0\\
0&0&y
\end{pmatrix}. \label{mat1}
\eea
For the sake of simplicity, we assume $\beta=\gamma$ and take out the phase $\alpha$ through the rotation $m_D^0\rightarrow e^{-i\alpha}m_D^0$.
Thus $m_D^0$ takes the form
\begin{equation}
 \begin{pmatrix}
a &\frac{1}{2}(b-c)e^{i\theta/2}&\frac{1}{2}(c-b)e^{i\theta/2}\\
a&be^{i\theta/2}&ce^{i\theta/2}\\
-a&ce^{i\theta/2}&be^{i\theta/2} \label{t1}
\end{pmatrix},
\end{equation}
where $\frac{\theta}{2}=\beta-\alpha$ and $a$, $b$, $c$ are all real positive parameters.
An arbitrary $G_3^{\mu\tau}$ breaking perturbation matrix could be added to $M_R^0$, 
since the overall $G_1^{TBM}$ invariance of the effective $M_\nu$  is independent of the form of the RH Majorana mass matrix. 
We choose the perturbation matrix to be 
\bea
M_R^{\epsilon}=\begin{pmatrix}
0&0&0\\
0&\epsilon_4 & 0\\
0&0&\epsilon_6
\end{pmatrix}. \label{dgpt}
\eea
Thus the effective $M_\nu$ is calculated (using phase rotated $m_D^0$ of eq.(\ref{t1}) and broken symmetric $M_R(=M_R^0+M_R^{\epsilon})$) as 
\bea
M_{\nu 2}^{G_1^{TBM}}=-m_D^0 M_R^{-1} (m_D^0)^T .\label{G2m}
\eea
Using the redefinition of the parameters as 
\bea
\frac{a}{\sqrt{x}}\rightarrow p,~\frac{b}{\sqrt{y}}\rightarrow q,~\frac{c}{\sqrt{y}}\rightarrow r,\epsilon_4^\prime\rightarrow\frac{\epsilon_4}{y},
\epsilon_6^\prime\rightarrow\frac{\epsilon_6}{y}
\eea
(with $p$, $q$, $r$,$\epsilon_4$, $\epsilon_6$ being real) the elements of $M_{\nu 2}^{G_1^{TBM}}$ matrix can be expressed as functions of $p,q,r,\theta,\epsilon_4^\prime,\epsilon_6^\prime$.
Explicit functional forms for the elements matrix $M_{\nu 2}^{G_1^{TBM}}$ can be found in Appendix \ref{appen2}.\\

In this case also, due to the $G_1^{TBM}$ invariance of $M_\nu$, the relation between $\theta_{12}$ and $\theta_{13}$ 
is same as that of (\ref{1strl}).  Another interesting point is that $m_D^0$ of (\ref{mat1}) is of determinant zero due 
to the imposed $G_1^{TBM}$ symmetry. Thus the matrix $M_{\nu 2}^{G_1^{TBM}}$  has one vanishing eigenvalue. Since 
$G_1^{TBM}$ also fixes the first column of the mixing matrix, the vanishing eigenvalue has to be $m_1$ which is 
allowed by the current oscillation data. \\

\noindent
{ Case 2. {\it $G_2^{TBM}$ invariance of $M_\nu$: breaking of $G_3^{\mu\tau}$ in $M_R$ }}\\

\noindent
The effective $M_\nu$ to be invariant under $G_2^{TBM}$ and $G_3^{\mu\tau}$ at the leading order, 
the constituent mass matrices follow the invariance equations
\bea
(G_{2}^{TBM})^T m_D^0 =-m_D^0, \hspace{1mm}(G_{3}^{\mu\tau})^T m_D^0, G_{3}^{\mu\tau}=m_D^0, \hspace{1mm}(G_{3}^{\mu\tau})^T M_R^0 G_{3}^{\mu\tau}=M_R^0.\label{G2m0}
\eea
The most general Dirac mass matrix $m_D^0$ and the Majorana mass matrix $M_R^0$ that satisfy (\ref{G2m0}) are of the forms
\bea
m_D^0=\begin{pmatrix}
-2a&-b+c&-c+b\\
a&b&c\\
-a&c&b
\end{pmatrix}, \hspace{2mm} M_R^0=\begin{pmatrix}
x&0&0\\
0&y&0\\
0&0&y
\end{pmatrix}, \label{mat2}
\eea
where we have used (\ref{G2h}) and (\ref{G3f}) to find these forms.
Similar to the previous case the effective $M_\nu $ can be calculated as 
\bea
M_{\nu 2}^{G_2^{TBM}}=-m_D^0 M_R^{-1} (m_D^0)^T, \label{G2m}
\eea
where
\bea
M_R=M_R^0+M_R^{\epsilon}
\eea
with $M_R^{\epsilon}$ being an arbitrary perturbation matrix. Here also  due to the imposed $G_2^{TBM}$ symmetry, the matrix $m_D^0$ of (\ref{mat2}) has zero 
determinant  which imply the $M_{\nu 2}^{G_2^{TBM}}$ matrix has one vanishing 
eigenvalue. Since $G_2^{TBM}$ fixes the second column of the mixing matrix, the vanishing eigenvalue has to be $m_2$ 
which is not allowed due to a positive definite value of the solar mass squared difference  ($\Delta m_{21}^2=m_2^2-m_1^2$) for both normal and inverted hierarchy. 
Therefore we discard this case in our analysis. \\
\subsection{$G_{1,2}^{TBM}$ and $G_3^{\mu\tau}$ on  $\nu_L$ only}
In this case the leading order transformations are
\bea
(G_{1}^{TBM})^T m_D^0 =-m_D^0, \hspace{1mm}(G_{3}^{\mu\tau})^T m_D^0 =-m_D^0, \label{G1m00}
\eea
\bea
(G_{1}^{TBM})^T m_D^0 =-m_D^0, \hspace{1mm}(G_{3}^{\mu\tau})^T m_D^0 =-m_D^0. \label{G2m00}
\eea
The most general effective $M_\nu$ for both the cases lead to two vanishing eigenvalues. 
Due to this degeneracy in masses, one can not fix the leading order mixing as the TBM mixing matrix, 
thus the residual symmetry approach breaks down (Due to the arbitrariness of the mixing matrix one cannot 
reconstruct the corresponding $\mathbb{Z}_2$ generators; the $G_i$ matrices). Therefore both of these cases 
are  discarded in our analysis.
\section{Baryogenesis through leptogenesis}\label{sc4}
Baryogenesis via leptogenesis is an excellent mechanism to understand the observed excess of 
baryonic matter over anti matter. The amount of baryon asymmetry is expressed by the parameter: ratio of
difference in number densities of baryons $(n_B)$ and anti baryons $(n_{\bar{B}})$ to the entropy density of the universe.
The experimentally observed value\cite{Ade:2013zuv,Ade:2015xua} of this baryon asymmetry parameter $(Y_B)$ is given by
\bea
Y_B=(n_B-n_{\bar{B}})/s=8.55\times 10^{-11}<Y_B<8.77\times 10^{-11}
\eea
with $s$ being the entropy density of the universe. In this mechanism, the CP violating and out of equilibrium 
decays of heavy RH neutrinos\cite{Fukugita:1986hr} create an excess lepton asymmetry which is further converted 
in to baryon asymmetry by nonperturbative  sphalerons\cite{Kolb:1990vq}. 
\subsection{Calculation of CP asymmetry parameter} \label{s3}
The part of our  Lagrangian relevant to the generation of a CP asymmetry is
\bea
 -\mathcal{L}=f_{i\alpha}^N \overline{\slashed{L}_{L\alpha}}\tilde{\phi} N_{Ri}+\frac{1}{2}\overline{{N_{Ri}}^C}(M_R)_{ij}N_{Rj}+{\rm h.c.},\label{cpa0}
 \eea
where $\slashed{L}_\alpha =(\nu_{L_\alpha}~\ell^{-}_{L\alpha})^T$ is the left-chiral SM lepton doublet of flavour $\alpha$, while   $\tilde{\phi}=(\phi^{0*}~-\phi^{-})^T$ is the charge conjugated Higgs scaler
doublet. It is evident from  (\ref{cpa0}) that the decay products of $N_i$ can be $\ell_\alpha^-\phi^+,\nu_\alpha\phi^0,\ell_\alpha^+\phi^-$ and $\nu_\alpha^C\phi^{0*}$. 
We are interested in the flavour dependent CP asymmetry parameter $\varepsilon^\alpha_i $ which is given by 
\begin{eqnarray}
 \varepsilon^\alpha_i &=&\frac{\Gamma({N}_i\rightarrow
    \slashed{L}_\alpha \phi)-\Gamma({N}_i\rightarrow
   \slashed{L}^C_\alpha \phi^{\dagger})}{\Gamma({N}_i\rightarrow
    \slashed{L} \phi)+\Gamma({N}_i\rightarrow
   \slashed{L}^C \phi^{\dagger})},\label{cpa}
\end{eqnarray}
where $\Gamma$ is  the corresponding  partial decay width and in the denominator a sum over the flavour index $\alpha$ has been considered. 
  A nonvanishing value of $\varepsilon^\alpha_i$ requires the interference between the tree level 
 and one loop decay contributions of $N_i$, since the tree level decay width is given by
 \bea
 \Gamma^{tree}(N_i\rightarrow \slashed{L}_\alpha \phi)=\Gamma^{tree}(N_i\rightarrow \slashed{L}^C_\alpha \phi^{\dagger})
 \eea
 and thus leads to a vanishing CP violation.
Before presenting the rigorous formulas of partial decay width and the CP asymmetry parameter, let us point out a subtle issue. Since the computations related to leptogenesis require the physical masses of the RH neutrinos, a nondiagonal RH neutrino mass matrix should be rotated to its diagonal basis.
A  Majorana type RH neutrino mass matrix $M_R$ could be put into diagonal form with a unitary matrix $V$ as
\begin{equation}
V^\dagger M_R V^\ast= {\rm diag}~(M_1,~M_2,~M_3), 
\end{equation}
where $M_i(i=1,2,3)$ are the  eigenvalues of $M_R$. Thus in the diagonal basis of $M_R$, the Dirac neutrino mass matrix (the neutrino Yukawa couplings) also
gets rotated as 
\begin{equation}
{f^N}^\prime=f^N V^\ast ~\Rightarrow~{m_D}^\prime=m_D V^\ast, 
\end{equation}
where $m_D$ is the Dirac neutrino mass matrix in the nondiagonal basis of $M_R$ and is given by $m_D=\frac{f^N v}{\sqrt{2}}$ with $v$ being the VEV of the SM Higgs. Accordingly, the tree 
level decay width can now be calculated as
\begin{equation}
 \Gamma^{tree}(N_i\rightarrow \slashed{L}_\alpha \phi)=\Gamma^{tree}(N_i\rightarrow \slashed{L}^C_\alpha \phi^{\dagger})=
 (16\pi)^{-1}({f_{i\alpha}^N}^{\prime\dagger}{f_{i\alpha}^N}^\prime)M_i.\label{tree}
\end{equation}
Along with (\ref{tree}), taking into account the contributions from one loop vertex and self energy diagrams  and without assuming any hierarchy of the right handed neutrinos the most general expression (keeping upto fourth order of Yukawa coupling) of the flavour dependent CP asymmetry parameter \cite{Pilaftsis:2003gt} 
can be calculated as
\bea
\varepsilon^\alpha_i
&=&\frac{1}{4\pi v^2 \mathcal{ H}^\prime_{ii}}\sum_{j\ne i} {\rm Im}\{\mathcal{ H}^\prime_{ij}({m_D^\prime}^\dagger)_{i\alpha} (m_D^\prime)_{\alpha j}\}
\left[f(x_{ij})+\frac{\sqrt{x_{ij}}(1-x_{ij})}
{(1-x_{ij})^2+\frac{{\mathcal{H}^\prime}_{jj}^2}{16 \pi^2 v^4}}\right]\nonumber\\
&+&\frac{1}{4\pi v^2 {\mathcal{H}^\prime}_{ii}}\sum_{j\ne i}\frac{(1-x_{ij}){\rm Im}\{{\mathcal{ H}^\prime}_{ji}({m_D^\prime}^\dagger)_{i\alpha} (m_D^\prime)_{\alpha j}\}}
{(1-x_{ij})^2+\frac{{\mathcal{H}^\prime}_{jj}^2}{16 \pi^2 v^4}},
\label{epsi_intro_h}
\eea
where $\mathcal{H}^\prime={m_D^\prime}^\dagger m_D^\prime$
\footnote{It is to be noted that if the right handed neutrino mass matrix is taken to be diagonal then
$V$ is a unit matrix and we would have ${m_D}^\prime=m_D$ and $\mathcal{H}^\prime=H$, where $\mathcal{H}={m_D}^\dagger m_D$.}, 
$x_{ij}=\frac{M_j^2}{M_i^2}$ and $f(x_{ij})$ is the loop function given by
\begin{equation}
f(x_{ij})=\sqrt{x_{ij}}\{1-(1+x_{ij})\ln(\frac{1+x_{ij}}{x_{ij}})\}.
\end{equation}
In the expression of $\varepsilon^\alpha_i$, the term proportional to $f(x_{ij})$ arises from the one loop vertex term 
interfering with the tree level contribution. The rest are originating from  interference of  the one loop self energy 
diagram with the tree level term. It is also worth clarifying the reason behind the explicit flavour index `$\alpha$' on 
the CP asymmetry parameter $\varepsilon_i$ in (\ref{epsi_intro_h}). Depending upon the temperature regime in which 
leptogenesis occurs, lepton flavours may be fully distinguishable, partly distinguishable or 
indistinguishable\cite{Abada:2006ea}. Assuming leptogenesis takes place at $T\sim M_{i}$, lepton flavours cannot be
treated separately if the concerned  process occurs above a temperature $T\sim M_{i}> 10^{12}~{\rm GeV}$. 
If the said temperature is lower, two possibilities might arise.  When $T\sim M_{i}< 10^9$ GeV, all three ($e,\mu,\tau$) 
flavours are individually active and we need three CP asymmetry parameters $\varepsilon^e_i,\varepsilon^\mu_i,\varepsilon^\tau_i$ 
for each generation of RH neutrinos. On the other hand when we have $10^9~{\rm GeV}<T\sim M_{i}< 10^{12}~{\rm GeV}$, only 
the $\tau$-flavour can be identified  while the $e$ and $\mu$ act indistinguishably. Here we need two CP asymmetry
parameters $\varepsilon^{(2)}_i=\varepsilon^e_i+\varepsilon^\mu_i$ and
$\varepsilon^\tau_i$ for each  of the RH neutrinos. As an aside, let us point out a simplification  for  
unflavoured leptogenesis which is relevant for the
regime $T\sim M_{i}> 10^{12}~{\rm GeV}$. Summing over all $\alpha$,
\begin{equation}
\sum\limits_{\alpha} {\rm Im} \{\hspace{1mm}\mathcal{H}_{ji}(m_D)_{i \alpha }({m_D}^*)_{j \alpha }\}
={\rm Im} \{\hspace{1mm}{\mathcal{H}}_{ji} {\mathcal{H}}_{ij}\}={\rm Im} \{\hspace{1mm}{\mathcal{H}}_{ji} {{\mathcal{H}}^*_{ji}}\}={\rm Im} \hspace{1mm}|{\mathcal{H}}_{ji}|^2=0,
\end{equation}
i.e. the second term in the RHS of (\ref{epsi_intro_h}) vanishes. The flavour-summed CP asymmetry parameter is 
therefore given by the simplified expression
\begin{eqnarray}
\varepsilon_i &=& \sum\limits_{\alpha}\varepsilon^\alpha_i \nonumber\\
              &=& 
\frac{1}{4\pi v^2 \mathcal{H}^\prime_{ii}}\sum_{j\ne i} {\rm Im}\{{\mathcal{H}^\prime_{ij}}^2\}
\left[f(x_{ij})+\frac{\sqrt{x_{ij}}(1-x_{ij})}
{(1-x_{ij})^2+\frac{{\mathcal{H}^\prime}_{jj}^2}{16 \pi^2 v^4}}\right].
\label{sum_epsi_intro_h2}
\end{eqnarray}
\vspace{5mm}
\paragraph{}
It is evident from the thorough discussion of different types of $\mathbb{Z}_2^{\mu\tau}$ breaking schemes 
presented in Sec.\ref{pt_tbm}, that only two of those may be compatible with the constraints of the neutrino
oscillation data. Therefore while performing the computations of leptogenesis, we should take into account only
those symmetry breaking patterns which at least have the potential to satisfy oscillation data and in present work those two theoretically relevant options are Case I of both Sec.\ref{brk_pt_tbm} and Sec.\ref{brk_pt_tbm1}.
For the Case I in  Sec.\ref{brk_pt_tbm}, the RH neutrino mass matrix is  diagonal in the  TBM limit. 
However, for the realistic scenario, i.e., in the broken TBM frame work, $M_R$ is off diagonal. 
On the other hand, for the Case I in Sec.\ref{brk_pt_tbm1}, $M_R$ is always diagonal. For the time being, let us leave the latter case for the numerical section and discuss here the former, i.e., Case I belonging to Sec.\ref{brk_pt_tbm}. Since right handed neutrino mass matrix is nondiagonal in this case, at first we have to  diagonalize the $M_R$ matrix and find out the corresponding diagonalizing matrix $V$ which would be used thereafter to rotate the Dirac matrix $m_D$.
After a straight forward diagonalization of the $M_R$ matrix in (\ref{tot_mr}), the eigenvalues come out to be
\bea
&&M_1=\frac{y}{2} \left(-\epsilon _4'-\epsilon _6'+2\right), \nonumber\\
&&M_2=\frac{y}{4} \left(3 \epsilon _4'+3 \epsilon _6'-\sqrt{7 \left(\epsilon _4'\right){}^2-10 \epsilon _6' \epsilon _4'+7 \left(\epsilon _6'\right){}^2}+4\right), \nonumber\\
&&M_3=\frac{y}{4} \left(3 \epsilon _4'+3 \epsilon _6'+\sqrt{7 \left(\epsilon _4'\right){}^2-10 \epsilon _6' \epsilon _4'+7 \left(\epsilon _6'\right){}^2}+4\right).\label{Rhe}
\eea
It is clear from (\ref{Rhe}) that the RH neutrino masses are very close to each other, separated only by the breaking parameters. This opens up a possibility of resonant enhancement of the CP asymmetry parameter which may yield the required value of baryon
asymmetry $Y_B$ at a very low mass scale. We have checked the condition for resonance\footnote{resonance in CP asymmetry 
parameters is achieved when $1-x_{ij} \simeq \frac{\mathcal{H}^\prime_{jj}}{4\pi v^2}$} very carefully and found that even for the lowest allowed (by oscillation data) value 
of the breaking parameters it is not possible the meet the resonant condition. This is since, given the neutrino oscillation data, the resonance condition in our model can be translated approximately to 
\bea
\Delta\sim 10^{-15} ~M~ {\rm GeV}^{-1},\label{reso}
\eea
where $\Delta\sim (M_j-M_i)/M_i$ and $M$ is the mass scale of the RH neutrino while unperturbed. If one wants to achieve resonance, say at $M\sim 10^{6}$ GeV, clearly, the scenario is inconsistent since in that case $\Delta\sim10^{-9}$ while as we shall see in the numerical section,  the LHS of (\ref{reso}) is $\sim 10^{-1}$. Interestingly, in this way, we could also circumvent the effect of RH neutrino flavour oscillation\cite{Dev:2014laa} where the same resonance condition leads to an additional CP asymmetry produced by the RH neutrino flavour oscillation\cite{Dev:2017wwc}. Nevertheless, since the RH neutrino masses are close to each other, we can not treat this scenario to be hierarchical where the asymmetries generated from the RH neutrinos of higher masses can be safely neglected\cite{Buchmuller:2003gz}.
Therefore we opt for the rigorous method of quasidegenerate leptogenesis where the contribution from all three right handed neutrinos are taken into account\cite{Pilaftsis:2003gt} and show how the produced asymmetry from each RH neutrino is affected by the other.
\paragraph{}
Since $M_R$ is a real matrix, the diagonalization matrix $V$ will also be a real. Thus $V$ is now an orthogonal matrix with its different elements in terms of the breaking parameters as
\bea
V_{11}&=&\sqrt{\frac{2}{3}},\nonumber\\
V_{12}&=&\frac{\sqrt{\frac{1}{2}} \left(-3 \epsilon _4'+\epsilon _6'+ \epsilon^{\prime\prime}\right)}{ \sqrt{35 \left(\epsilon _4'\right){}^2-\left(50 \epsilon _6'+
13  \epsilon^{\prime\prime}\right)
\epsilon _4'+\epsilon _6' \left(35 \epsilon _6'+11  \epsilon^{\prime\prime}\right)}},\nonumber\\\nonumber\\
V_{13}&=&\frac{-\frac{3}{\sqrt{2}} \epsilon _4'+\frac{1}{\sqrt{2}} \epsilon _6'-\frac{1}{\sqrt{2}}  \epsilon^{\prime\prime}}{ \sqrt{35 \left(\epsilon _4'\right){}^2+
\left(13  \epsilon^{\prime\prime}-50 \epsilon _6'\right)
\epsilon _4'+\epsilon _6' \left(35 \epsilon _6'-11 \epsilon^{\prime\prime}\right)}},\nonumber\\\nonumber\\
V_{21}&=&-\frac{1}{\sqrt{6}},\nonumber\\\nonumber\\
V_{22}&=&\frac{-\frac{5}{\sqrt{2}} \epsilon _4'+\frac{5}{\sqrt{2}} \epsilon _6'+\sqrt{2}  \epsilon^{\prime\prime}}{ \sqrt{35 \left(\epsilon _4'\right){}^2-
\left(50 \epsilon _6'+13  \epsilon^{\prime\prime}\right)
\epsilon _4'+\epsilon _6' \left(35 \epsilon _6'+11  \epsilon^{\prime\prime}\right)}},\nonumber\\\nonumber\\
V_{23}&=&\frac{-\frac{5}{\sqrt{2}} \epsilon _4'+\frac{5}{\sqrt{2}} \epsilon _6'-\sqrt{2}  \epsilon^{\prime\prime}}{ \sqrt{35 \left(\epsilon _4'\right){}^2+
\left(13  \epsilon^{\prime\prime}-50 \epsilon _6'\right)
\epsilon _4'+\epsilon _6' \left(35 \epsilon _6'-11  \epsilon^{\prime\prime}\right)}},\nonumber\\\nonumber\\
V_{31}&=&\frac{1}{\sqrt{6}},\nonumber\\\nonumber\\
V_{32}&=&\frac{\frac{1}{\sqrt{2}}\left(\epsilon _4'+3 \epsilon _6'\right)}{\sqrt{35 \left(\epsilon _4'\right){}^2-\left(50 \epsilon _6'+13  \epsilon^{\prime\prime}\right) \epsilon _4'+\epsilon _6' \left(35 \epsilon _6'+11  \epsilon^{\prime\prime}\right)}},\nonumber\\\nonumber\\
V_{33}&=&  \frac{\frac{1}{\sqrt{2}}\left(\epsilon _4'+3 \epsilon _6'\right)}{\sqrt{35 \left(\epsilon _4'\right){}^2+\left(13  \epsilon^{\prime\prime}-50 \epsilon
   _6'\right) \epsilon _4'+\epsilon _6' \left(35 \epsilon _6'-11  \epsilon^{\prime\prime}\right)}},\nonumber\\
\eea
where
\bea
 \epsilon^{\prime\prime}= \sqrt{7 \left(\epsilon _4'\right){}^2-10 \epsilon _6' 
   \epsilon _4'+7 \left(\epsilon _6'\right){}^2}.
 \eea
\paragraph{Unflavoured CP asymmetry parameter:} The expression of unflavoured CP asymmetry parameter in (\ref{sum_epsi_intro_h2}) involves the matrix ${\mathcal{H}^\prime}$ which can further be written as
\bea
{\mathcal{H}^\prime}={{m_D}^\prime}^\dagger {m_D}^\prime
                    =V^T {m_D}^\dagger {m_D} V^\ast 
                    = V^T {m_D}^\dagger {m_D} V.
\eea
Taking into account the explicit form of  $m^0_D$  given in (\ref{g1_md}),  different elements of the ${m^0_D}^\dagger {m^0_D}$  can be written as
\bea
&& ({m^0_D}^\dagger {m^0_D})_{11}=3 a^2 +(b-c)^2-2\cos (\beta^\prime) a(b-c),\nonumber\\
&& ({m^0_D}^\dagger {m^0_D})_{12}=2\cos (\beta^\prime) a(b-c)-a^2, \nonumber\\
&& ({m^0_D}^\dagger {m^0_D})_{13}=2\cos (\beta^\prime) a(b-c)+a^2, \nonumber\\
&& ({m^0_D}^\dagger {m^0_D})_{22}=a^2+b^2+c^2,\nonumber\\
&& ({m^0_D}^\dagger {m^0_D})_{23}=a^2+2 b c, \nonumber\\
&& ({m^0_D}^\dagger {m^0_D})_{33}=a^2+b^2+c^2.
\eea
It is clear from the above set of equations that ${m^0_D}^\dagger {m^0_D}$ is completely real matrix which in turn dictates 
that ${\mathcal{H}^\prime}$ is also a real matrix due to the real nature of the matrix  $V$. Since the CP asymmetry parameter in (\ref{sum_epsi_intro_h2}) is proportional to ${\rm Im}\{{\mathcal{H}^\prime}_{ij}^2\}$, it can easily be inferred that  generation of lepton asymmetry is not at all possible in the unflavoured regime. Now our task is to examine whether we can have nonvanishing values of flavour dependent CP asymmetry parameters
so that we can get generate baryon asymmetry in the fully flavoured or partly flavoured regime.

\paragraph{Flavour dependent CP asymmetry parameters:} The flavoured CP asymmetry parameters of (\ref{epsi_intro_h}) can be represented in a little bit simpler form as
\begin{eqnarray}
 \varepsilon^\alpha_i
&=&\frac{1}{4\pi v^2 \mathcal{ H}^\prime_{ii}}\sum_{j\ne i} {\rm Im}\{\mathcal{ H}^\prime_{ij}({m_D^\prime}^\dagger)_{i\alpha} (m_D^\prime)_{\alpha j}\}
g(x_{ij})\nonumber\\
&+&\frac{1}{4\pi v^2 {\mathcal{H}^\prime}_{ii}}\sum_{j\ne i}{\rm Im}\{\mathcal{ H}^\prime_{ji}({m_D^\prime}^\dagger)_{i\alpha} (m_D^\prime)_{\alpha j}\}g^\prime(x_{ij}),\label{flav_cp}
\end{eqnarray}
where $g(x_{ij})=\left[f(x_{ij})+\frac{\sqrt{x_{ij}}(1-x_{ij})}{(1-x_{ij})^2+\frac{{\mathcal{H}^\prime}_{jj}^2}{16 \pi^2 v^4}}\right]$ and
$g^\prime(x_{ij})=\frac{(1-x_{ij})}{(1-x_{ij})^2+\frac{{\mathcal{H}^\prime}_{jj}^2}{16 \pi^2 v^4}}$.
The $\varepsilon_i^\alpha$ in (\ref{flav_cp})  can  further be simplified as
\begin{equation}
 \varepsilon^\alpha_i=\frac{1}{4\pi v^2 \mathcal{ H}^\prime_{ii}}\sum_{j\ne i} {\rm Im}\{({m_D^\prime}^\dagger)_{i\alpha} (m_D^\prime)_{\alpha j}\}
 \left[\mathcal{ H}^\prime_{ij} g(x_{ij}) +\mathcal{ H}^\prime_{ji} g^\prime (x_{ij}) \right]. 
\end{equation}
Although $\mathcal{H}^\prime$ matrix is a real matrix, ${m^0_D}^\prime$ matrix contains  complex parameters. Since  $\varepsilon^\alpha_i$ contains individual elements of $m_D$ matrix, we would have a nonzero imaginary part which leads to a nonvanishing flavoured CP asymmetry parameters and thus nonzero $Y_B$. The full functional forms of these flavoured CP asymmetry
parameters ($\varepsilon^\alpha_i,\alpha=e,\mu,\tau~{\rm and~}i=1,2,3$) in terms of parameters of $m^0_D$ matrix and the symmetry breaking parameters
would be too cumbersome to present here. We calculate all nine of them and use in the  Boltzmann equations suitably. Nevertheless, to realize the significance of the phase parameter $(\beta^\prime)$, it is useful to simplify further the  expression of $\varepsilon^\alpha_i$. For this, let us focus on the first term (the second term would be treated in the same manner) in the RHS of (\ref{flav_cp})
\begin{equation}
(\varepsilon^\alpha_i)_1
=\frac{1}{4\pi v^2 \mathcal{ H}^\prime_{ii}}\sum_{j\ne i} {\rm Im}\{\mathcal{ H}^\prime_{ij}({m_D^\prime}^\dagger)_{i\alpha} (m_D^\prime)_{\alpha j}\}
g(x_{ij}). 
\end{equation}
It is to be noted that the phase parameter $\beta^\prime$ is contained in the unprimed matrices ($m_D ~{\rm and}~\mathcal{H}$).
The diagonalization matrix $V$ doesn't depend on $\beta^\prime$. Now $\mathcal{H}$ matrix possesses the phase in its off-diagonal
elements in form of $\cos \beta^\prime$.
Therefore inversion of sign of $\beta^\prime$ will not affect
sign of $\mathcal{ H}^\prime_{ij}$. A closer inspection of the elements of the $m_D$ matrix reveals that 
${\rm Im}\{({m_D^\prime}^\dagger)_{i\alpha} (m_D^\prime)_{\alpha j}\}$ (for $i \neq j$) is
bound to be function of $\sin \beta^\prime$ which  will appear in the expression of CP asymmetry parameters as an overall multiplicative factor. Thus we can say that phase dependence of the flavoured CP asymmetry parameters is of the form : 
$\varepsilon^\alpha_i \sim \sin \beta^\prime f(\cos\beta^\prime)$. So $\varepsilon^\alpha_i$ is an odd function of the phase $\beta^\prime$, i.e
$\varepsilon^\alpha_i(-\beta)=-\varepsilon^\alpha_i(\beta)$.

\subsection{Boltzmann equations and baryon asymmetry in different mass regimes}\label{s4}
For  an  evolution down to the electroweak scale,  one  needs  to
solve the corresponding Boltzmann Equations (BEs) for the number density $n_a$ of a particle type `a' (in our context, 
either a right-chiral heavy neutrino $N_i$ or a left-chiral lepton doublet $\slashed{L}$). For this purpose it is 
convenient to define $\eta_{a}(z)=n_a(z)/n_\gamma (z)$ with $z=M_i/T$, $n_\gamma (z)=2M_i^3/\pi^2z^3$. 
We  follow here the treatment given in Ref.\cite{Pilaftsis:2003gt}. The equations involve decay transitions 
between $N_i$ and $\slashed{L}_\alpha \phi$ as well as $\slashed{L}_\alpha^C \phi^\dagger$ plus scattering 
transitions  $Q u^C\leftrightarrow N_i\slashed{L}_\alpha,\slashed{L}_\alpha Q^C\leftrightarrow N_i u^C,\slashed{L}_\alpha u \leftrightarrow N_i Q,\slashed{L}_\alpha \phi \leftrightarrow N_i V_\mu, \phi^\dagger V_\mu\leftrightarrow N_i\slashed{L}_\alpha,\slashed{L}_\alpha V_\mu \leftrightarrow N_i \phi^\dagger $. Here $Q$ represents the left-chiral quark doublet with $Q^T=(u_L\hspace{2mm}d_L)$ and $V_\mu$  stands for either $B$ or $W_{1,2,3}$. The number density of a particle of species $a$ and mass $m_a$ with $g_a$ internal degrees of freedom is given by\cite{Edsjo:1997bg} 
\bea
n_a (T)=  \frac{g_a\, m^2_a\,T\ e^{\mu_a (T)/T}}{2\pi^2}\
K_2\bigg(\frac{m_a}{T}\bigg)\;,
\eea
 $K_2$ being the modified Bessel function of the second kind with order 2. The corresponding equilibrium density is given by
\bea
n^{\rm eq}_a (T)=  \frac{g_a\, m^2_a\,T\ }{2\pi^2}\
K_2\bigg(\frac{m_a}{T}\bigg).
\eea
Stage is now set up for the usage of the Boltzmann evolution equations given in Ref.\cite{Pilaftsis:2003gt} -- generalized
with flavour\cite{Adhikary:2014qba}. In making this generalization, one comes across a subtlety: the active flavour in the 
mass regime (given by the  leptogenesis scale $T\sim  M_i$) under consideration may not be individually $e$, $\mu$ or $\tau$
but some combination thereof. So instead of $\alpha$ we use a general active flavour index $\lambda$ for the lepton asymmetry. 
Now we write the relevant BEs as
\begin{eqnarray}
  \label{BEN} 
\frac{d \eta_{N_i}}{dz} &=& \frac{z}{H(z=1)}\ \bigg[\,\bigg( 1
\: -\: \frac{\eta_{N_i}}{\eta^{\rm eq}_{N_i}}\,\bigg)\,\sum\limits_{\beta=e,\mu,\tau} \bigg(\,
\Gamma^{\beta Di} \: +\: \Gamma^{\beta Si}_{\rm Yukawa}\: +\:
\Gamma^{\beta Si}_{\rm Gauge}\, \bigg)\nonumber\\ &&-\frac{1}{4}\sum\limits_{\beta=e,\mu,\tau}\eta_L^\beta\varepsilon^\beta_i \bigg(\,
\Gamma^{\beta Di} \: +\: \tilde{\Gamma}^{\beta Si}_{\rm Yukawa}\: +\:
\tilde{\Gamma}^{\beta Si}_{\rm Gauge}\ \bigg) \bigg],\nonumber\\
\frac{d \eta^\lambda_L}{dz} &=& -\, \frac{z}{H(z=1)}\, \bigg [\,
\sum\limits_{i=1}^3\,\varepsilon^\lambda_i \ 
\bigg( 1 \: -\: \frac{\eta_{N_i}}{\eta^{\rm eq}_{N_i}}\,\bigg)\,\sum\limits_{\beta=e,\mu,\tau} \bigg(\,
\Gamma^{\beta Di} \: +\: \Gamma^{\beta S i}_{\rm Yukawa}\: +\:
\Gamma^{\beta S i}_{\rm Gauge}\, \bigg) \nonumber\\ 
&&+\, \frac{1}{4}\, \eta^\lambda_L\, \bigg \{\, \sum\limits_{i=1}^3\, 
\bigg(\, \Gamma^{\lambda D i} \: +\: 
\Gamma^{\lambda Wi}_{\rm Yukawa}\: 
+\: \Gamma^{\lambda Wi}_{\rm Gauge}\,\bigg)\: +\:
\Gamma^{\lambda \Delta L =2}_{{\rm Yukawa}} \bigg \}\,\bigg ]\,.\label{BEL}
\end{eqnarray} 
In each RHS of (\ref{BEL}), apart from the Hubble rate of expansion $H$ at the decay temperature, there are various 
transition widths $\Gamma$  which are linear combinations (normalized to the photon density) of different CP conserving
collision terms $\gamma_Y^X$  for the transitions $X\rightarrow Y$ and $\bar{X}\rightarrow \bar{Y}$. Here $\gamma^X_Y$ is 
defined as
\begin{equation}
  \label{CT}
\gamma^X_Y\ \equiv \ \gamma ( X\to Y)\: +\: \gamma ( \overline{X}
\to \overline{Y} )\;,
\end{equation}
with 
\begin{equation}
\gamma ( X\to Y)\ =\ \int\! d\pi_X\, d\pi_Y\, (2\pi )^4\,
\delta^{(4)} ( p_X - p_Y )\ e^{-p^0_X/T}\, |{\cal M}( X \to Y )|^2\; . \label{gmint}
\end{equation}
In (\ref{gmint}) a short hand notation has been used for the phase space 
\bea
d\pi_x=\frac{1}{S_x} \prod\limits_{i=1}^{n_x} \frac{d^4 p_i}{(2\pi)^3}\delta(p_i^2-m_i^2)\theta(p_i^0)
\eea
with $S_X =n_{id}!$ being a symmetry factor in case the initial state $X$ contains a number $n_{id}$ of identical particles.
In addition, the squared matrix element in  (\ref{gmint}) is summed  (not averaged) over the internal degrees of freedom 
of the initial and final states.\\ 

The transition widths $\Gamma$ in (\ref{BEL}) are given as follows:
\begin{eqnarray}
  \label{GD}
&&\Gamma^{\lambda Di}  =  \frac{1}{n_\gamma}\ \gamma^{N_i}_{\slashed{L}_\lambda \phi^{\dagger}}\;, \\
  \label{GSY}
&&\Gamma^{\lambda Si}_{\rm Yukawa}  = \frac{1}{n_\gamma}\
\bigg(\, \gamma^{N_i \slashed{L}_\lambda}_{Q u^C}\: +\:  \gamma^{N_i u^C}_{\slashed{L}_\lambda Q^C}\: 
+\: \gamma^{N_i Q}_{\slashed{L}_\lambda u}\, \bigg)\; ,\\
&&\widetilde{\Gamma}^{\lambda Si}_{\rm Yukawa} = \frac{1}{n_\gamma}\
\bigg(\, \frac{\eta_{N_i}}{\eta^{\rm eq}_{N_i}}\, \gamma^{N_i \slashed{L}_\lambda}_{Q u^C}\: 
+\: \gamma^{N_i u^C}_{\slashed{L}_\lambda Q^C}\: +\: \gamma^{N_i Q}_{\slashed{L}_\lambda u}\, \bigg)\;,\\
  \label{GSG}
&&\Gamma^{\lambda Si}_{\rm Gauge}  =  \frac{1}{n_\gamma}\ 
\bigg(\, \gamma^{N_i V_\mu}_{\slashed{L}_\lambda ~\phi}\: +\: 
\gamma^{N_i \slashed{L}_\lambda}_{\phi^\dagger V_\mu}\: +\: 
\gamma^{N_i\phi^\dagger }_{\slashed{L}_\lambda V_\mu}\, \bigg)\;,\\
&&\widetilde{\Gamma}^{\lambda Si}_{\rm Gauge} = \frac{1}{n_\gamma}\ 
\bigg(\, \gamma^{N_i V_\mu}_{\slashed{L}_\lambda \phi}\: +\: 
\frac{\eta_{N_i}}{\eta^{\rm eq}_{N_i}}\, 
\gamma^{N_i \slashed{L}_\lambda}_{\phi^\dagger V_\mu}\: +\: 
\gamma^{N_i\phi^\dagger }_{\slashed{L}_\lambda V_\mu}\, \bigg)\; ,\\
  \label{GWY}
&&\Gamma^{\lambda Wi}_{\rm Yukawa}  =  \frac{2}{n_\gamma}\
\bigg(\, \gamma^{N_i\slashed{L}_\lambda}_{Q u^C}\: +\:  \gamma^{N_i u^C}_{\slashed{L}_\lambda Q^C}\: 
+\: \gamma^{N_i Q}_{\slashed{L}_\lambda u}\: +\: \frac{\eta_{N_i}}{2\eta^{\rm eq}_{N_i}}\,
\gamma^{N_i \slashed{L}_\lambda}_{Q u^C}\, \bigg)\; ,\\
  \label{GWG}
&&\Gamma^{\lambda Wi}_{\rm Gauge}  =  \frac{2}{n_\gamma}\ 
\bigg(\, \gamma^{N_i V_\mu}_{\slashed{L}_\lambda \phi}\: +\: 
\gamma^{N_i \slashed{L}_\lambda}_{\phi^\dagger V_\mu}\: +\: 
\gamma^{N_i\phi^\dagger }_{\slashed{L}_\lambda V_\mu}\: +\:
\frac{\eta_{N_i}}{2\eta^{\rm eq}_{N_i}}\, 
\gamma^{N_i \slashed{L}_\lambda}_{\phi^\dagger V_\mu}\, \bigg)\;,\\
  \label{GDL2}
&&\Gamma^{\lambda \Delta L =2}_{\rm Yukawa} = \frac{2}{n_\gamma}\sum_{\beta=e,\mu\tau}\ 
\bigg(\, \gamma^{\,\prime \slashed{L}_\lambda\phi}_{\,L^{ C}_\beta\phi^\dagger} 
+\:  2\gamma^{\slashed{L}_\lambda \slashed{L}_\beta}_{\phi^\dagger\phi^\dagger}\, \bigg)\;. 
\end{eqnarray}
The explicit expressions for  $\gamma$ and $\gamma^{\prime}$  have been considered here from the Appendix B of Ref.\cite{Pilaftsis:2003gt}.
The subscripts $D$, $S$ and $W$ stand for decay, scattering and washout respectively. We rewrite the Boltzmann equations 
in terms of $Y_{N_i}(z)=\eta_{N_i}(z)s^{-1}\eta_\gamma$ and certain $D$-functions of $z$ as given in the following.\\

Consider the first  equation in (\ref{BEL}) to start with. Its second RHS term has been neglected for our assumed scenario  
leptogenesis due to quasi degenerate RH neutrinos. Unlike the pure resonant leptogenesis\cite{Pilaftsis:2003gt,Dev:2017wwc},
here  both  $\eta_L^\beta$ and $\varepsilon_i^\beta$ are each quite small and their product much smaller\footnote{In order of
magnitude this product is $ 10^{-6}\times10^{-5}\sim 10^{-11}$, as compared with the first term which is $\mathcal{O}(1)$.}.
Using some shorthand notation, as explained in Eqs. (\ref{ki1}) - (\ref{ki2}) below, we can now write 
\begin{equation}
\frac{d Y_{N_i(z)}}{d z}=\{D_i(z)+D^{\rm SY}_i(z)+D^{\rm SG}_i(z)\}\{(Y^{\rm eq}_{N_i}(z)-Y_{N_i}(z)\}\label{BEN_Y1},
\end{equation}
where
\begin{eqnarray}
&&D_i(z)=\sum \limits_{\beta=e,\mu,\tau} D^\beta_i(z) = \sum \limits_{\beta=e,\mu,\tau} \frac{z}{H(z=1)}\frac{\Gamma^{\beta Di}}{\eta^{\rm eq}_{N_i}(z)},\label{ki1}\\
&&D^{\rm SY}_i(z)=\sum \limits_{\beta=e,\mu,\tau} \frac{z}{H(z=1)}\frac{\Gamma^{\beta S i}_{\rm Yukawa}}{\eta^{\rm eq}_{N_i}(z)},\\
&&D^{\rm SG}_i(z)=\sum \limits_{\beta=e,\mu,\tau} \frac{z}{H(z=1)}\frac{\Gamma^{\beta S i}_{\rm Gauge}}{\eta^{\rm eq}_{N_i}(z)}.\label{ki2}
\end{eqnarray}
Turning to the second equation in (\ref{BEL}) and neglecting the $\Delta L=2$ scattering terms, we rewrite it as
\begin{eqnarray}
\frac{d \eta^\lambda_L(z)}{dz} = & -& \sum\limits_{i=1}^3\,\varepsilon^\lambda_i \{D_i(z)+D^{\rm SY}_i(z)+D^{\rm SG}_i(z))(\eta^{\rm eq}_{N_i}(z)-\eta_{N_i}(z)\}
\nonumber\\& - & \frac{1}{4}\eta^\lambda_L\sum\limits_{i=1}^3 
\{\frac{1}{2}D^\lambda_i(z)z^2 K_2(z)+D^{\lambda \rm YW}_i(z)+ D^{\lambda \rm GW}_i(z) )\}  \label{BEL1}
\end{eqnarray}
with
\begin{eqnarray}
&&D^{\rm YW}_i(z)=\sum \limits_{\beta=e,\mu,\tau} \frac{z}{H(z=1)}\Gamma^{\beta Wi}_{\rm Yukawa},\\
&&D^{\rm GW}_i(z)=\sum \limits_{\beta=e,\mu,\tau} \frac{z}{H(z=1)}\Gamma^{\beta Wi}_{\rm Gauge}.
\end{eqnarray}
A major simplification (\ref{BEL1}) occurs in our model when one considers a sum over the active 
flavour $\lambda$  since $\Sigma_\lambda\varepsilon_i^\lambda=0$ and only the second RHS term contributes to
the evolution of $\Sigma_\lambda\eta^\lambda$. Then the solution of the equation 
becomes \cite{Buchmuller:2004nz}
\bea
\Sigma_\lambda\eta_L^\lambda(z)=\Sigma_\lambda\eta_L^\lambda(z=0)\exp[{-\frac{1}{4}\int_{0}^{z}W(z^\prime)dz^\prime }],\label{ete}
\eea
where 
\bea
W(z)=
\frac{1}{2}\Sigma_\lambda \left[ D^\lambda_i(z)z^2 K_2(z)+D^{\lambda \rm YW}_i(z)+ D^{\lambda \rm GW}_i(z)\right].
\eea
Thus any lepton asymmetry cannot be dynamically produced unless we assume a pre-existing lepton asymmetry 
at $z\rightarrow 0$.\\

To calculate the baryon asymmetry from the lepton asymmetry for the flavoured regimes, it is first convenient to
define the variable 
\bea
Y_\lambda=\frac{n^\lambda_L-n^\lambda_{\bar{L}}}{s}=\frac{n_\gamma}{s}\eta_L^\lambda,
\eea
i.e. the leptonic minus the antileptonic number density of the active flavour $\lambda$ normalized to the entropy density. 
The factor  $s/\eta_\gamma$ is  equal to $1.8g_{\ast s}$ and is a function of temperature. 
For $T>10^2~{\rm GeV}$, $g_{\ast s}$  remains nearly constant with temperature at a value (with three right chiral neutrinos)
of about $112$\cite{Kolb:1990vq}. Sphaleronic processes convert the lepton asymmetry created by the decay of the right chiral
heavy neutrinos into a baryon asymmetry by keeping $\Delta_\lambda=\frac{1}{3}B-L^\lambda$ conserved.  
$Y_{\Delta_\lambda}$, defined as $s^{-1}\{1/3(n_B-n_{\bar{B}})-(n_L-n_{\bar{L}})\}$, and $Y_\lambda$  are linearly 
related, as under
 \bea 
Y_\lambda=\sum\limits_{\rho}A_{\lambda\rho}Y_{\Delta_\rho}, \label{yalph}
\eea
where $A_{\lambda\rho}$ are a set of numbers whose values depend on certain chemical equilibrium conditions for different
mass regimes. These are discussed in brief later in the section. Meanwhile, we can rewrite (\ref{BEL1}) as
\begin{eqnarray}
\frac{d Y_{\Delta_\lambda}}{dz}& = & \sum\limits_{i=1}^3[\varepsilon^\lambda_i \{D_i(z)+D^{\rm SY}_i(z)+D^{\rm SG}_i(z)\}\{Y^{\rm eq}_{N_i}(z)-Y_{N_i}(z)
\} ] \nonumber\\ & + & \frac{1}{4}\sum\limits_{\rho}A_{\lambda\rho}Y_{\Delta_\rho} \sum\limits_{i=1}^3 
\{\frac{1}{2}D^\lambda_i(z)z^2 K_2(z)+D^{\lambda ~\rm YW}_i(z)+ D^{\lambda ~\rm GW}_i(z)\}. \label{BEL_Y1}
\end{eqnarray}
We need to solve (\ref{BEN_Y1}) and (\ref{BEL_Y1}) and evolve $Y_{N_i}$ as well as $Y_{\Delta_\lambda}$ upto a value 
of $z$ where the quantities $Y_{\Delta_\lambda}$ saturate to constant values. The final baryon asymmetry $Y_B$ is 
obtained \cite{Abada:2006ea} linearly in terms $Y_{\Delta_\lambda}$, the coefficient depending on the mass regime in
which $M_{i}$ is located, as explained in what follows. Let us then discuss three mass regimes separately.\\

\noindent
 {\bf $\bf {M_{i}>{10}^{12}}$ GeV (One flavour regime):} In this case all the lepton flavours are out of equilibrium and thus act indistinguishably
 leading to a single CP asymmetry parameter $\varepsilon_i=\sum\limits_{\lambda}\varepsilon^\lambda_i$.
 \begin{figure}[H]
\begin{center}
\includegraphics[scale=.34]{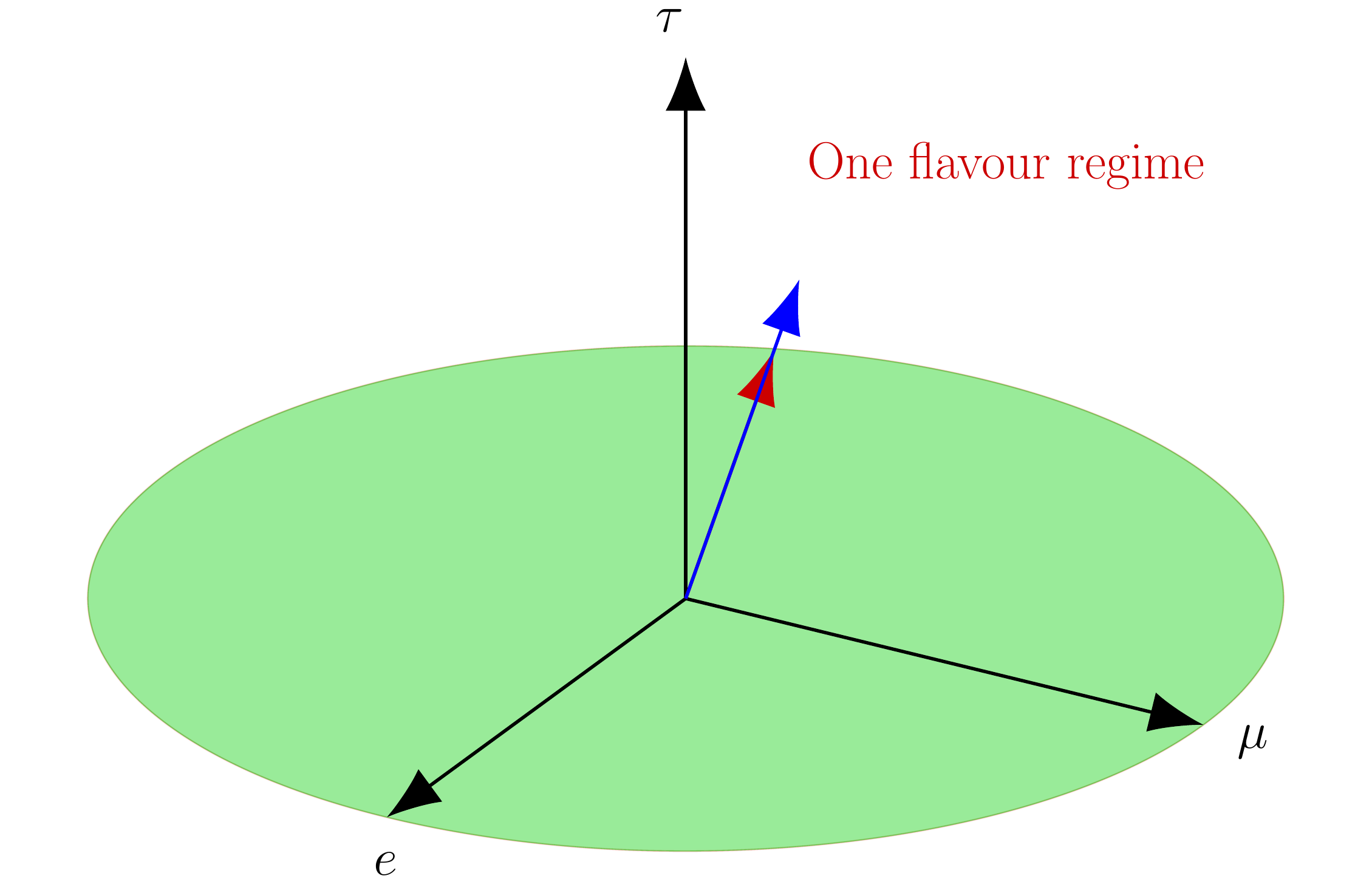}\includegraphics[scale=.34]{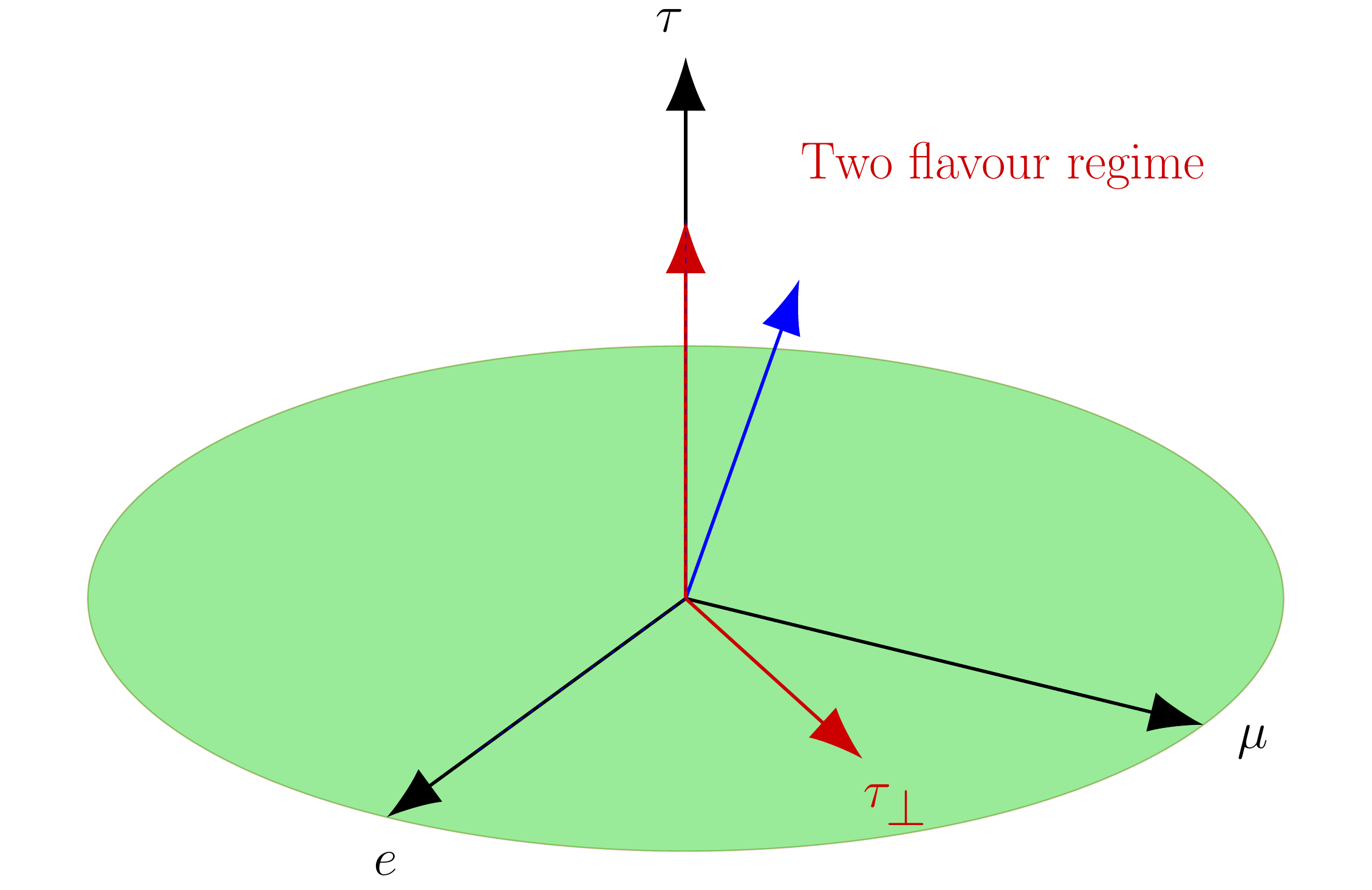}\\\includegraphics[scale=.34]{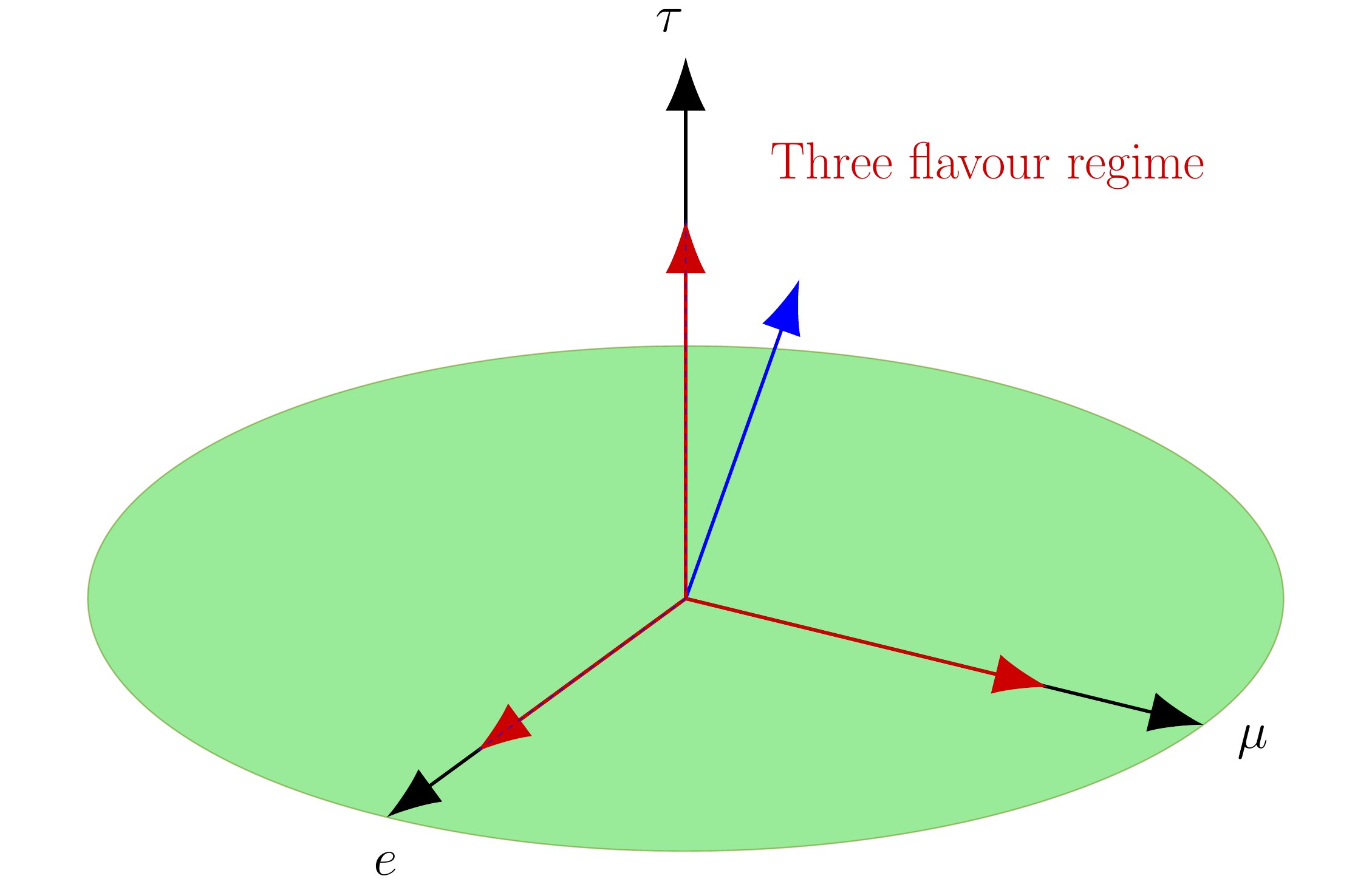}
\caption{Pictorial view of the flavoured regimes. The blue arrow indicates the direction of a state produced by  a heavy RH neutrino. The red arrows indicate the direction of the flavoured asymmetries.}
\label{p_p3}
\end{center} 
\end{figure}
 As mentioned
 earlier, $\sum\limits_{\lambda}\eta_L^\lambda=0$, therefore $Y_B=0$ and no baryogenesis is possible in this mass regime.\\  
 
 \noindent
{\bf ${\bf{{10}^{9}\,\,{\rm{\bf{{\rm \bf GeV}}}}<M_{i}<{10}^{12}}}$ GeV (Two flavour regime):} In this regime the $\tau$ flavour is in equilibrium 
and hence distinguishable but  $e$ and $\mu$ flavours cannot be distinguished since they are not in equilibrium. 
It is therefore convenient to define two sets of CP asymmetry parameters  $\varepsilon^\tau$ and 
$\varepsilon^{(2)}=\varepsilon^e+\varepsilon^\mu$, the index $\lambda$ takes the values $\tau$ and $2$ ($\tau_\perp$, Fig.\ref{p_p3}). 
The Boltzmann equations lead to the two asymmetries $Y_{\Delta_\tau}$ and $Y_{\Delta_2}$. 
They are related to $Y_\tau$ and $Y_2=Y_e+Y_\mu$  by a $2\times2$ flavour coupling A-matrix  given by \cite{Abada:2006ea}
\begin{equation}
A=\left(\begin{array}{cc}
  -417/589  & 120/589\\
   30/589 & -390/589
\end{array}\right). 
\end{equation}
The final baryon asymmetry $Y_B$ is then calculated as
\begin{equation}
Y_B=\frac{28}{79}(Y_{\Delta_2}+Y_{\Delta_\tau}) \label{pf_yb}.
\end{equation}
{\bf ${\bf{M_{i}<{10}^{9}}}$ GeV (Three flavour regime):} Here  muon flavour comes to an equilibrium thus three lepton flavours are separately 
distinguishable. Now the flavour index  $\lambda$ can just be $\lambda=e$ or $\mu$ or $\tau$. 
In this case the $3\times3$ $A$ matrix, whose $\lambda,\rho$ element relates $Y_\lambda$ and $Y_{\Delta_\rho}$, 
is given by\cite{Abada:2006ea}
\begin{equation}
A=\left(\begin{array}{ccc}
-151/179 & 20/179 & 20/179\\
25/358 & -344/537 & 14/537\\
25/358 &  14/537 &  -344/537
\end{array}\right).
\end{equation}
Now the final baryon asymmetry  normalized to the entropy density, is given by
\begin{equation}
Y_B=\frac{28}{79}( Y_{\Delta_e}+ Y_{\Delta_\mu}+Y_{\Delta_\tau}).  \label{ff_yb}
\end{equation}
\section{Numerical discussion}\label{sc5}
Before going into  detail of the numerical analysis, let us address an important issue first. 
Unlike the other literatures which also deal with perturbation to the effective light neutrino 
mass matrix $M_\nu$, here we use an exact diagonalization\cite{Adhikary:2013bma} procedure for the effective $M_\nu$. 
This in turn allows us to take large\footnote{Here by `large' we mean a number whose square order can not be neglected} 
values of the perturbation parameters (this is not allowed in 
the perturbative diagonalization procedure). Obviously in our numerical analysis we do not go beyond 
$|\epsilon^\prime|=1$ which implies a full breaking of the leading order symmetry. Here the numerical analysis  
is basically a two step process in which at first we constrain the primed parameters (e.g. Eq.(\ref{prm}) )  by
the $3\sigma$ experimental limits on the neutrino oscillation observables and then explore the  related low energy phenomenology. 
\paragraph{}
As it is mentioned earlier that we should carry out the numerical analysis only for the 
theoretically viable cases, we proceed to constrain the Lagrangian parameters with the 
$3\sigma$ experimental bounds on the oscillation data for $M_{\nu 1}^{G_1^{TBM}}$ in (\ref{G1m}) and
$M_{\nu 1}^{G_2^{TBM}}$ in (\ref{G2m}). Despite the fact that both the mass orderings for the matrix $M_{\nu 1}^{G_1^{TBM}}$ in (\ref{G1m}) and only the normal
mass ordering of $M_{\nu 1}^{G_2^{TBM}}$ in (\ref{G2m}) are theoretically allowed, given the present global fit 
oscillation constraints\cite{Capozzi:2017ipn}, the upper bound 0.17 eV\cite{Ade:2015xua} on the sum of the light 
neutrino masses $\Sigma_im_i$ and non zero values of both the breaking parameters, the latter case is disfavoured along with the inverted ordering for the former.
Therefore  the detailed discussions on numerical analysis is based entirely on the
phenomenological consequences and outcomes for the normal mass ordering case of $M_{\nu 1}^{G_1^{TBM}}$.
\paragraph{}
We then turn to the computation of baryogenesis via leptogenesis. Note that the calculation of 
the CP asymmetry parameters as well as the other involved decay and scattering process require full information of the Lagrangian
elements, i.e., the parameters of $m_D$ as well as $M_R$. For this purpose, we first fix the elements of $m_D^\prime$ 
that correspond to the lowest values of the breaking  parameters $\epsilon_{4,6}^\prime$ consistent with the oscillation data.
Then varying the unperturbed values of RH neutrino masses (here the relevant unperturbed parameter is $y$ only) we generate 
the parameters of $m_D$ using the relation $m_D=m_D^\prime\sqrt{y}$. Thus for the fixed values of the primed parameters 
and the corresponding breaking parameters, we are able to calculate $Y_B$ for each of the chosen values of unperturbed 
RH neutrino masses  and hence the elements of $m_D$. Since $Y_B$ has an upper and a lower limit we end up with an upper and a lower
limit on the RH neutrino masses also. Again, a  subtle point should be noted that since there are large number values 
for $|\epsilon_i|<1$ consistent with the oscillation data, we are obliged to take the simultaneous minimum values of 
the breaking parameters for the computation of baryon asymmetry.  However, we also discuss how the large values of the breaking parameters could affect the final asymmetry. A  detail numerical discussion is now given in what follows.\\
\subsection{Fit to neutrino oscillation data and predictions on low energy neutrino parameters}
\begin{figure}[!h]
\begin{center}
\includegraphics[scale=.3]{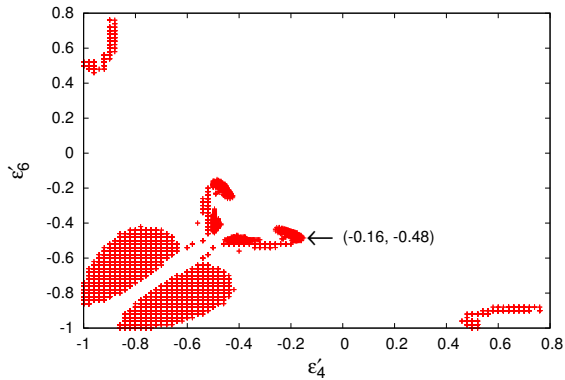}
\caption{Constrained values of breaking parameters.(the point indicated by the arrow is the lowest allowed pair of breaking parameters) }
\label{p_p}
\end{center} 
\end{figure}

\begin{figure}[H]
\begin{center}
\includegraphics[scale=.5]{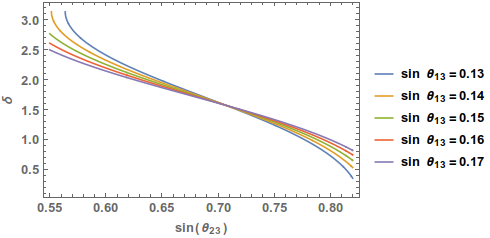}\includegraphics[scale=.5]{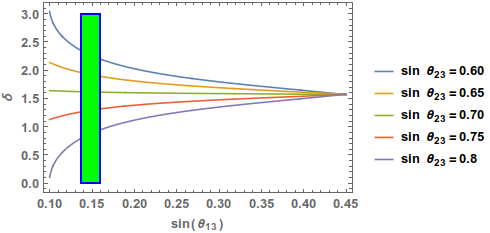}
\caption{Plot in the left side: Variation of $\delta$ with $\theta_{23}$ for different values of $\theta_{13}$. Plot in 
the right side: Variation of $\delta$ with $\theta_{13}$ for different values of $\theta_{23}$ where the green band 
represents the latest $3\sigma$ range for $\theta_{13}$. }\label{dvth}
\end{center}
\end{figure}

\begin{figure}[H]
\centering
\includegraphics[scale=.6]{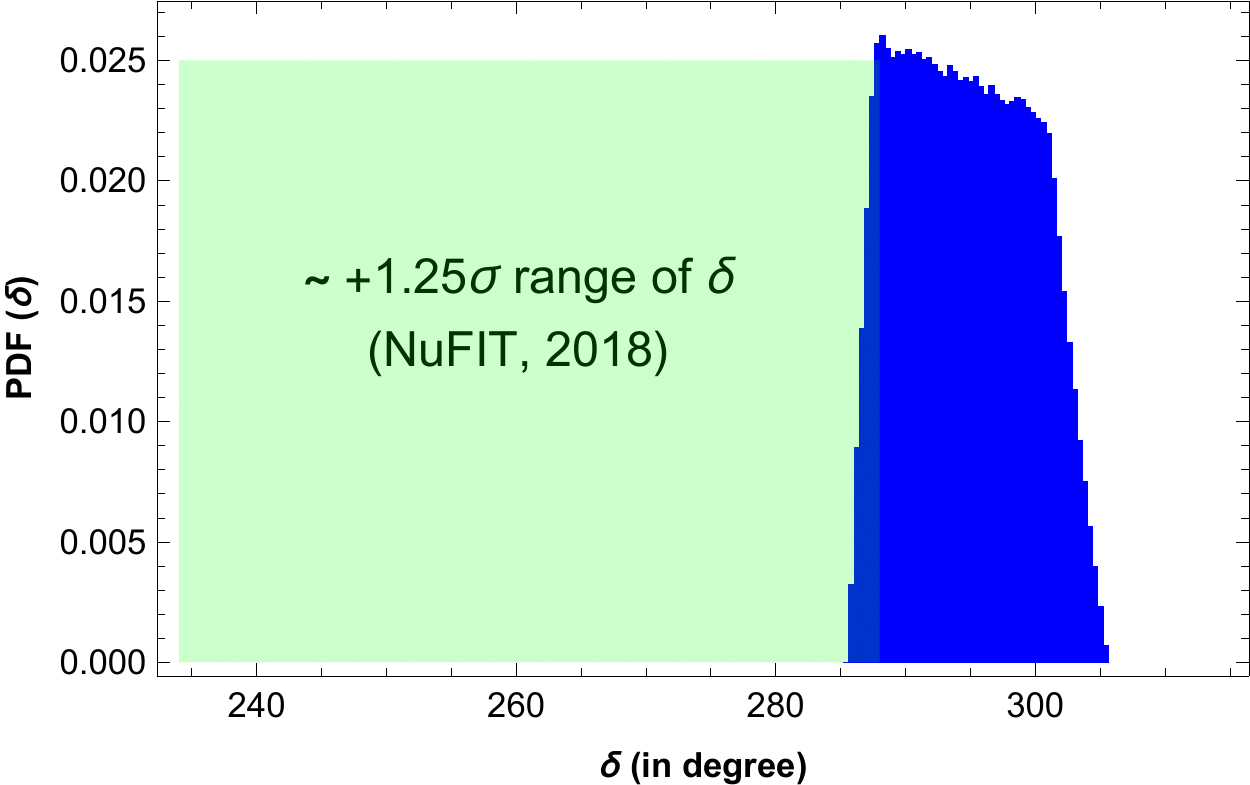}~\includegraphics[scale=.6]{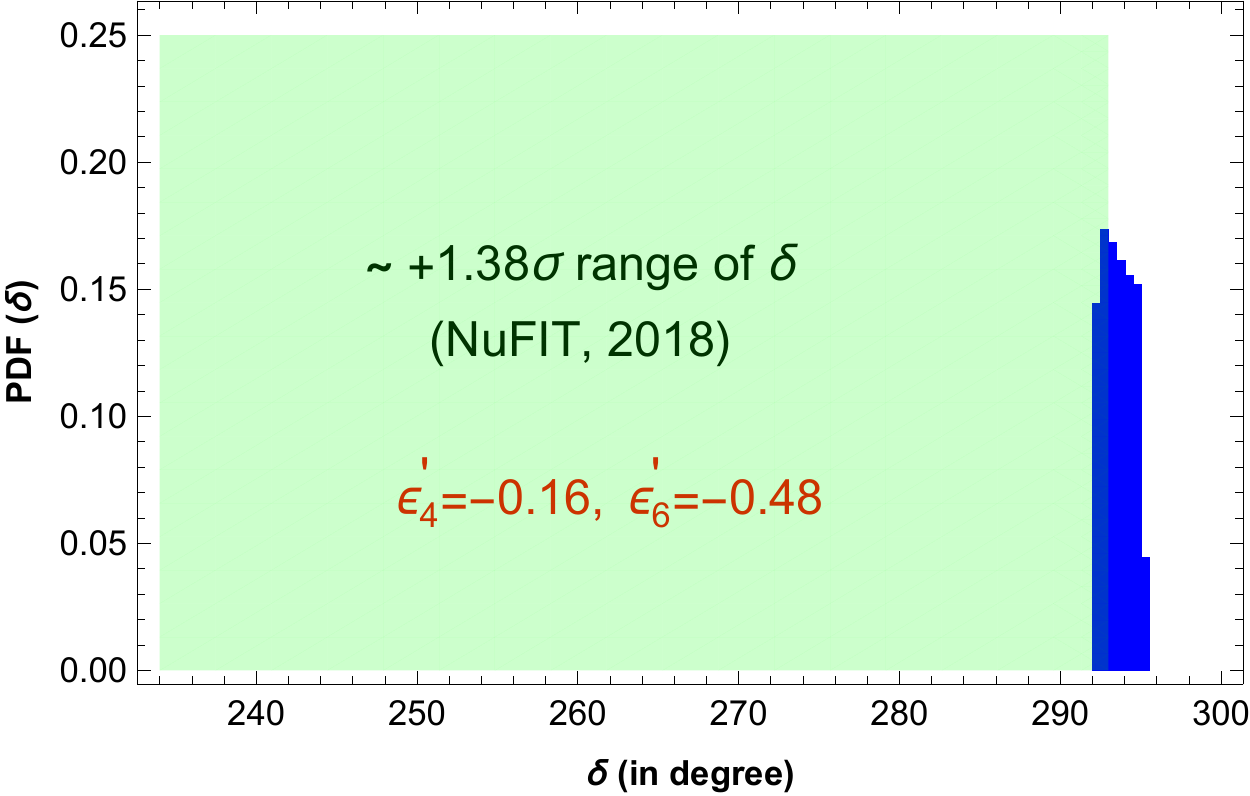}
\caption{Left: Probability distribution of the Dirac CP phase $\delta$ for normal mass ordering allowing the breaking parameters upto $\pm 0.5$.  Probability of most probable value of $\delta$ $\sim 288^\circ$ is found upon a numerical integration :  $\int_{288}^{288\pm 0.2}PDF(\delta) ~d\delta=0.025$. Most probable value $\sim 288^\circ$, is disfavoured at $\sim 1.25\sigma$ (NuFIT, 2018). Right: Probability distribution of the Dirac CP phase $\delta$ for normal mass ordering with most simultaneous values of the breaking parameters. Probability of most probable value of $\delta$ $\sim 292.5^\circ$ is found upon a numerical integration :  $\int_{295}^{295 \pm 0.2}PDF(\delta) ~d\delta=0.142$. Most probable value $\sim 292.5^\circ$, is disfavoured at $\sim 1.38 \sigma$ (NuFIT, 2018).}\label{disd}
\end{figure}

For numerical computation, we use the explicit analytic formulae for the light neutrino masses and mixing angles   
originally obtained in Ref.\cite{Adhikary:2013bma} for a general $3\times 3$ complex symmetric matrix. 
Thus after knowing the neutrino mass matrix elements $(M_\nu)_{ij}$ in terms of the model parameters $(a^\prime,b^\prime,c^\prime,\epsilon_4^\prime,\epsilon_6^\prime)$
we insert them in to the generalised formula for masses and mixing angles obtained in Ref.\cite{Adhikary:2013bma} (which in priciple enable us to express those oscillation observables in terms
of the model parameters)
and do a random scanning of the parameters using the experimental constraints on masses and mixing angles. We do not present here the  explicit forms 
of the equations since they are quite lengthy due  the complicated structure of the mass matrix as shown in Appendix A.1. In any case, as we said,  
given the general forms of those equations in Ref.\cite{Adhikary:2013bma}, one  has just to insert the mass matrix parameters in those equations and do
a random scanning of those parameters. In our case,  we vary the model parameters $(a^\prime,b^\prime,c^\prime)$ within a  range $\sim 0.01\rightarrow 0.1$, the 
phase parameter $\beta$ within the  interval as $-\pi< \beta^\prime< \pi$ and the breaking parameters in the range $-1<\epsilon^\prime_{4,6}<1$.
Then all the parameters are constrained using the $3\sigma$ global fit of neutrino data and $\sum m_i$. 
\begin{figure}[H]
\begin{center}
\includegraphics[scale=.22]{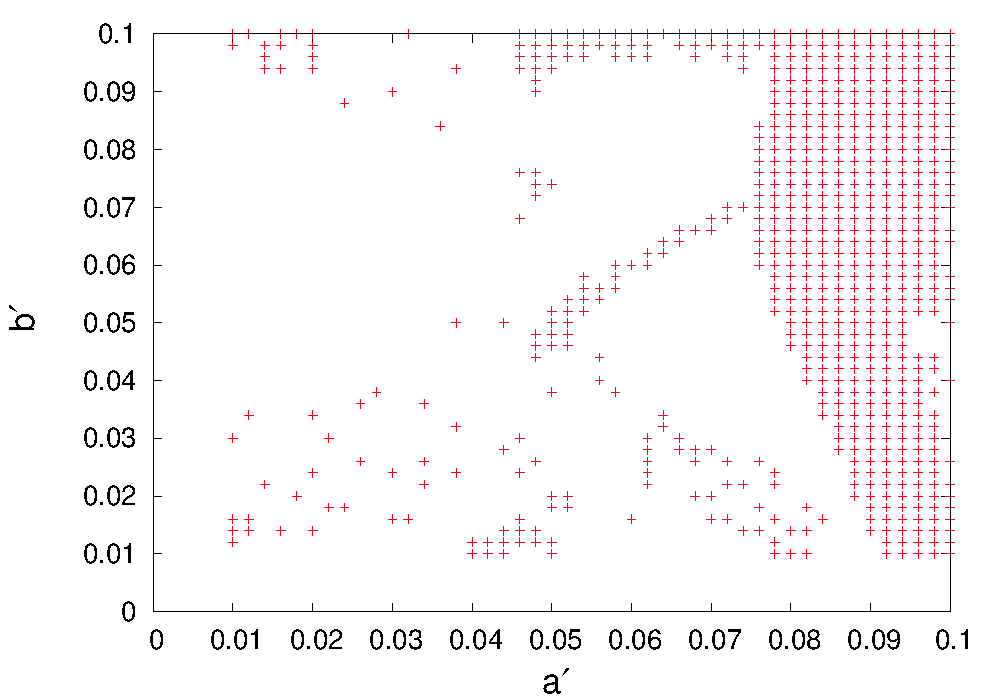}\includegraphics[scale=.22]{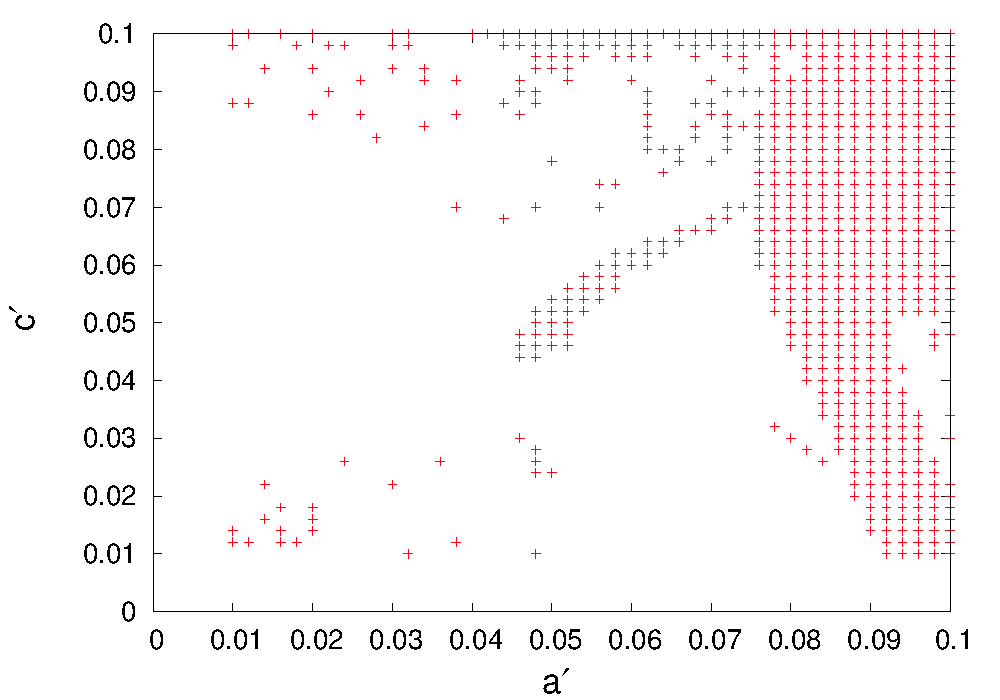}
\includegraphics[scale=.22]{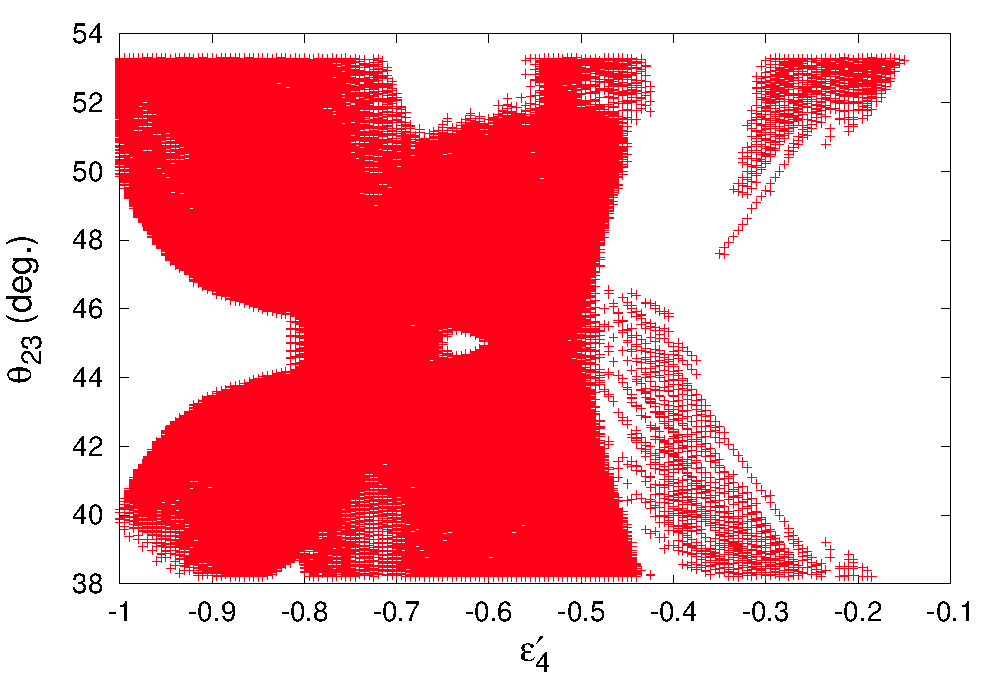}\includegraphics[scale=.22]{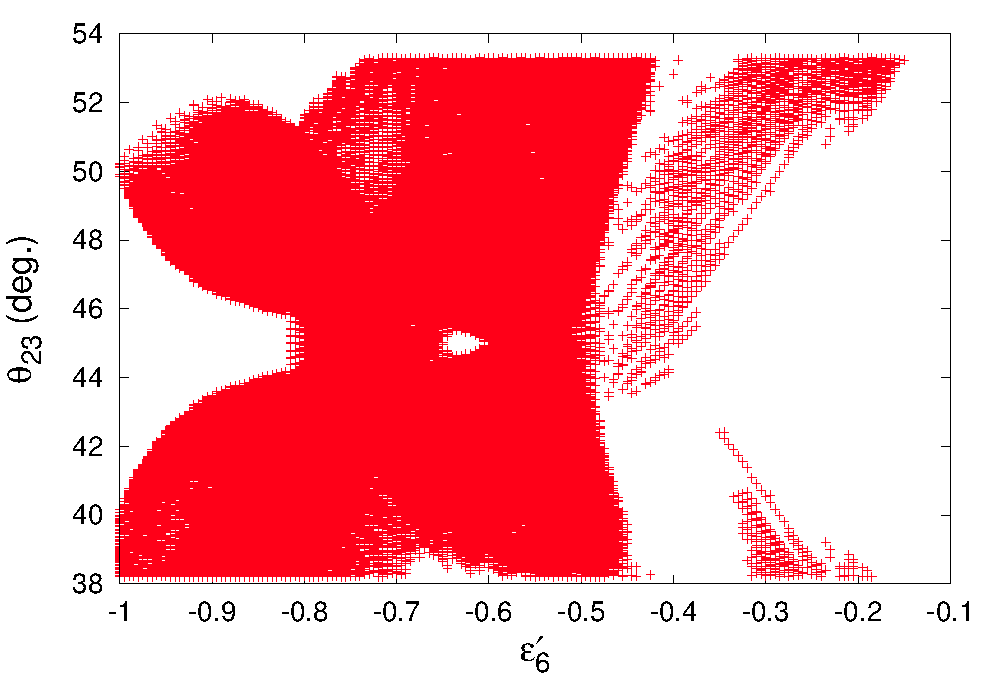}
\caption{Upper panel: Constrained range of the parameters $a^\prime$, $b^\prime$ and $c^\prime$. Lower panel: $\theta_{23}$ vs $\epsilon_4^\prime$ and $\epsilon_6^\prime$. }
\label{p_p1}
\end{center} 
\end{figure}
As shown in the upper panel of Fig.\ref{p_p1}, although
$a^\prime,b^\prime$ and $c^\prime$  are allowed almost throughout the given range $(0.01\rightarrow 0.1)$, their all possible 
combinations in the given range are not allowed\footnote{Notice that  all the plots in Fig.\ref{p_p1} and Fig.\ref{p_p} are two dimensional projections of the six dimensional coupled parameter space allowed by the oscillation data. It is difficult to infer an one to one analytic correlation due to the random shapes of the projections. However, the narrow disallowed strip in Fig.\ref{p_p} could easily be understood, since that corresponds to $\epsilon_4^\prime\simeq \epsilon_6^\prime$ which means, we essentially have one independent breaking parameter and one can not fit all the oscillation constraints with one  breaking parameter in this model.}. The phase $\beta^\prime$ remains unconstrained, i.e. all possible values
in the interval $(-\pi \rightarrow \pi)$ are allowed. The breaking parameters get significant restriction which is depicted in Fig.\ref{p_p}.
One crucial observation regrading the allowed parameter space should be mentioned here. 
The constrained parameter space is totally symmetric with respect to the sign of phase $\beta^\prime$, 
i.e., in other words, if the constrained parameter space contains a certain
set of points $(a^\prime,b^\prime,c^\prime,\epsilon^\prime_4,\epsilon^\prime_6,\beta^\prime)$ then the set 
$(a^\prime,b^\prime,c^\prime,\epsilon^\prime_4,\epsilon^\prime_6,-\beta^\prime)$ must belong to the same 
constrained parameter space. Given the constraints from the neutrino oscillation global fit data, we find it
very difficult to fit  maximal or  near maximal values of $\theta_{23}$ with small breaking parameters. For example,
to fit $\theta_{23}$ to 48$^\circ$, one needs $54\%$ and $38\%$ breaking in $\epsilon_4^\prime$ and $\epsilon_6^\prime$ for $\beta^\prime=\pm 104^\circ$. On the other hand a maximal value of $\theta_{23}$ requires $55\%$ and $43\%$ breaking on the same breaking 
parameters for $\beta^\prime=\pm 90^\circ$. However, if we restrict ourselves to consider breaking in one of  the parameters up to $25\%$ while keeping the other more than $40\%$ but less than $50\%$,  we can fit the value of $\theta_{23}$ between, e.g., $49.4^\circ-53^\circ$ (lower panel, Fig.\ref{p_p1}). For example, the most 
simultaneous minimal values of $\epsilon_4^\prime$ and $\epsilon_6^\prime$ correspond to $16\%$ and $48\%$ breaking
in the respective parameters ($\epsilon_4^\prime$, $\epsilon_6^\prime$). With this choice of values we can fit $\theta_{23}$ to a value $\sim 53^\circ$. 
We have used these minimal values of the breaking parameters in the leptogenesis calculation also.
Nevertheless, as one can see, to fit $\theta_{23}$ to its maximality or near maximality, this spacial 
case of TM1 mixing  under consideration requires large breaking. On the other hand, a sizeable departure from maximality in the second octant or in the first octant, could be fitted better with relatively small breaking parameters. The matrix $M_{\nu 1}^{G_1^{TBM}}$ also possesses testable prediction on the neutrinoless double beta decay parameter $|M_{ee}|$ (here $(M_{\nu 1}^{G_1^{TBM}})_{11}$) 3 meV-35 meV. Significant upper limits 
on $|M_{ee}|$ are available from  ongoing search experiments for $\beta\beta 0\nu$ decay. KamLAND-Zen \cite{Asakura:2015ajs}
and EXO \cite{Auger:2012ar} had earlier constrained this value to be $<0.35$ eV. Nevertheless, the most impressive upper bound
till date  is provided by  GERDA phase-II data\cite{Majorovits:2015vka}: $|M_{ee}|<0.098$ eV. Though the aforementioned experiments cannot test this model, predictions of our model could be probed by the combined GERDA + MAJORANA experiments \cite{Abgrall:2013rze}. The sensitivity reach of  other promising 
experiments such as LEGEND-200 (40 meV), LEGEND-1K (17 meV) and nEXO (9 meV)\cite{Agostini:2017jim} are also exciting to 
probe our predictions.
One of the significant result of the matrix $M_{\nu 1}^{G_1^{TBM}}$ is its prediction on the Dirac CP violating 
phase $\delta$. Since the $G_1^{TBM}$ symmetry fixes the first column of $U_{PMNS}$ to the first column 
of $U^{TBM}$ (cf. Eq.\ref{utbm}), using the equality $|{U_{\mu 1}}^{TBM}|=|{U_{\tau 1}}^{TBM}|$ and the relation
in (\ref{1strl}) one can calculate 
\bea
\cos\delta=\frac{\left(1-5 \sin\theta_{13}^2\right) \left(2 \sin\theta_{23}^2-1\right)}{4 \sqrt{2} \sin\theta_{13} \sin\theta_{23}\sqrt{\left(1-3
   \sin\theta_{13}^2\right) \left(1-\sin\theta_{23}^2\right)}}.\label{dd}
\eea
This is clear from (\ref{dd}) that the indicated maximality in the Dirac CP phase from T2K\cite{Abe:2017uxa} would arise 
for a maximal  atmospheric mixing. One can also track $\delta$ for nonmaximal values of $\theta_{23}$ (this has recently 
been hinted by NO$\nu$A at 2.6$\sigma$\cite{Adamson:2017qqn}) as shown in the Fig.\ref{dvth}.  
However, please note that, this prediction comes from the residual unbroken $\mathbb{Z}_2$ symmetry thus this is also present in a pure TM1 mixing. But, as we indicate, in our model, for small values of breaking parameters, it is not possible to reproduce the entire 3$\sigma$ range of $\theta_{23}$.  In  left panel of Fig.\ref{disd}, we present a probability distribution of $\delta$ allowing one the breaking parameters to vary upto $- 0.5$ while the other one to $-0.25$. In this case the most probable value of $\delta$, say $288^\circ$ is disfavoured at 1.25$\sigma$ by the present best fit of $\delta$\cite{nufit}. In the right panel, the distribution for $\delta$ is presented for the most simultaneous minimal values of $\epsilon_{4,6}^\prime$. Here the most probable value $\sim 292.5^\circ$ is disfavoured at 1.38$\sigma$. Though the statements on $\delta$ in the global fit is not very precise due to poor statistics, however, future measurements of $\delta$ would be an excellent  test of  the goodness of the framework under consideration.
\subsection{Numerical results for baryogenesis via leptogenesis}
Now we turn to the calculations  related to baryogenesis via leptogenesis. The main ingredients to evaluate baryon 
asymmetry are the flavoured CP asymmetry parameters and decay/scattering terms that appear in the  relevant Boltzmann equations.
Computation of these quantities require explicit values of the unprimed parameters of $m_D$ and the mass scale of $M_R$. 
Note that only the primed parameters $(a^\prime,b^\prime,c^\prime,\beta^\prime)$ along with two breaking 
parameters $\epsilon_{4,6}^\prime$ have been constrained by the oscillation data. Therefore to get the unprimed 
parameters $a,b,c$ from the primed ones, we need to  vary the mass scale $y$ which is a free parameter. Since there 
are huge number of sets of primed parameters that are consistent with the 3$\sigma$ global fit constraints, it is 
therefore impractical to numerically solve the BEs for each of the sets. For this, we take a fixed set of primed parameters 
and vary $y$ through a wide range of masses form $10^5$ GeV to $10^{12}$ GeV to study the phenomena baryogenesis via 
leptogenesis for each of the mass regimes. Question might arise, which set of the primed parameters should be taken into 
account for the computation related to baryogenesis? In principle each of the data set for the primed parameters are allowed. 
Nevertheless, we choose that set of the primed elements which corresponds to the minimum values for the breaking parameters 
required to fit the oscillation data. In other words, we have readily opted for the minimal symmetry breaking scenario (as already pointed out, the values are $\epsilon_4^\prime=-0.16$ and $\epsilon_6^\prime=-0.48$)
to compute the baryon asymmetry in our model. The set of primed parameters  are displayed in
Table \ref{osc_para}. 
\begin{table}[H]
\caption{primed parameters corresponding to lowest value of breaking parameters allowed by oscillation data}
\begin{center}
\begin{tabular}{ |c|c|c|c|c|c| } 
\hline
 $a^\prime$ & $b^\prime$  & $c^\prime$ & $\beta^\prime$ & $\epsilon_4^\prime$ & $\epsilon_6^\prime$    \\ \hline
 $0.066$ & $0.064$  & $0.1$ & $-98^\circ$ & $-0.16$ & $-0.48$ \\ \hline
\end{tabular}
\label{osc_para}
\end{center}
\end{table}
As explained in the theoretical section, the unflavoured leptogenesis scenario ($M_i>10^{12}$ GeV) is disfavored for our scheme.
We therefore present our numerical results for the other two regimes in what follows.\\

\noindent
{\bf $\tau$-flavoured regime ($10^9$ GeV $<M_i<10^{12}$ GeV):} As explained in the previous section, in this regime 
the $\tau$ flavour is in equilibrium and thus it has a separate identity but $e$ and $\mu$ are indistinguishable. 
So practically here we have two lepton flavours $(e+\mu)$ (denoted by 2 or $\tau_\perp$) and $\tau$ and correspondingly we have 
CP asymmetry parameters $\varepsilon^2_i(=\varepsilon^e_i+\varepsilon^\mu_i)$
and $\varepsilon^\tau_i$. In the flavoured Boltzmann equations (\ref{BEL_Y1}), lepton flavour index $\lambda$ can take 
only two values $e+\mu=2$ and $\tau$. Thus we get two equations involving the differentials of 
flavoured asymmetries $Y_{\Delta_2}$ and $Y_{\Delta_\tau}$ which have to be solved simultaneously (using $2\times2$ $A$ matrix)
to get the values of those asymmetries at very low temperature or equivalently at fairly large value of $z(=y/T)$ where these
asymmetries get frozen. Those final values of asymmetries are then added up  and multiplied by a suitable Sphaleronic 
conversion factor (cf. (\ref{pf_yb})) to arrive at the observed range of $Y_B$.
\paragraph{}
Now for the set of the primed parameters given in Table \ref{osc_para} we generate the unprimed parameters by varying the mass
scale parameter $y$ over the entire range $10^9$ GeV to $10^{12}$ GeV. For every value of $y$ within this range $Y_B$ indeed 
freezes to a positive value at high $z$, but the correct order of $Y_B$ ($\sim 8\times10^{-11}$) is achieved 
when $y \sim 10^{11}$ GeV. We present only few such values of $y$ and corresponding $Y_B$ in the Table \ref{Yb_m} for 
which $Y_B$ is mostly within the experimentally observed range $(8.55<Y_B\times10^{11}<8.77)$.
\begin{table}[h!]
\caption{$Y_B$ for different values of the mass scale $y$}
\begin{center}
\begin{tabular}{ |c|c|c|c|c|c|c|c|c|c|c| } 
\hline
$\frac{M_1}{10^{11}}$ (GeV) &  $3.84$ & $3.88$ & $3.92$ & $3.96$ & $4.00$ & $4.04$ & $4.08$ & $4.12$ & $4.16$ & $4.20$ \\ \hline
$Y_B\times10^{11}$          & $8.36$  & $8.44$ & $8.53$& $8.62$& $8.70$& $8.79$& $8.88$& $8.96$& $9.05$& $9.14$\\ \hline
\end{tabular}
\label{Yb_m}
\end{center}
\end{table}

\begin{center}
\begin{figure}[H]
\includegraphics[scale=0.071]{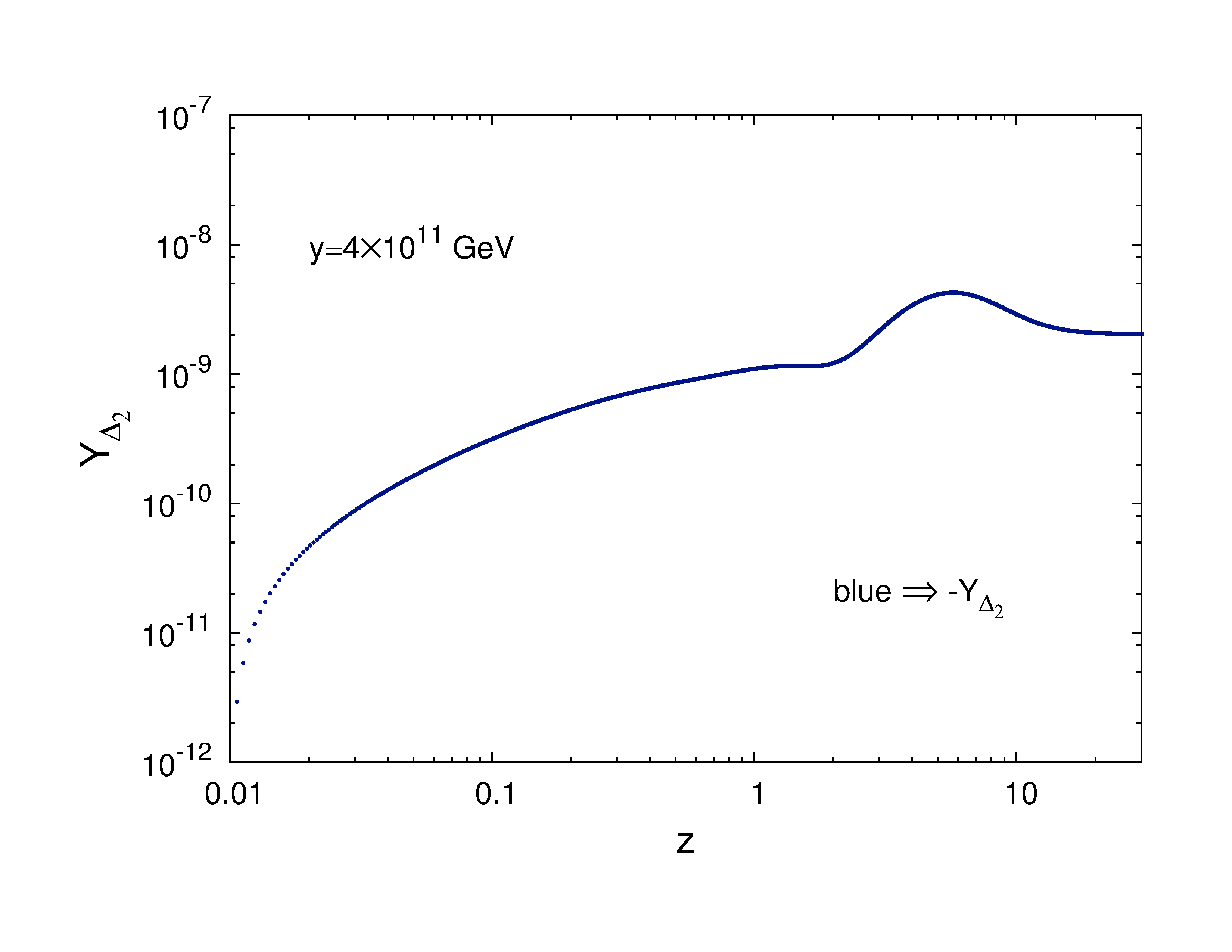}\includegraphics[scale=0.07]{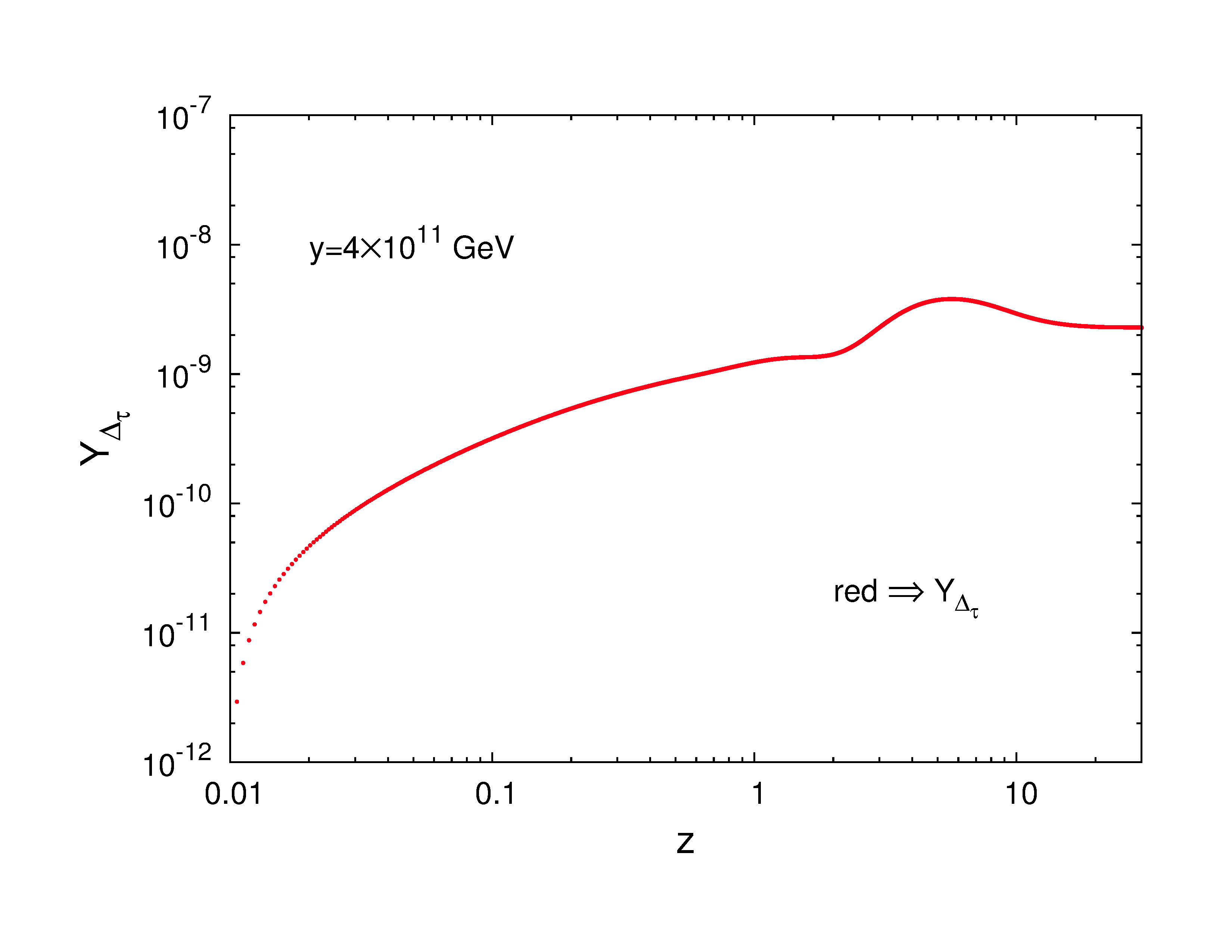}\\
\includegraphics[scale=0.07]{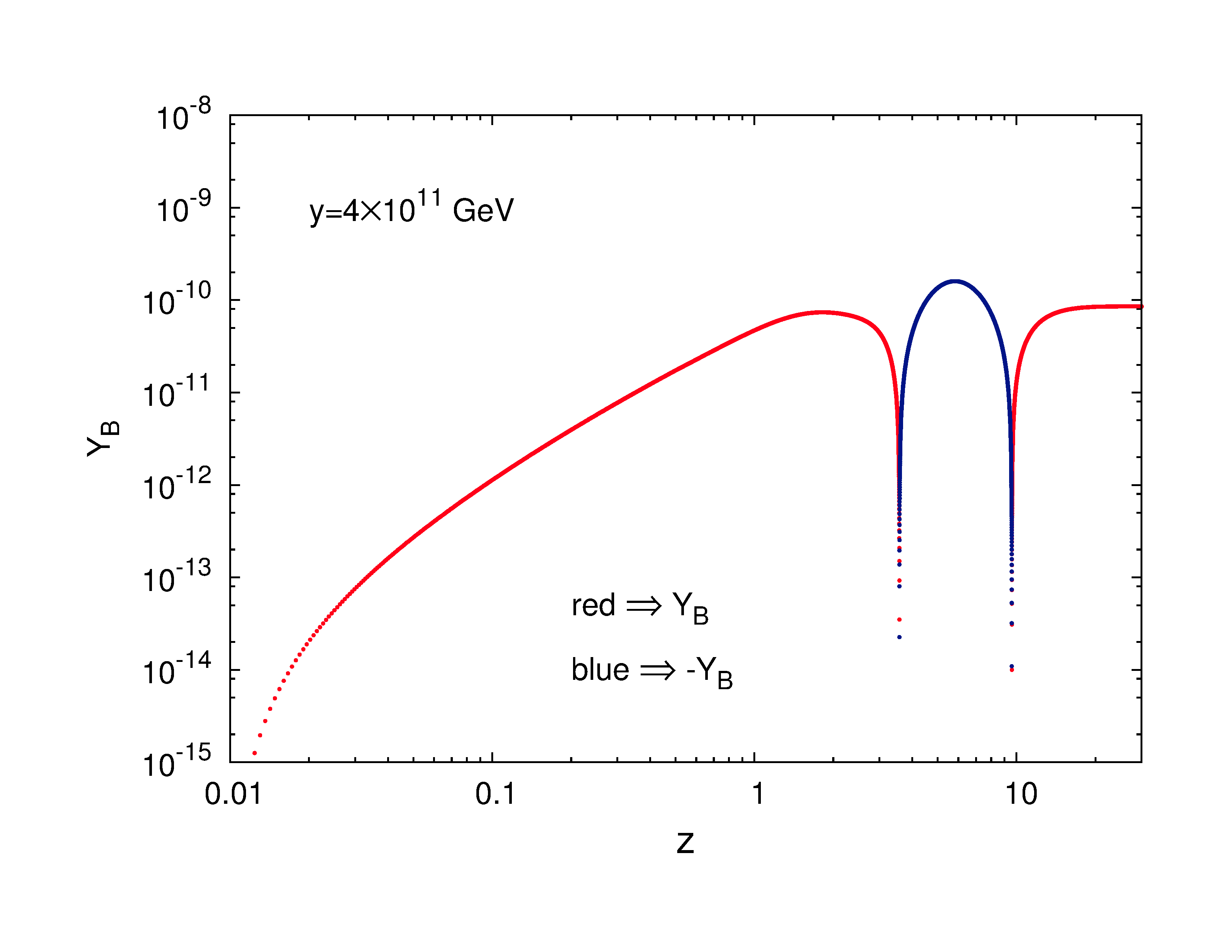}\includegraphics[scale=0.07]{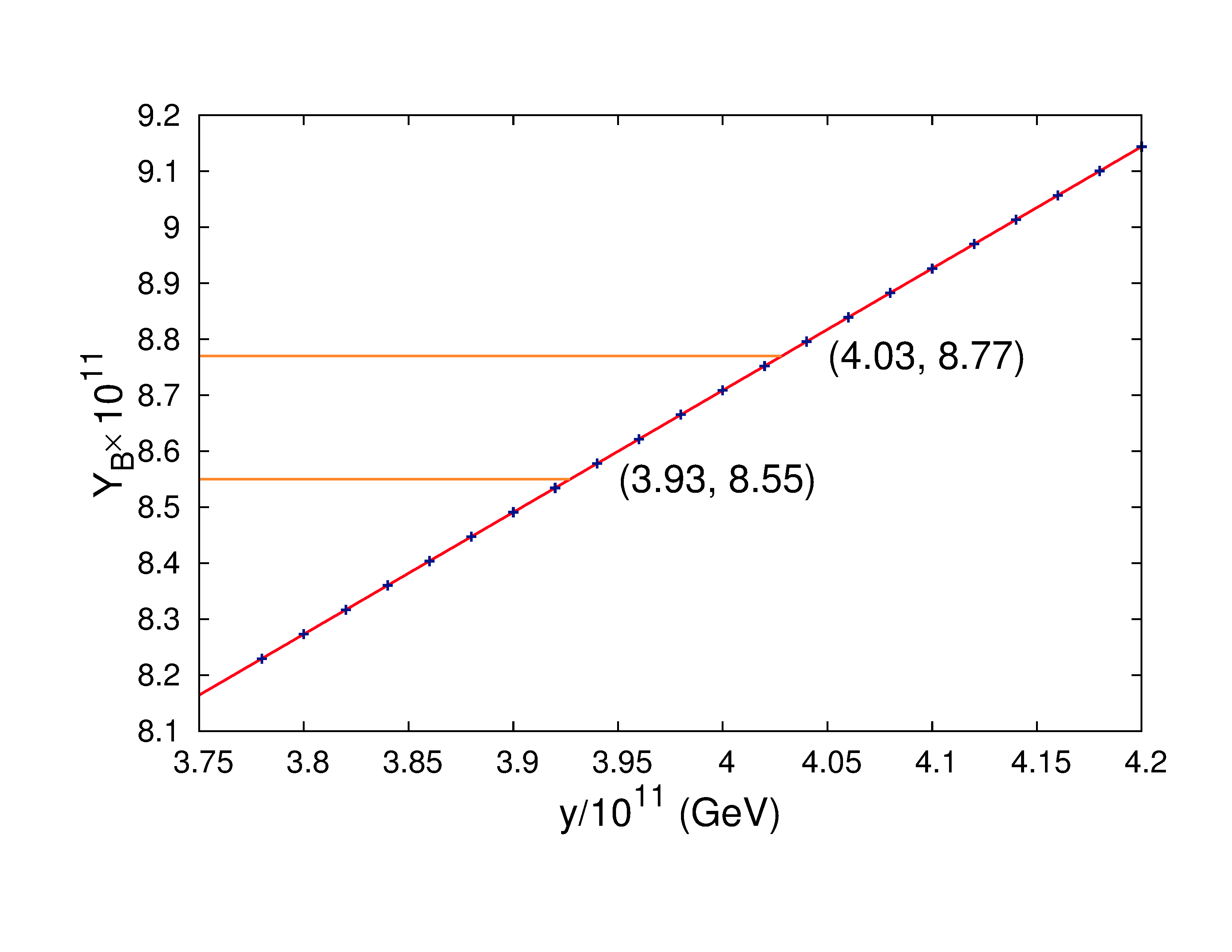}
\caption{Upper panel: variation of $Y_{\Delta_2}$ (left), $Y_{\Delta_\tau}$ (right) with $z$. Lower panel: variation of $Y_B$ with $z$ (left)in the mass regime (2) for
a definite value of $y$, variation of final value of $Y_B$ for different values of $y$ (right).
N.B. since these asymmetry parameters ($Y_{\Delta_2},Y_{\Delta_\tau},Y_B$) become negative for certain values of $z$, their negatives have been plotted on
the log scale for those values of $z$. A normal mass ordering for the light neutrinos has been assumed.}
\label{asy_m}
\end{figure}
\end{center}

Among all these values we choose $y=4\times10^{11}$ GeV and show the variation of $Y_\Delta$ asymmetries and finally $Y_B$ with $z$
in Fig.\ref{asy_m}. It can be understood from Table \ref{Yb_m} that the value of final baryon asymmetry parameter more or less increases linearly 
with the mass scale parameter $y$. Thus it is clear that for the observed range of $Y_B$, we should have a lower and an upper
bound on  $y$. The figure in the right side in the lower panel of Fig.\ref{asy_m} represents the variation of $Y_B$ with $y$. 
Two straight lines parallel to the abscissa have been drawn respectively at $Y_B=8.55\times10^{-11}$ and $Y_B=8.77\times10^{-11}$ 
corresponding to the upper and lower bounds on the observed $Y_B$.
The values of $y$ where these two straight lines touches the curve give the highest and lowest allowed value for the
mass scale parameter $y$. These values are found to be 
$y_{\rm low}=3.93\times10^{11}$ GeV and $y_{\rm high}=4.03\times10^{11}$ GeV.\\


\noindent
{\bf Fully flavoured regime ($M_i<10^9$ GeV):} In this case all three lepton flavours  can be distinguished from one 
another and consequently there are $9$ different CP asymmetry parameters $(\varepsilon^\alpha_i, \alpha=e,\mu,\tau ~{\rm and}~i=1,2,3)$ 
which have to be inserted in suitable places of fully flavour dependent Boltzmann equations (cf. (\ref{BEL_Y1})). 
These Boltzmann equations are then solved to obtain the flavoured asymmetry parameters $Y_{\Delta_\alpha}$ which
are required to obtain the final asymmetry parameter $Y_B$. 
As expected, the CP asymmetry parameters and the final baryon asymmetry parameter are found to be significantly less 
for the lower masses of right handed neutrinos $(10^7 ~-~10^8)$ GeV. So we try with the highest value of the 
right handed neutrino mass $(9\times10^8)$ GeV allowed in this regime. With $y=9\times10^8$ GeV and primed set of parameters
as given in Table \ref{osc_para}, we calculate CP asymmetry and thereafter solving the full set of coupled Boltzmann equations, we compute
$Y_B$. It is found that final value of $Y_B$ at high $z$ attains a negative value. It has already been made
clear in the second paragraph of numerical discussion that we have a similar set of points (Table.\ref{osc_para}) with
every primed parameters unaltered except $\beta^\prime \rightarrow -\beta^\prime$. Following the discussion of the last
paragraph in Sec.\ref{s3}, it is easy to understand that the sign of the CP asymmetries will be reversed while
they are computed with $-\beta^\prime$ instead of $\beta^\prime$. Therefore as a result the parameter set 
of Table.\ref{osc_para} with $\beta^\prime=98^\circ$ yields a positive value of baryon asymmetry parameter 
at high $z$, but still it is one order lower than the experimentally observed value of $Y_B$. To be precise, the value of $Y_B$ (at $z \geq 20$) for $y=9\times10^8$
is $Y_B \sim 6.5 \times10^{-12}.$
\vspace{1cm}\\
{\bf Few remarks on the effect of the two heavier neutrinos ($N_{2,3}$) on the final baryon asymmetry:} As already mentioned in Sec.\ref{s3}, the 
resonance enhancement and heavy neutrino flavour oscillation are not significant in our scenario. However, due to the small mass splitting between the
RH neutrinos, it is expected that dynamics of the heavy neutrinos are not decoupled since washout due to a particular species of RH neutrino affects
the production of the asymmetry due the lighter RH neutrinos, also the asymmetry produced by a heavier one is not fully washed out by the lighter one. 
This is why we have solved the network of  Boltzmann equations where `all production' is affected by `all washout'\cite{Pilaftsis:2003gt}.
This could qualitatively be understood by a simple two RH neutrino scenario considering the simplest form of the Boltzmann equations 
where only the decays and inverse decays are involved (however for a realistic three RH neutrino scenario, where  all the other effects, e.g., effects of 
scattering, charged lepton flavour effect, flavour coupling etc. are involved, the qualitative picture does not change). In this simplest scenario the solution 
for the lepton asymmetry $Y_L$ is given by 
\bea
Y_L=-\sum_i\varepsilon_i\kappa_i,
\eea
where $\kappa_i$ is the efficiency of production of lepton asymmetry due to `$i$'th RH neutrino. An explicit analytical  expression for $\kappa_i$ is given by
\bea
\kappa_i=-\int_{0}^\infty \frac{dY_i}{dz}e^{-\sum_i \int_{z}^\infty {\rm WID_i(z^\prime) dz^\prime}}dz,\label{effi}
\eea
where `${\rm WID}$' means the inverse decay and has the standard expression\cite{Pilaftsis:2003gt} in the hot early universe.
If the RH neutrinos are strongly hierarchical, two standard expressions for the efficiency factors are given by\cite{Buchmuller:2004nz}
\bea
\kappa_1^f(K_{N_1})\equiv \kappa_1(K_{N_1})= \frac{2}{K_{N_1}z_B}(1-e^{-\frac{K_{N_1}z_B}{2}}),~ \kappa_2^f(K_{N_2})=\kappa_2(K_{N_2})e^{-(3\pi/8) K_{N_1}},\label{eff2}
\eea
where 
\bea
z_B=2+4K_{N_1}^{0.13}e^{\frac{-2.5}{K_{N_1}}},~ K_{N_i}=\frac{|(m_{D})_{\alpha i}|^2}{M_im_*}
\eea
with $m_*$ as an equilibrium neutrino mass $\simeq 10^{-3}$ eV.
\begin{figure}[H]
\centering
\includegraphics[scale=.5]{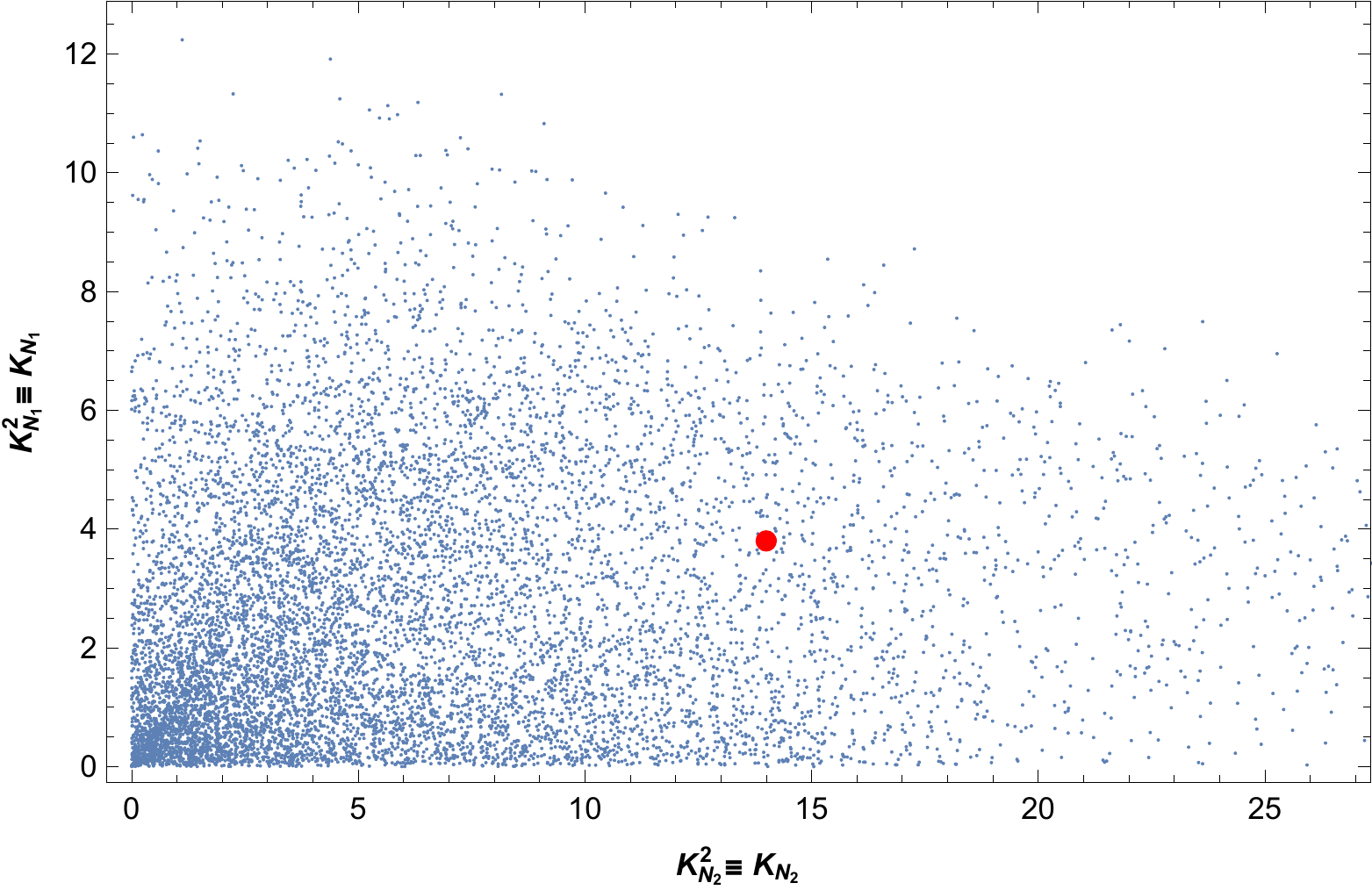}\includegraphics[scale=.8]{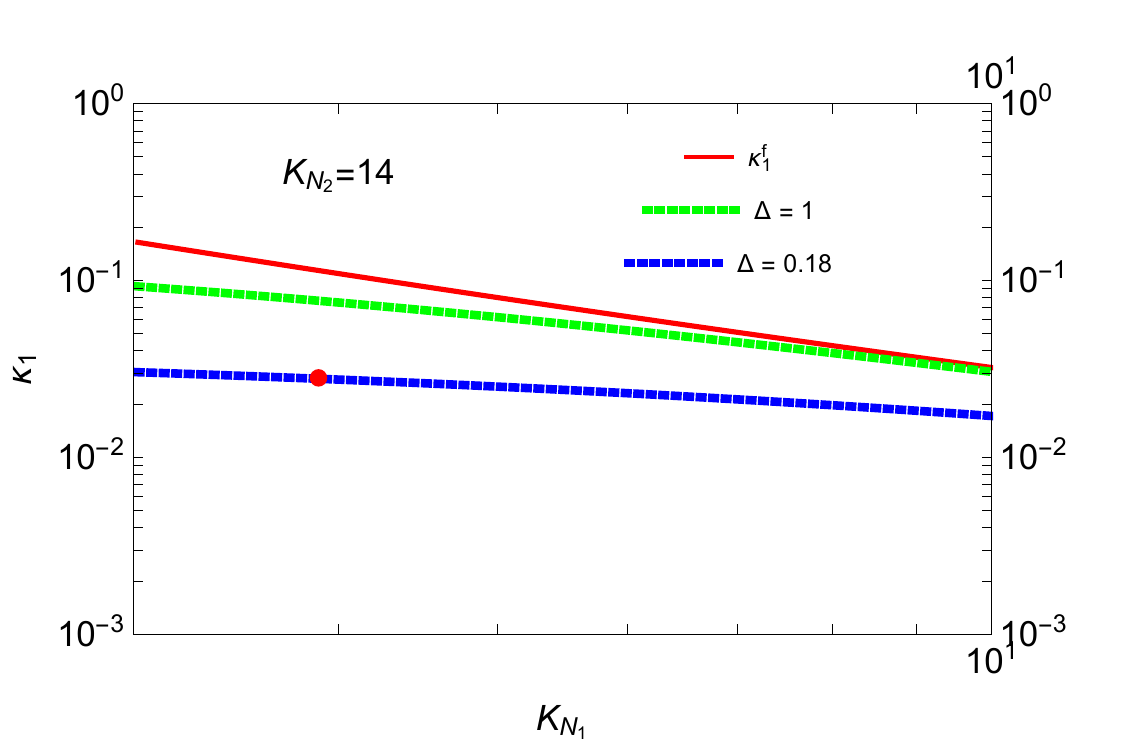}\\
\includegraphics[scale=.8]{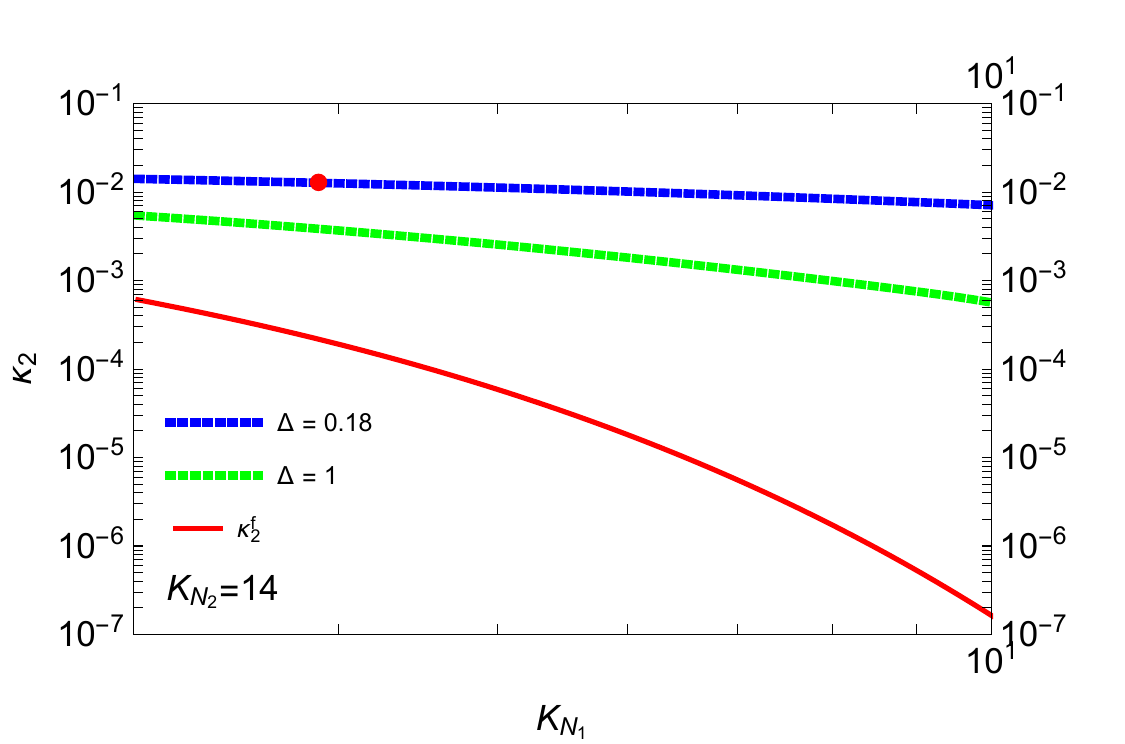}
\caption{Top left: Allowed values of the decay parameters. The red point represents value of the decay parameters which correspond to the minimal pair of breaking parameters. Top right. A comparison between the efficiency factor $\kappa_1$ due to the standard $N_1$ dominated scenario and in our model.  Bottom:  A comparison between the efficiency factor $\kappa_2$ due to the standard $N_1$ dominated scenario and in our model. }\label{effi_d}
\end{figure}
Our task is to show, given our model, whether a numerical integration of (\ref{effi}) is consistent with (\ref{eff2}) or not. If these two equations match, then contribution from the heavier neutrinos are irrelevant, i.e, we are in standard $N_1$ dominated scenario. On the contrary, if they are inconsistent, then we can conclude that contributions from the heavier neutrinos are not negligible.\\

 Let us choose the $\tau_\perp$ or the `2' flavour for a specific example and denote the decay parameters as
$ K_{N_1}^2\equiv  K_{N_1}$ and $ K_{N_2}^2\equiv  K_{N_2}$ for $N_1$ and $N_2$ respectively. It is clear from the figure on the RHS (top) of Fig.\ref{effi_d}, that the efficiency factor of the standard $N_1$ dominated scenario and our model do not match at all. In fact,  the efficiency factor $\kappa_1$ decreases as we consider smaller breaking of $\mathbb{Z}_2^{\mu\tau}$.  This is simply due to the fact, that as the $\mathbb{Z}_2^{\mu\tau}$ breaking parameters which are also related to the mass splitting of the RH neutrinos, become smaller, in  addition to the $N_1$-washout, the washout by  $N_2$ also affects the asymmetry production due to $N_1$ decays (In a typical $N_1$ dominated scenario, $N_2$ inverse decay goes out of equilibrium before the $N_1$ decay or inverse decay reaches equilibrium\cite{Blanchet:2006dq}). The blue line represents the $\kappa_1$ for the minimal set of breaking parameters $\epsilon_{4}'=-0.16$ and  $\epsilon_{6}'=-0.48$ which corresponds to a mass splitting $\Delta=(M_2-M_1)/M_1=0.18$ (cf Eq.\ref{Rhe}) where we allow $ K_{N_1}$ to vary within the obtained range $\sim 3-10 $ and take $ K_{N_2}=14$. The red dot represents  a particular set  ($ K_{N_1}=3.89,~ K_{N_2}=14$) which corresponds to the earlier mentioned simultaneous minimal pair of breaking parameters. The green line corresponds to $\Delta=1$. As one can see, starting from our scenario which corresponds to small breaking parameters, as one approaches to a pure TM1 mixing which requires complete breaking of the $\mu\tau$ symmetry and hence large breaking parameters, the effect of the next to the lightest of the heavy neutrinos decreases so that the efficiency factor $\kappa_1$ tends to match with  its standard expression in a pure $N_1$ dominated or strongly hierarchical case. On the other hand, as one can see from the figure at the bottom in Fig.\ref{effi_d}, for small values of the breaking parameters, the efficiency factor $\kappa_2$ escapes from the exponential washout (cf. Eq.\ref{eff2}) due to $N_1$ and increases from its standard value $\kappa_2^f$, thus leaves a non-negligible contribution to the final asymmetry.
\subsection{Prediction of flux ratios at neutrino telescopes} \label{sec4d} 
Recently IceCube \cite{Aartsen:2013bka} has discovered long expected Ultra High Energy (UHE) neutrinos events and  thus  opened  a new era in the neutrino astronomy.  IceCube has reported 82 high-energy starting events (HESE) (Including track +shower)   which constitute more than 7$\sigma$ excess  over the atmospheric background and thus points towards an extraterrestrial origin of the UHE neutrinos (for a latest updated result, please see \cite{ice2018}). Also, no significant spatial clustering has been found\cite{Adrian-Martinez:2015ver} and the recent data seems to be consistent with isotropic neutrino flux from uniformly distributed point sources and points towards extra galactic nature of the observed events. Nevertheless, the origin of these UHE neutrinos still remains unknown. Although the HESE events are not consistent\footnote{Using a flavour composition 1:1:1 at the earth and deposited energy 60 TeV-10 PeV, 6-years HESE best fit to the spectral index is $\gamma=2.92^{+0.29}_{-0.33}$. However, the 8-years through going muon (TG) data which corresponds to 1000  extraterrestrial neutrinos above 10 TeV, corresponds to a best fit $\gamma=2.19\pm 0.1$ which is close to the theoretically preferred $E^{-2}$ spectrum.} with the standard astrophysical `one component' unbroken isotropic power-law spectrum 
\bea
\Phi (E_\nu)=\Phi_0\left( \frac{ E_\nu}{100 ~{\rm TeV}}\right)^{-\gamma}
\eea
with $\gamma\simeq 2$ (much harder spectrum than the HESE best fit) and also suffer constraints from multi-messenger gamma-ray observation\cite{Becker:2007sv}, `two component'  explanation of the observed neutrino flux from purely astrophysical sources is still a plausible scenario \cite{Sui:2018bbh}. Thus with enhanced statistics at the neutrino telescopes and future determination of the flavour composition of UHE neutrinos at the earth would pin point the viability of the astrophysical sources  as the origin of the UHE neutrinos. In our model, without going into any fit to the present data, we predict the the flavour flux ratios at the earth, assuming the conventional $pp$ and $\gamma p$ sources. The dominant source of ultra high energy cosmic neutrinos are $pp$ (hadro-nuclear) collisions in cosmic ray reservoirs such as galaxy clusters  and $p\gamma$ (photo-hadronic) collisions in cosmic ray accelerators such as gamma-ray bursts, active galactic nuclei and blazars\cite{Ahlers:2015lln,Hummer:2010vx}. In  $pp$ collisions, protons of TeV$-$PeV range produce neutrinos via the processes $\pi^+\to \mu^+\nu_\mu, \pi^-\to \mu^-\bar{\nu}_\mu, \mu^{+}\to e^+\nu_e\bar{\nu}_\mu$ and $\mu^-\to e^{-}\bar{\nu}_e\nu_\mu.$ Therefore, the ratio of the normalized flux distributions over flavour is 
\begin{equation}\phi^S_{\nu_e}:\phi^S_{\bar{\nu}_e}:\phi^S_{\nu_\mu}:\phi^S_{\bar{\nu}_\mu}:\phi^S_{\nu_\tau}:\phi^S_{\bar{\nu}_\tau}=\phi_0\Big\{\frac{1}{6}:\frac{1}{6}:\frac{1}{3}:\frac{1}{3}:0:0\Big\}, \label{stan}
\end{equation}
 where the superscript $S$ denotes `source' and $\phi_0$ denotes the overall flux normalization. For $p\gamma$ collisions, one has either $\gamma p\to X \pi^{\pm} $ leading to the same flux ratios in Eq.\ref{stan} or the resonant production $\gamma p\to \Delta^+\to \pi^+ n$ and $\pi^+\to \mu^+\nu_\mu, \mu^+\to e^+\nu_e\bar{\nu}_\mu.$  corresponding normalized flux distributions over flavour 
 \begin{equation}
\phi^S_{\nu_e}:\phi^S_{\bar{\nu}_e}:\phi^S_{\nu_\mu}:\phi^S_{\bar{\nu}_\mu}:\phi^S_{\nu_\tau}:\phi^S_{\bar{\nu}_\tau}=\phi_0\Big\{\frac{1}{3}:0:\frac{1}{3}:\frac{1}{3}:0:0\Big\}.\end{equation} 
In either case, since the Icecube does not distinguish between neutrino and antineutrinos (other than the Glashow resonance: $\bar{\nu}_ee^-\to W^-$ at $E_\nu\simeq 6.3$ PeV)  we take $\phi^S_l=\phi^S_{\nu_l}+\phi^S_{\bar{\nu}_l}$ with $l=e,\mu,\tau$ as
 \begin{equation}
\phi_e^S:\phi_\mu^S:\phi_\tau^S=\phi_0\Big\{\frac{1}{3}:\frac{2}{3}:0\Big\}.
\end{equation}
Since the source-to-telescope distance is much greater than the oscillation length, the flavour oscillation probability averaged over many oscillations is given by 
\begin{equation}
P(\nu_m\to\nu_l)=P(\bar{\nu}_m\to
\bar{\nu}_l)\approx \sum\limits_{i}|U_{l i}|^2|U_{m i}|^2.\end{equation}
Thus the flux reaching the telescope is given by
 \begin{equation}
\phi_l^T=\sum\limits_{i}\sum\limits_{m}\phi_m^S|U_{l i}|^2|U_{m i}|^2=\frac{\phi_0}{3}\sum\limits_{i}|U_{l i}|^2(|U_{ei}|^2+2|U_{\mu i}|^2).
\end{equation}
which simplifies to 
  \begin{equation}
  \phi_l^T=\frac{\phi_0}{3}[1+\sum\limits_{i}|U_{l i}|^2(|U_{\mu i}|^2-|U_{\tau i}|^2)]=\frac{\phi_0}{3}[1+\sum\limits_{i}|U_{l i}|^2\Delta_i].
  \end{equation} 
\begin{figure}[H]
\centering
\includegraphics[scale=.5]{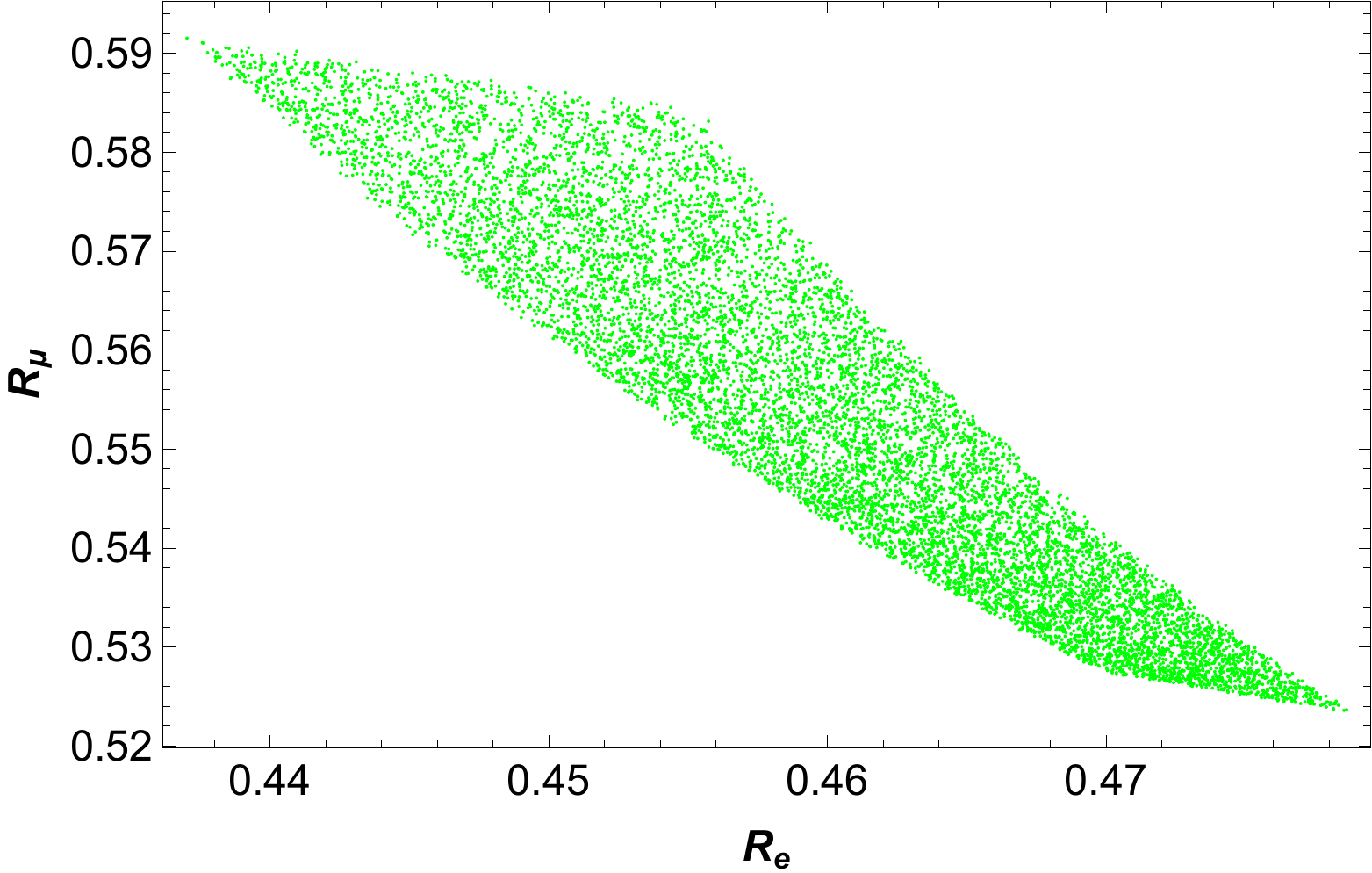}\includegraphics[scale=.5]{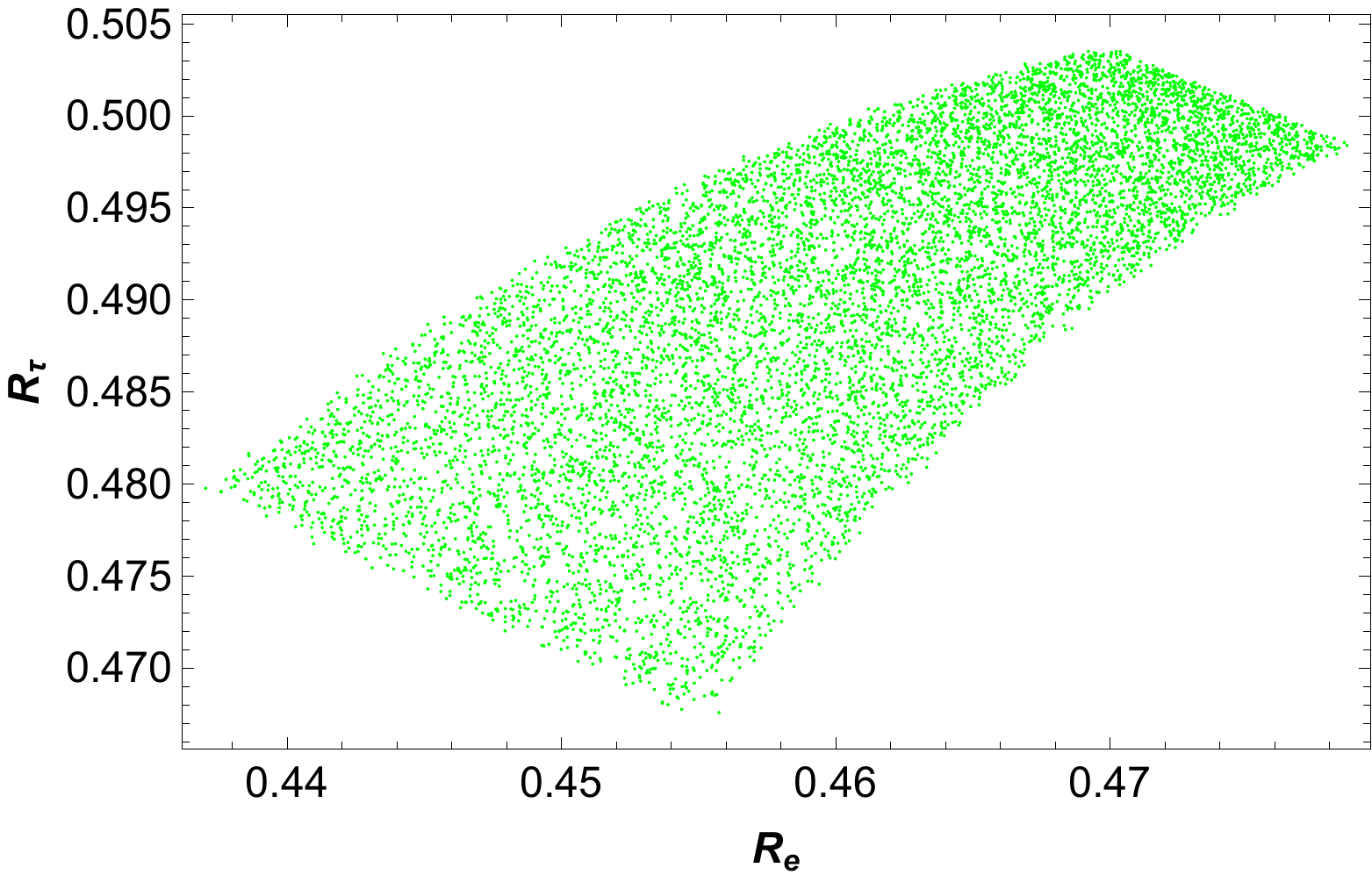}\\

\caption{ $R_e$ vs. $R_{\mu,\tau}$ in our model allowing  one of the  breaking parameters up to $-0.50$ keeping the other one upto $-0.25$. }\label{flux}
\end{figure}
where $\Delta_i=|U_{\mu i}|^2-|U_{\tau i}|^2$ and we have used the unitarity of the PMNS matrix i.e., $|U_{ei}|^2+|U_{\mu i}|^2+|U_{\tau i}|^2=1$.
With the above background, one can define  flavour flux ratios $R_l$ ($l=e,\mu,\tau$) at the neutrino telescope as
\begin{equation}
 R_l\equiv\frac{\phi_l^T}{\sum\limits_{m}\phi_m^T-\phi_l^T}=\frac{1+\sum\limits_{i}|U_{l i}|^2\Delta_i}{2-\sum\limits_{i}|U_{l i}|^2\Delta_i},
\end{equation}
 where $m=e,\mu,\tau$ and $U$ is as in \eqref{eu}. 
 Note that for the exact TBM,  $\Delta_i=0$ and thus $R_e:R_\mu:R_\tau=1:1:1$ -- this is well known\cite{Learned:1994wg,Pakvasa:2007dc,Rodejohann:2006qq,Xing:2008fg}. In our model,  we find interesting deviation from this democratic flavour distribution at the telescopes.  As we show in the Fig. {\ref{flux} whilst $R_e$ and $R_\tau$ prefers the values less than the standard value 0.5, $R_\mu$ prefers values greater than 0.5 in this model. In Fig.\ref{tria} we present a Ternary plot for a better visualization of the flavour compositions. Here $\alpha_l=R_l/\Sigma_lR_l$. The red `$\ast$' represents the TBM democratic prediction 1:1:1. The green area represents the allowed range of the flavours for 3$\sigma$ interval of the mixing parameters. The blue region (within the green one) is our model prediction. The red `$+$' is the HESE best fit 0.29:0.50:0.21\cite{ice2018}. Clearly the present HESE best fit and the flavour composition allowed by standard neutrino oscillation as well as the composition predicted in our model are in tension. Though these could be reconciled well within the HESE  $68\%$ CL\cite{ice2018}. 
\begin{figure}[H]
\centering
\includegraphics[scale=.6]{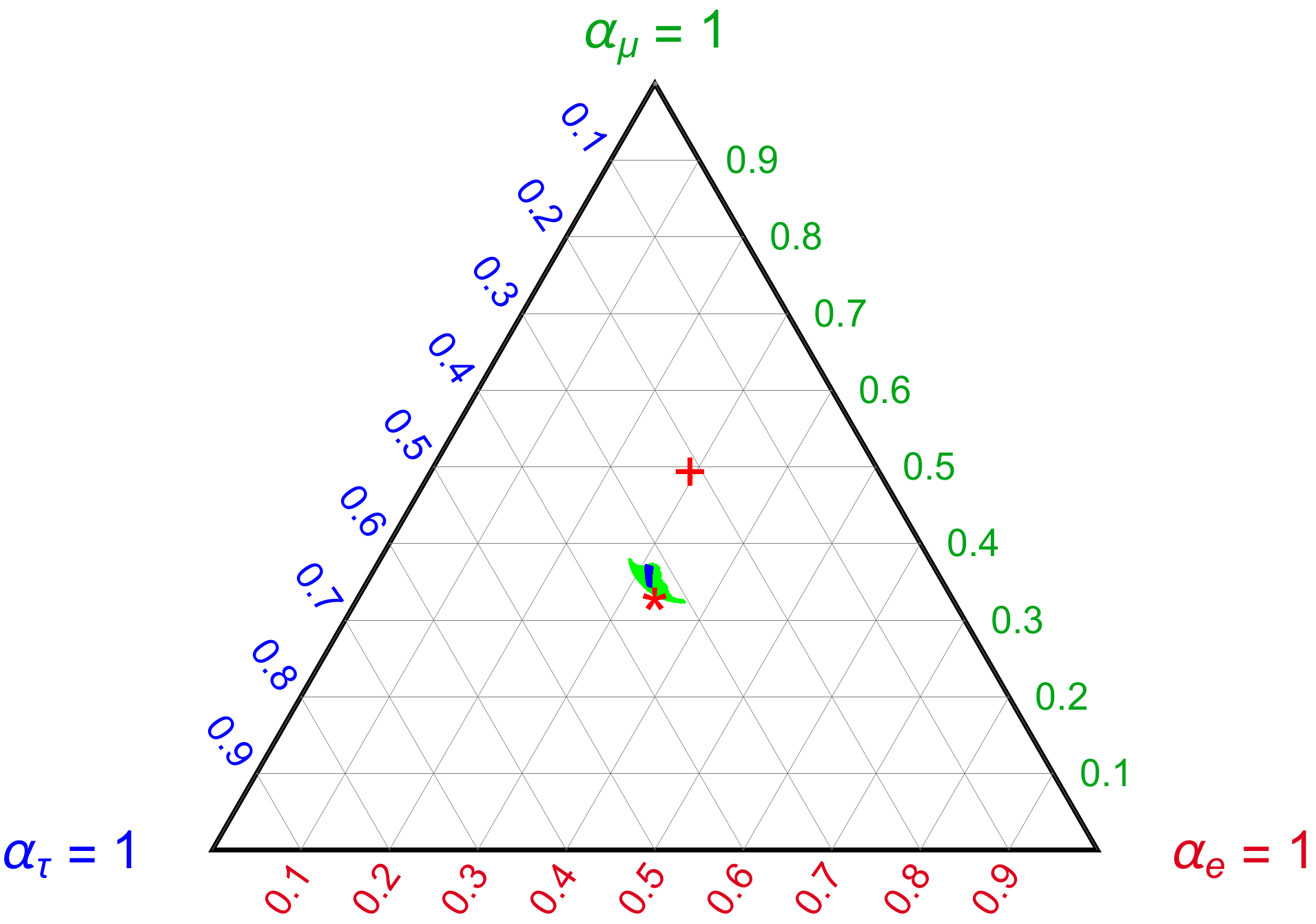}

\caption{The red `$\ast$' represents the TBM democratic prediction 1:1:1. The green area represents the allowed range of the flavours for 3$\sigma$ interval of the mixing parameters. The blue region (within the green one) is our model prediction. The red `$+$' is the HESE best fit 0.29:0.50:0.21\cite{ice2018}. }\label{tria}
\end{figure}
\subsection{A comparative study with the other works and few final remarks}
Though at the end of the introduction section we have tried to focus on the novelty and new results of our work, for the sake of completeness 
and a quantitative comparison, we would like to expense few lines in this subsection also. As already pointed out in the introduction, 
in a bottom up approach, starting from the  residual symmetry framework, we have studied the goodness of the $\mu\tau$ symmetry under the
lamppost of a TM1 symmetry in quite a general way.  Though there is sizeable amount work devoted to TBM mixing as we cited in the 
introduction, e.g., Ref.\cite{He:2006qd,Albright:2008rp,Xing:2006ms,tm1,tm2} etc, most of them discuss either completely broken TBM
or a pure TM1 symmetry. Thus we feel, the results obtained in our work (apart from the correlations which are also present in a pure TM1 mixing)
are entirely novel and more testable. As we have already pointed out, we are motivated  by Ref.\cite{joshi} Ref. \cite{rode}, where, keeping the TM1
generator unbroken, alteration of the $\mu\tau$ symmetry has been studied. Though the alterations have been done by the usage of the $\mu\tau$ generator
as a CP generator; instead of an exact $\mu\tau$ interchange symmetry. However, here we have studied the modification of the $\mu\tau$ interchange by
breaking it explicitly. Thus, though the philosophy behind our work is same as that of Ref.\cite{joshi} Ref. \cite{rode}, phenomenological outcomes are 
different due different treatment of the $\mu\tau$ symmetry. Since the underlying philosophy for handling the symmetry is different,  our  low energy predictions such as the Dirac CP phase as well as the estimated range of $\theta_{23}$ are   also different from a comprehensive analysis \cite{Albright:2008rp} which discusses different variants of TBM mixing. From leptogenesis perspective, our 
work and the Ref.\cite{Mohapatra:2004hta}  share some common ground. To be more precise, the common origin of nonzero $\theta_{13}$ and quasi 
degeneracy of the RH neutrino masses. However, the analysis is done only with $\mu\tau$ interchange symmetry  with one small breaking parameter 
(so that the second order terms could be neglected)  whereas in our case we have done braking of $\mu\tau$ keeping TM1 symmetry intact and
this in turn leads the structure of the breaking pattern  not be arbitrary. In addition, our leptogenesis analysis is rigorous and  also include flavour effects
as well as theoretical uncertainties such as flavour couplings. We also show, how starting from our scenario, as one approaches to a pure TM1 mixing, 
the effects of  the heavy RH neutrinos become weak. We would like to draw the same conclusion for Ref.\cite{Ge} as we do for Ref.\cite{Mohapatra:2004hta}.
Though Ref.\cite{Xing:2006ms} studied leptogenesis under a TM1 symmetry, the framework is minimal seesaw and thus one of the mass eigenvalue is zero plus
due to the typical structure of the mass matrices, the prediction is $\theta_{23}=\pi/4$ and $\delta=\pm\pi/2$. In addition, the authors assume the RH 
masses are arbitrarily close so that resonant condition could be satisfied in a low RH mass scale whereas in our case as we already point out,
one cannot choose the arbitrary mass splitting between the RH neutrinos. \\

Some  final remarks: In this precision era of the low energy neutrino phenomenology,   it is a high time for the rigorous computation in any neutrino mass model so that it could be tested in the experiments unambiguously. In this work, we have tried to be as concise as possible in the computation while rigorously  studying an unexplored scenario related to the modification to a TBM scheme.    We report that, the framework under consideration is not compatible with very small breaking parameters and the neutrino oscillation data. The minimal pair corresponds to $\epsilon_4^\prime=-0.16$ and $\epsilon_6^\prime=-0.48$. Even if we allow breaking in both the parameters upto $40\%$, our work disfavours maximal mixing. Thus in future, strong statements on $\theta_{23}$ would be an excellent probe to test the goodness of our idea. In addition, whilst an in-depth computation (with the best of our expertise) of baryogenesis via leptogenesis seeks the  RH mass scale to be more than $10^{11}$ GeV, validity of this framework with relatively small breaking parameters is also testable via its very sharp predictions on the Dirac CP phase as well as UHE neutrino flavour ratios.\\
\section{Conclusion}\label{sc6}
We have analyzed the broken TBM mass matrices which are invariant under a residual $\mathbb{Z}_2\times\mathbb{Z}_2^{\mu\tau}$
(TBM-Klein) symmetry at the leading order. To explore a predictive scenario, we have opted for the minimal breaking 
scheme where only $\mathbb{Z}_2^{\mu\tau}$ is broken to generate a nonzero reactor mixing angle $\theta_{13}$. We started with
the Type-I seesaw mechanism which contains the Dirac type $m_D$ and Majorana type $M_R$ as the constituent matrices. 
In the diagonal basis of the charged lepton as well as the RH neutrino mass matrix $M_R$, the implemented residual 
TBM-Klein symmetry leads to degenerate RH neutrino masses. The $\mathbb{Z}_2^{\mu\tau}$ is then broken in $M_R$ to lift 
the mass degeneracy as well as to generate nonvanishing value of $\theta_{13}$. Thus the observed small  
value of $\theta_{13}$  restricts the level of degeneracy in the RH neutrino masses. Phenomenologically allowed case 
in our analysis gives rise to a TM1 type mixing and predicts a normal mass ordering for the light neutrinos. 
Testable predictions on the Dirac CP phase $\delta$ and the neutrinoless double beta decay parameter $|(M_\nu)_{11}|$ have
also been obtained. Our analysis is also interesting from leptogenesis perspective. Unlike the standard hierarchical
$N_1$-leptogenesis scenario, here due to the implemented symmetry and the phenomenologically viable breaking pattern 
of that symmetry, the baryogenesis via leptogenesis scenario is realized due to quasi degenerate RH neutrinos. 
It has been clarified by a brief mathematical calculation that other two RH neutrinos $(N_2,N_3)$ have sizeable
contribution in generating lepton asymmetry.
For  computation of the final baryon asymmetry we make use of the flavour dependent coupled Boltzmann Equations to 
track the evolution of the produced lepton asymmetry down to the low temperature scale. Only $\tau$-flavoured leptogenesis 
scheme  is allowed in our analysis. Consistent with the observed range of $Y_B$ a lower and an upper bound on the RH 
neutrino masses have  also been obtained. We also estimate the   testable flux ratios of three UHE neutrino flavours (detected at Icecube).
At the end we elucidate the novelty and importance of this present work through a comparative study with the existing literature.
\section*{Acknowledgement}
RS would like to thank Prof. Pasquale Di Bari for a very useful discussion regarding leptogenesis with quasi degenerate neutrinos. RS would like to thank the Royal Society (UK) and SERB (India) for the Newton International Fellowship (NIF). For financial support from Siksha ’O’ Anusandhan (SOA), Deemed to be University, M. C. acknowledges a Post-Doctoral fellowship.
\appendix
\section{Appendix}
\subsection{Explicit algebraic forms of elements of $M_{\nu 1}^{G_{1}^{TBM}}$}\label{appen1}
Since $M_{\nu 1}^{G_{1}^{TBM}}$ is symmetric matrix, it has only six independent complex parameters (namely
$f_{11},f_{12},f_{13},f_{22},f_{23},f_{33}$) each of which contains a common factor $z$ in the denominator given by
\begin{equation}
z=\left(\epsilon _4'+\epsilon _6'-2\right) \left(\left(\epsilon _4'\right){}^2+2 \left(7 \epsilon _6'+6\right) \epsilon _4'+\epsilon _6' \left(\epsilon _6'+12\right)+8\right).
\end{equation}
The explicit functional forms of the six independent elements of the $M_{\nu 1}^{G_{1}^{TBM}}$ are as follows
\small{
\begin{eqnarray}
f_{11} &=&(1/z)[-8 a' e^{i \beta '} \left(\left(\epsilon _4'\right){}^2+2 \left(5 \epsilon _6'+4\right)
\epsilon _4'+\epsilon _6' \left(\epsilon _6'+8\right)+4\right) \left(b'-c'\right)\nonumber\\
&+&
4\left(a'\right)^2 e^{2 i \beta '} \left(\left(\epsilon _4'\right){}^2+2 \left(9 \epsilon _6'+
8\right) \epsilon _4'+\epsilon _6' \left(\epsilon _6'+16\right)+12\right)\nonumber \\ &+&16 \left(\epsilon _4'+1\right)
   \left(\epsilon _6'+1\right) \left(b'-c'\right)^2] \\\nonumber\\
f_{12}&=&(1/z)[ -a' e^{i \beta '} (b' \left(\left(\epsilon _4'\right){}^2+2 \left(8 \epsilon _6'+7\right) 
\epsilon _4'-\left(\epsilon_6'-18\right) \epsilon _6'+16\right)\nonumber\\ &+& c' (\left(\epsilon _4'\right)^2-  
2 \left(8 \epsilon _6'+9\right) \epsilon_4'-\epsilon _6' \left(\epsilon _6'+14\right)-16))\nonumber\\
&+&2 \left(a'\right)^2 e^{2 i \beta '} \left(\left(\epsilon_4'\right){}^2+2 \left(5 \epsilon _6'+4\right) \epsilon _4'+
\epsilon _6' \left(\epsilon _6'+8\right)+4\right)\nonumber\\ &+& 2\left(b'-c'\right) \left(b' \left(\epsilon _6'+1\right) 
\left(3 \epsilon _4'+\epsilon _6'\right)-c' \left(\epsilon _4'+1\right)
   \left(\epsilon _4'+3 \epsilon _6'\right)\right)]\\\nonumber\\
 f_{13}&=&(1/z)[a' e^{i \beta '} (b' ((\epsilon _4'){}^2-2 (8 \epsilon _6'+9)
 \epsilon _4'-\epsilon _6'(\epsilon _6'+14)-16)\nonumber\\ &+& c' ((\epsilon _4'){}^2
 2(8 \epsilon _6'+7) \epsilon _4'-(\epsilon _6'-18) \epsilon _6'+16))\nonumber\\
&+& 2 \left(a'\right)^2 e^{2 i \beta '} \left(\left(\epsilon_4'\right){}^2+2 \left(5 \epsilon _6'+4\right) 
 \epsilon _4'+\epsilon _6' \left(\epsilon _6'+8\right)+4\right)\nonumber\\ &+& 2\left(b'-c'\right) 
 \left(b' \left(\epsilon _4'+1\right) \left(\epsilon _4'+3 \epsilon _6'\right)-c' \left(\epsilon _6'+1\right)
 \left(3 \epsilon _4'+\epsilon _6'\right)\right)]\\\nonumber\\
 f_{22}&=&(1/z)[-8 a' e^{i \beta '} \left(b' \left(\epsilon _6'+1\right) \left(3 \epsilon _4'+\epsilon _6'\right)-
 c' \left(\epsilon _4'+1\right)\left(\epsilon _4'+3 \epsilon _6'\right)\right)\nonumber \\ &+&16 \left(a'\right)^2 e^{2 i \beta '} 
 \left(\epsilon _4'+1\right) \left(\epsilon_6'+1\right)-2 b' c' \left(3 \epsilon _4'+\epsilon _6'\right)\nonumber\\ 
&& \left(\epsilon _4'+3 \epsilon _6'\right)-\left(b'\right)^2\left(\left(\epsilon _4'+3 \epsilon _6'\right){}^2-
 16 \left(\epsilon _6'+1\right)\right)\nonumber\\ &-&\left(c'\right)^2 \left(3 \epsilon
   _4'+\epsilon _6'\right){}^2+16 \left(c'\right)^2 \left(\epsilon _4'+1\right)]\\\nonumber\\
 f_{23}&=&(1/z)[4 a' e^{i \beta '} \left(\left(\epsilon _4'\right){}^2+\left(6 \epsilon _6'+4\right) \epsilon _4'+
 \epsilon _6' \left(\epsilon_6'+4\right)\right) \left(b'-c'\right)\nonumber \\ &-& 16 \left(a'\right)^2 e^{2 i \beta '} 
 \left(\epsilon _4'+1\right) \left(\epsilon_6'+1\right)- \nonumber\\
&& 2 b' c' \left(5 \left(\epsilon _4'\right){}^2+ 
\left(6 \epsilon _6'-8\right) \epsilon _4'+5 \left(\epsilon_6'\right){}^2-8 \epsilon _6'-16\right)\nonumber \\ &-&
 \left(b'\right)^2 \left(3 \epsilon _4'+\epsilon _6'\right) \left(\epsilon _4'+3
 \epsilon _6'\right)-\left(c'\right)^2 \left(3 \epsilon _4'+\epsilon _6'\right) \left(\epsilon _4'+3 \epsilon _6'\right)]\\\nonumber\\
f_{33} &=&(1/z)[-8 a' e^{i \beta '} \left(b' \left(\epsilon _4'+1\right) \left(\epsilon _4'+3 \epsilon _6'\right)-
c' \left(\epsilon _6'+1\right)\left(3 \epsilon _4'+\epsilon _6'\right)\right)\nonumber\\ &+& 16 \left(a'\right)^2 e^{2 i \beta '} 
\left(\epsilon _4'+1\right) \left(\epsilon _6'+1\right)-2 b' c' \left(3 \epsilon _4'+\epsilon _6'\right)\nonumber\\ 
&&\left(\epsilon _4'+3 \epsilon _6'\right)-\left(b'\right)^2\left(9 \left(\epsilon _4'\right){}^2+
2 \left(3 \epsilon _6'-8\right) \epsilon _4'+\left(\epsilon_6'\right){}^2-16\right)\nonumber \\ &-&\left(c'\right)^2 
\left(\left(\epsilon _4'+3 \epsilon _6'\right){}^2-16 \left(\epsilon_6'+1\right)\right)].
\end{eqnarray}
}
\subsection{Explicit algebraic forms of elements of $M_{\nu 1}^{G_{2}^{TBM}}$}\label{appen2}
The elements of the matrix $M_{\nu 2}^{G_1^{TBM}}$ can be parametrized as
\bea
(M_{\nu 2}^{G_1^{TBM}})_{11}&=&-p^2-\frac{r^2+q^2-2qr}{4}e^{i\theta}\left(\frac{2+\epsilon_4^\prime+\epsilon_6^\prime}{(1+\epsilon_6^\prime)(1+\epsilon_4^\prime)}\right),\nonumber\\
(M_{\nu 2}^{G_1^{TBM}})_{12}&=&-p^2-e^{i\theta}\left(\frac{r^2}{2(1+\epsilon_6^\prime)}+\frac{q^2}{2(1+\epsilon_4^\prime)}-\frac{qr(2+\epsilon_4^\prime+\epsilon_6^\prime)}{2(1+\epsilon_6^\prime)(1+\epsilon_4^\prime)}\right),\nonumber\\
\nonumber
\eea
\bea
(M_{\nu 2}^{G_1^{TBM}})_{13}&=&p^2+e^{i\theta}\left(\frac{q^2}{2(1+\epsilon_6^\prime)}+\frac{r^2}{2(1+\epsilon_4^\prime)}-\frac{qr(2+\epsilon_4^\prime+\epsilon_6^\prime)}{2(1+\epsilon_6^\prime)(1+\epsilon_4^\prime)}\right),\nonumber\\
(M_{\nu 2}^{G_1^{TBM}})_{22}&=&-p^2-e^{i\theta}\left(\frac{q^2}{(1+\epsilon_4^\prime)}+\frac{r^2}{(1+\epsilon_6^\prime)}\right),\nonumber\\
(M_{\nu 2}^{G_1^{TBM}})_{23}&=&p^2-e^{i\theta}qr\left(\frac{1}{(1+\epsilon_4^\prime)}+\frac{1}{(1+\epsilon_6^\prime)}\right),\nonumber\\
(M_{\nu 2}^{G_1^{TBM}})_{33}&=&-p^2-e^{i\theta}\left(\frac{q^2}{(1+\epsilon_6^\prime)}+\frac{r^2}{(1+\epsilon_4^\prime)}\right),\label{m2g1}
\eea
where we define the parameters in Eq.(\ref{m2g1}) as
\bea
\frac{a}{\sqrt{x}}\rightarrow p,~\frac{b}{\sqrt{y}}\rightarrow q e^{i \theta/2},~\frac{c}{\sqrt{y}}\rightarrow r e^{i \theta/2},\epsilon_4^\prime\rightarrow\frac{\epsilon_4}{y},\epsilon_6^\prime\rightarrow\frac{\epsilon_6}{y}
\eea
with $p$, $q$, $r$,$\epsilon_4$, $\epsilon_6$ being real.\\

{}

\begin{thebibliography}{}
\bibitem{An:2012eh} 
  F.~P.~An {\it et al.} [Daya Bay Collaboration],
  Phys.\ Rev.\ Lett.\  {\bf 108}, 171803 (2012)
 [\xlink{1203.1669} [hep-ph]].  
\bibitem{Ahn:2012nd} 
  J.~K.~Ahn {\it et al.} [RENO Collaboration],
  Phys.\ Rev.\ Lett.\  {\bf 108}, 191802 (2012)
  [\xlink{1204.0626} [hep-ph]].

\bibitem{Harrison:2002er} 
  P.~F.~Harrison, D.~H.~Perkins and W.~G.~Scott,
  Phys.\ Lett.\ B {\bf 530}, 167 (2002)
  [\xlink{hep-ph/0202074} [hep-ph]]. P.~F.~Harrison and W.~G.~Scott,
  Phys.\ Lett.\ B {\bf 535}, 163 (2002)
  [\xlink{hep-ph/0203209} [hep-ph]].

\bibitem{Xing:2002sw} 
  Z.~z.~Xing,
  Phys.\ Lett.\ B {\bf 533}, 85 (2002)
 [\xlink{hep-ph/0204049} [hep-ph]]. X.~G.~He and A.~Zee,
  Phys.\ Lett.\ B {\bf 560}, 87 (2003)
  [\xlink{hep-ph/0301092} [hep-ph]].  E.~Ma,
  Phys.\ Rev.\ Lett.\  {\bf 90}, 221802 (2003)
   [\xlink{hep-ph/0303126} [hep-ph]].C.~I.~Low and R.~R.~Volkas,
  Phys.\ Rev.\ D {\bf 68}, 033007 (2003)
   [\xlink{hep-ph/0305243} [hep-ph]]. E.~Ma,
  Phys.\ Rev.\ D {\bf 70}, 031901 (2004)
  [\xlink{hep-ph/0404199} [hep-ph]].
   G.~Altarelli and F.~Feruglio,
  Nucl.\ Phys.\ B {\bf 720}, 64 (2005)
 [\xlink{hep-ph/0504165} [hep-ph]].
    S.~F.~King,
  JHEP {\bf 0508}, 105 (2005)
  [\xlink{hep-ph/0506297} [hep-ph]].
   E.~Ma,
  Phys.\ Rev.\ D {\bf 73}, 057304 (2006)
  [\xlink{hep-ph/0511133} [hep-ph]].
  G.~Altarelli and F.~Feruglio,
  Nucl.\ Phys.\ B {\bf 741}, 215 (2006)
  [\xlink{hep-ph/0512103} [hep-ph]].
  E.~Ma,
  Mod.\ Phys.\ Lett.\ A {\bf 22}, 101 (2007)
  [\xlink{hep-ph/0610342} [hep-ph]]. 
 K.~S.~Babu and X.~G.~He,
  hep-ph/0507217. S.~Pakvasa, W.~Rodejohann and T.~J.~Weiler,
  Phys.\ Rev.\ Lett.\  {\bf 100}, 111801 (2008)
  [\xlink{0711.0052} [hep-ph]]. 
  G.~Altarelli, F.~Feruglio and L.~Merlo,
  JHEP {\bf 0905}, 020 (2009)
  [\xlink{0903.1940} [hep-ph]]. 
  E.~Ma and D.~Wegman,
  Phys.\ Rev.\ Lett.\  {\bf 107}, 061803 (2011)
 [\xlink{1106.4269} [hep-ph]].  R.~de Adelhart Toorop, F.~Feruglio and C.~Hagedorn,
  Nucl.\ Phys.\ B {\bf 858}, 437 (2012)
  [\xlink{1112.1340} [hep-ph]].
 E.~Ma,
  Phys.\ Lett.\ B {\bf 583}, 157 (2004)
 [\xlink{hep-ph/0308282} [hep-ph]]. 
  \bibitem{Ishimori:2010au} 
  G.~Altarelli and F.~Feruglio,
  Rev.\ Mod.\ Phys.\  {\bf 82}, 2701 (2010)
  doi:10.1103/RevModPhys.82.2701
  [\xlink{1002.0211} [hep-ph]].
   H.~Ishimori, T.~Kobayashi, H.~Ohki, Y.~Shimizu, H.~Okada and M.~Tanimoto,
  Prog.\ Theor.\ Phys.\ Suppl.\  {\bf 183}, 1 (2010)
  doi:10.1143/PTPS.183.1
  [\xlink{1003.3552} [hep-ph]].
  S.~F.~King,
  Prog.\ Part.\ Nucl.\ Phys.\  {\bf 94}, 217 (2017)
  doi:10.1016/j.ppnp.2017.01.003
  [\xlink{1701.04413} [hep-ph]].
   S.~T.~Petcov
  [\xlink{1711.10806}[hep-ph]].

\bibitem{He:2006qd} 
 F.~Plentinger and W.~Rodejohann,
  Phys.\ Lett.\ B {\bf 625}, 264 (2005)
 [\xlink{hep-ph/0507143} [hep-ph]].  
  X.~G.~He and A.~Zee,
  Phys.\ Lett.\ B {\bf 645}, 427 (2007)
 [\xlink{hep-ph/0607163} [hep-ph]].  
 Z.~z.~Xing, H.~Zhang and S.~Zhou,
  Phys.\ Lett.\ B {\bf 641}, 189 (2006)
  [\xlink{hep-ph/0607091} [hep-ph]].  R.~N.~Mohapatra and H.~B.~Yu,
  Phys.\ Lett.\ B {\bf 644}, 346 (2007)
  [\xlink{hep-ph/0610023} [hep-ph]]. M.~Hirsch, E.~Ma, J.~C.~Romao, J.~W.~F.~Valle and A.~Villanova del Moral,
  Phys.\ Rev.\ D {\bf 75}, 053006 (2007)
  [\xlink{hep-ph/0606082} [hep-ph].  K.~A.~Hochmuth, S.~T.~Petcov and W.~Rodejohann,
  Phys.\ Lett.\ B {\bf 654}, 177 (2007)
 [\xlink{0706.2975} [hep-ph]. 
 Y.~Koide and H.~Nishiura,
  Phys.\ Lett.\ B {\bf 669}, 24 (2008)
 [\xlink{0808.0370} [hep-ph].  S.~F.~King,
  Phys.\ Lett.\ B {\bf 659}, 244 (2008)
 [\xlink{0710.0530} [hep-ph]]. 
 C.~H.~Albright, A.~Dueck and W.~Rodejohann,
  Eur.\ Phys.\ J.\ C {\bf 70}, 1099 (2010)
  [\xlink{1004.2798} [hep-ph]].X.~G.~He and A.~Zee,
  Phys.\ Rev.\ D {\bf 84}, 053004 (2011). R.~Samanta, M.~Chakraborty and A.~Ghosal,
  Nucl.\ Phys.\ B {\bf 904}, 86 (2016)
 [\xlink{1502.06508} [hep-ph]]. R.~Samanta and M.~Chakraborty,
 [\xlink{1703.09579} [hep-ph]]. 

  \bibitem{models}  
 B.~Adhikary, B.~Brahmachari, A.~Ghosal, E.~Ma and M.~K.~Parida,
  Phys.\ Lett.\ B {\bf 638}, 345 (2006).  B.~Adhikary and A.~Ghosal,
  Phys.\ Rev.\ D {\bf 78}, 073007 (2008)
 [\xlink{0803.3582} [hep-ph]]. 
 J.~Barry and W.~Rodejohann,
  Phys.\ Rev.\ D {\bf 81}, 093002 (2010)
  Erratum: [Phys.\ Rev.\ D {\bf 81}, 119901 (2010)]
 [\xlink{1003.2385} [hep-ph]]. 
  Y.~F.~Li and Q.~Y.~Liu,
  Mod.\ Phys.\ Lett.\ A {\bf 25}, 63 (2010)
  [\xlink{0911.2670} [hep-ph]]. 
 Y.~Lin,
  Nucl.\ Phys.\ B {\bf 824}, 95 (2010)
  [\xlink{0905.3534} [hep-ph]].
 [\xlink{1106.4359} [hep-ph]]. S.~F.~King and C.~Luhn,
  JHEP {\bf 1203}, 036 (2012)
 [\xlink{1112.1959} [hep-ph]].  W.~Rodejohann and H.~Zhang,
  Phys.\ Rev.\ D {\bf 86}, 093008 (2012)
  [\xlink{1207.1225} [hep-ph]]. S.~K.~Garg and S.~Gupta,
  JHEP {\bf 1310}, 128 (2013)
  [\xlink{1308.3054} [hep-ph]]. V.~V.~Vien and H.~N.~Long,
  Int.\ J.\ Mod.\ Phys.\ A {\bf 28}, 1350159 (2013)
 [\xlink{1312.5034} [hep-ph]]. 
   Z.~h.~Zhao,
  JHEP {\bf 1411}, 143 (2014)
  doi:10.1007/JHEP11(2014)143
   [\xlink{1405.3022} [hep-ph]]. Z.~z.~Xing and Z.~h.~Zhao,
  Rept.\ Prog.\ Phys.\  {\bf 79}, no. 7, 076201 (2016)
  [\xlink{1512.04207} [hep-ph]]. 
   J.~Talbert,
  JHEP {\bf 1412}, 058 (2014)
 [\xlink{1409.7310} [hep-ph]]. 
R.~Samanta and A.~Ghosal,
  Nucl.\ Phys.\ B {\bf 911}, 846 (2016)
  [\xlink{1507.02582} [hep-ph]]. 
   J.~Zhang and S.~Zhou,
  JHEP {\bf 1609}, 167 (2016)
  [\xlink{1606.09591} [hep-ph]].  S.~K.~Garg,
 [\xlink{1712.02212} [hep-ph]]. 
\bibitem{Albright:2008rp} 
  C.~H.~Albright and W.~Rodejohann,
  Eur.\ Phys.\ J.\ C {\bf 62}, 599 (2009)
 [\xlink{0812.0436} [hep-ph]]. 
\bibitem{Xing:2006ms} 
  Z.~z.~Xing and S.~Zhou,
  Phys.\ Lett.\ B {\bf 653}, 278 (2007)
  [\xlink{hep-ph/0607302} [hep-ph]].
\bibitem{Grimus:2008tt} 
  W.~Grimus and L.~Lavoura,
  JHEP {\bf 0809}, 106 (2008)
  [\xlink{0809.0226} [hep-ph]].
    W.~Grimus and L.~Lavoura,
  Phys.\ Lett.\ B {\bf 671}, 456 (2009)
  [\xlink{0810.4516} [hep-ph]].
\bibitem{tm1} 
 I.~de Medeiros Varzielas and L.~Lavoura,
  J.\ Phys.\ G {\bf 40}, 085002 (2013)
  [\xlink{1212.3247} [hep-ph]].
  Nucl.\ Phys.\ B {\bf 875}, 80 (2013)
  [\xlink{1306.2358} [hep-ph]].
   C.~C.~Li and G.~J.~Ding,
  Nucl.\ Phys.\ B {\bf 881}, 206 (2014)
  [\xlink{1312.4401} [hep-ph]].
\bibitem{tm2} 
  C.~Luhn, S.~Nasri and P.~Ramond,
  J.\ Math.\ Phys.\  {\bf 48}, 073501 (2007)
  doi:10.1063/1.2734865
  [\xlink{hep-th/0701188} [hep-ph]].
    J.~A.~Escobar and C.~Luhn,
  J.\ Math.\ Phys.\  {\bf 50}, 013524 (2009)
   [\xlink{0809.0639} [hep-ph]].
  \bibitem{Lam:2007qc} 
  C.~S.~Lam,
  Phys.\ Lett.\ B {\bf 656}, 193 (2007)
  [\xlink{0708.3665} [hep-ph]].
\bibitem{Lam:2008rs} 
  C.~S.~Lam,
  Phys.\ Rev.\ Lett.\  {\bf 101}, 121602 (2008),
  [\xlink{0804.2622} [hep-ph]].
\bibitem{Lam:2008sh} 
  C.~S.~Lam,
  Phys.\ Rev.\ D {\bf 78}, 073015 (2008).
  [\xlink{0809.1185} [hep-ph]].
  \bibitem{Ge:2011qn} 
   D.~A.~Dicus, S.~F.~Ge and W.~W.~Repko,
  Phys.\ Rev.\ D {\bf 83}, 093007 (2011)
  [\xlink{1012.2571} [hep-ph]]. 
S.~F.~Ge, D.~A.~Dicus and W.~W.~Repko,
  Phys.\ Lett.\ B {\bf 702}, 220 (2011)
  [\xlink{1104.0602} [hep-ph]].
   S.~F.~Ge, D.~A.~Dicus and W.~W.~Repko,
  Phys.\ Rev.\ Lett.\  {\bf 108}, 041801 (2012)
  [\xlink{1108.0964} [hep-ph]].
   A.~D.~Hanlon, S.~F.~Ge and W.~W.~Repko,
  Phys.\ Lett.\ B {\bf 729}, 185 (2014)
  doi:10.1016/j.physletb.2013.12.063
  [\xlink{1308.6522} [hep-ph]].
  \bibitem{He:2011kn} 
  H.~J.~He and F.~R.~Yin,
  Phys.\ Rev.\ D {\bf 84}, 033009 (2011)
 [\xlink{1104.2654} [hep-ph]]. 
 H.~J.~He and X.~J.~Xu,
  Phys.\ Rev.\ D {\bf 86}, 111301 (2012)
  doi:10.1103/PhysRevD.86.111301
  [\xlink{1203.2908} [hep-ph]].
 \bibitem{mutaus} R.N Mohapatra and S. Nussinov, Phys. Rev. {\bf D60}(1999)013002.  T.~Fukuyama and H.~Nishiura,
  [\xlink{hep-ph/9702253} [hep-ph]].
 C.~S.~Lam,
  Phys.\ Lett.\ B {\bf 507}, 214 (2001)
  [\xlink{0104116} [hep-ph]]. 
  E.~Ma and M.~Raidal,
  Phys.\ Rev.\ Lett.\  {\bf 87}, 011802 (2001)
  Erratum: [Phys.\ Rev.\ Lett.\  {\bf 87}, 159901 (2001)]
[\xlink{0102255} [hep-ph]]. 
  K.~R.~S.~Balaji, W.~Grimus and T.~Schwetz,
  Phys.\ Lett.\ B {\bf 508}, 301 (2001)
   [\xlink{0104035} [hep-ph]]. 
     A.~Ghosal,
 [\xlink{hep-ph/0304090.} [hep-ph]]. A.~Ghosal,
  Mod.\ Phys.\ Lett.\ A {\bf 19}, 2579 (2004).
  J.~C.~Gomez-Izquierdo and A.~Perez-Lorenzana,
  Phys.\ Rev.\ D {\bf 77}, 113015 (2008)
  [\xlink{0711.0045} [hep-ph]].
  J.~C.~Gomez-Izquierdo and A.~Perez-Lorenzana,
  Phys.\ Rev.\ D {\bf 82}, 033008 (2010)
  [\xlink{0912.5210} [hep-ph]].
 J.~C.~Gómez-Izquierdo, F.~Gonzalez-Canales and M.~Mondragón,
  Int.\ J.\ Mod.\ Phys.\ A {\bf 32}, no. 28-29, 1750171 (2017)
  [\xlink{1705.06324} [hep-ph]].
 Z.~z.~Xing and Z.~h.~Zhao,
  Rept.\ Prog.\ Phys.\  {\bf 79}, no. 7, 076201 (2016)
 [\xlink{1512.04207} [hep-ph]].
  \bibitem{CP1} 
  R.~Samanta, P.~Roy and A.~Ghosal,
  Eur.\ Phys.\ J.\ C {\bf 76}, no. 12, 662 (2016)
   [\xlink{1604.06731} [hep-ph]].   R.~Samanta, P.~Roy and A.~Ghosal,
  Acta Phys.\ Polon.\ Supp.\  {\bf 9}, 807 (2016)
[\xlink{1604.01206} [hep-ph]]. 
R.~Samanta, M.~Chakraborty, P.~Roy and A.~Ghosal,
  JCAP {\bf 1703}, no. 03, 025 (2017)
  doi:10.1088/1475-7516/2017/03/025
 [\xlink{1610.10081} [hep-ph]].  
 R.~Sinha, R.~Samanta and A.~Ghosal,
  JHEP {\bf 1712}, 030 (2017)
  [\xlink{1706.00946} [hep-ph]]
  \bibitem{mutau}
 P.~F.~Harrison and W.~G.~Scott,
  Phys.\ Lett.\ B {\bf 547}, 219 (2002)
  [\xlink{0210197} [hep-ph]].
   W. Grimus and L. Lavoura,   Phys.\ Lett.\ B {\bf 579}, 113 (2004)
  [\xlink{0305309} [hep-ph]].
  Fortsch.\ Phys.\  {\bf 61}, 535 (2013)
  [\xlink{1207.1678} [hep-ph]].
   R.~N.~Mohapatra and C.~C.~Nishi,
  JHEP {\bf 1508}, 092 (2015)
  [\xlink{1506.06788} [hep-ph]]. Also see: G. Ecker, W. Grimus, H. Neufeld, J.Phys. A20, L807 (1987); Int.J.Mod.Phys. A3, 603 (1988).
  W.~Grimus and M.~N.~Rebelo,
  Phys.\ Rept.\  {\bf 281}, 239 (1997)
  [\xlink{9506272} [hep-ph]].
  \bibitem{Ma:2015fpa} 
  E.~Ma,
  Phys.\ Lett.\ B {\bf 752}, 198 (2016)
   [\xlink{1510.02501} [hep-ph]].
\bibitem{CPt}
 R.~N.~Mohapatra and C.~C.~Nishi,
  Phys.\ Rev.\ D {\bf 86}, 073007 (2012)
  doi:10.1103/PhysRevD.86.073007
 [\xlink{1208.2875} [hep-ph]].
F.~Feruglio, C.~Hagedorn and R.~Ziegler,
  JHEP {\bf 1307}, 027 (2013)
  [\xlink{1211.5560} [hep-ph]].
  M.~Holthausen, M.~Lindner and M.~A.~Schmidt,
  JHEP {\bf 1304}, 122 (2013)
  [\xlink{1211.6953} [hep-ph]].
   M.~C.~Chen, M.~Fallbacher, K.~T.~Mahanthappa, M.~Ratz and A.~Trautner,
  Nucl.\ Phys.\ B {\bf 883}, 267 (2014)
   [\xlink{1402.0507} [hep-ph]].
    G.~J.~Ding, S.~F.~King, C.~Luhn and A.~J.~Stuart,
  JHEP {\bf 1305}, 084 (2013)
 [\xlink{1303.6180} [hep-ph]]. 
  G.~J.~Ding, S.~F.~King and A.~J.~Stuart,
  JHEP {\bf 1312}, 006 (2013)
 [\xlink{1307.4212} [hep-ph]].  
 F.~Feruglio, C.~Hagedorn and R.~Ziegler,
  Eur.\ Phys.\ J.\ C {\bf 74}, 2753 (2014)
  doi:10.1140/epjc/s10052-014-2753-2
  [\xlink{1303.7178} [hep-ph]].
P.~Chen, C.~Y.~Yao and G.~J.~Ding,
  Phys.\ Rev.\ D {\bf 92}, no. 7, 073002 (2015)
  [\xlink{1507.03419} [hep-ph]].
   H.~J.~He, W.~Rodejohann and X.~J.~Xu,
  Phys.\ Lett.\ B {\bf 751}, 586 (2015)
  [\xlink{1507.03541} [hep-ph]].
C.~C.~Nishi,
  Phys.\ Rev.\ D {\bf 93}, no. 9, 093009 (2016)
 [\xlink{1601.00977} [hep-ph]].
   C.~C.~Nishi and B.~L.~Sánchez-Vega,
  JHEP {\bf 1701}, 068 (2017)
  [\xlink{1611.08282} [hep-ph]].
   J.~T.~Penedo, S.~T.~Petcov and A.~V.~Titov,
  JHEP {\bf 1712}, 022 (2017)
 [\xlink{1705.00309} [hep-ph]]. R.~Samanta, P.~Roy and A.~Ghosal,
 [\xlink{1712.06555} [hep-ph]].  N.~Nath, Z.~z.~Xing and J.~Zhang,
 [\xlink{1801.09931} [hep-ph]].
A comprehensive review : S.~F.~King,
  Prog.\ Part.\ Nucl.\ Phys.\  {\bf 94}, 217 (2017)
  [\xlink{1701.04413} [hep-ph]].
  \bibitem{Abe:2017uxa} 
  K.~Abe {\it et al.} [T2K Collaboration],
  Phys.\ Rev.\ Lett.\  {\bf 118}, no. 15, 151801 (2017)
  [\xlink{1701.00432}[hep-ex]].
\bibitem{joshi}
 S.~Gupta, A.~S.~Joshipura and K.~M.~Patel,
  Phys.\ Rev.\ D {\bf 85}, 031903 (2012)
 [\xlink{1112.6113} [hep-ph]].
  \bibitem{rode}
  W.~Rodejohann and X.~J.~Xu,
  Phys.\ Rev.\ D {\bf 96}, no. 5, 055039 (2017)
 [\xlink{1705.02027} [hep-ph]].
  \bibitem{Capozzi:2017ipn} 
  F.~Capozzi, E.~Di Valentino, E.~Lisi, A.~Marrone, A.~Melchiorri and A.~Palazzo,
  Phys.\ Rev.\ D {\bf 95}, no. 9, 096014 (2017)
  doi:10.1103/PhysRevD.95.096014
[\xlink{1703.04471} [hep-ph]]. 
 \bibitem{Adamson:2017qqn} 
  P.~Adamson {\it et al.} [NOvA Collaboration],
  Phys.\ Rev.\ Lett.\  {\bf 118}, no. 15, 151802 (2017)
  doi:10.1103/PhysRevLett.118.151802
  [\xlink{1701.05891} [hep-ph]].
 \bibitem{Agashe:2014kda} 
  K.~A.~Olive {\it et al.} [Particle Data Group],
  Chin.\ Phys.\ C {\bf 38}, 090001 (2014).
  \bibitem{Sakharov:1967dj} 
  A.~D.~Sakharov,
  Pisma Zh.\ Eksp.\ Teor.\ Fiz.\  {\bf 5}, 32 (1967)
  [JETP Lett.\  {\bf 5}, 24 (1967)]
  [Sov.\ Phys.\ Usp.\  {\bf 34}, 392 (1991)]
  [Usp.\ Fiz.\ Nauk {\bf 161}, 61 (1991)].
  \bibitem{Aartsen:2013bka} 
  M.~G.~Aartsen {\it et al.} [IceCube Collaboration],
  Phys.\ Rev.\ Lett.\  {\bf 111}, 021103 (2013)
  doi:10.1103/PhysRevLett.111.021103
  [arXiv:1304.5356 [astro-ph.HE]].
  M.~G.~Aartsen {\it et al.} [IceCube Collaboration],
  Science {\bf 342}, 1242856 (2013)
  doi:10.1126/science.1242856
  [arXiv:1311.5238 [astro-ph.HE]].
   M.~G.~Aartsen {\it et al.} [IceCube Collaboration],
  Phys.\ Rev.\ Lett.\  {\bf 113}, 101101 (2014)
  doi:10.1103/PhysRevLett.113.101101
  [arXiv:1405.5303 [astro-ph.HE]].
  M.~G.~Aartsen {\it et al.} [IceCube Collaboration],
  arXiv:1510.05223 [astro-ph.HE].
   M.~G.~Aartsen {\it et al.} [IceCube Collaboration],
  arXiv:1710.01191 [astro-ph.HE].
  \bibitem{Aartsen:2016oji} 
  M.~G.~Aartsen {\it et al.} [IceCube Collaboration],
  Astrophys.\ J.\  {\bf 835}, no. 2, 151 (2017)
  doi:10.3847/1538-4357/835/2/151
  [arXiv:1609.04981 [astro-ph.HE]].
  arXiv:1710.01179 [astro-ph.HE].
  \bibitem{Mohapatra:2004hta} 
  R.~N.~Mohapatra and S.~Nasri,
  Phys.\ Rev.\ D {\bf 71}, 033001 (2005)
  doi:10.1103/PhysRevD.71.033001
  [hep-ph/0410369].
  R.~N.~Mohapatra, S.~Nasri and H.~B.~Yu,
  Phys.\ Lett.\ B {\bf 615}, 231 (2005)
  doi:10.1016/j.physletb.2005.03.082
  [hep-ph/0502026].
  \bibitem{Ge}
  S.~F.~Ge, H.~J.~He and F.~R.~Yin,
  JCAP {\bf 1005}, 017 (2010)
  [\xlink{1001.0940} [hep-ph]]. 
  \bibitem{Sui:2018bbh} 
  Y.~Sui and P.~S.~Bhupal Dev,
  JCAP {\bf 1807}, no. 07, 020 (2018)
  doi:10.1088/1475-7516/2018/07/020
  [arXiv:1804.04919 [hep-ph]].
  \bibitem{An:2015rpe} 
  F.~P.~An {\it et al.} [Daya Bay Collaboration],
  Phys.\ Rev.\ Lett.\  {\bf 115}, no. 11, 111802 (2015)
 [\xlink{1505.03456} [hep-ph]].
\bibitem{Karmakar:2014dva} 
  B.~Karmakar and A.~Sil,
  Phys.\ Rev.\ D {\bf 91}, 013004 (2015)
  [\xlink{1407.5826} [hep-ph]].
\bibitem{Ade:2013zuv} 
  P.~A.~R.~Ade {\it et al.} [Planck Collaboration],
  Astron.\ Astrophys.\  {\bf 571}, A16 (2014)
  doi:10.1051/0004-6361/201321591
  [\xlink{arXiv:1303.5076} [astro-ph.CO]]. 
\bibitem{Ade:2015xua} 
  P.~A.~R.~Ade {\it et al.} [Planck Collaboration],
  Astron.\ Astrophys.\  {\bf 594}, A13 (2016)
  doi:10.1051/0004-6361/201525830
  [\xlink{1502.01589}[astro-ph.CO]].
  N.~Aghanim {\it et al.} [Planck Collaboration],
  Astron.\ Astrophys.\  {\bf 596}, A107 (2016)
  doi:10.1051/0004-6361/201628890
  \xlink{arXiv:1605.02985} [astro-ph.CO]].
  \bibitem{Fukugita:1986hr} 
  M.~Fukugita and T.~Yanagida,
  Phys.\ Lett.\ B {\bf 174}, 45 (1986).
A.~Riotto and M.~Trodden,
  Ann.\ Rev.\ Nucl.\ Part.\ Sci.\  {\bf 49}, 35 (1999)
   [\xlink{hep-ph/9901362}].
S.~Davidson, E.~Nardi and Y.~Nir,
  Phys.\ Rept.\  {\bf 466}, 105 (2008)
  doi:10.1016/j.physrep.2008.06.002
 [\xlink{0802.2962} [hep-ph]].
\bibitem{Kolb:1990vq} 
  E.~W.~Kolb and M.~S.~Turner,
  Front.\ Phys.\  {\bf 69}, 1 (1990).
  \bibitem{Pilaftsis:2003gt} 
  A.~Pilaftsis and T.~E.~J.~Underwood,
  Nucl.\ Phys.\ B {\bf 692}, 303 (2004)
  [\xlink{hep-ph/0309342} ].

  \bibitem{Abada:2006ea} 
  A.~Abada, S.~Davidson, A.~Ibarra, F.-X.~Josse-Michaux, M.~Losada and A.~Riotto,
  JHEP {\bf 0609}, 010 (2006)
  [\xlink{hep-ph/0605281}].
 E.~Nardi, Y.~Nir, E.~Roulet and J.~Racker,
  JHEP {\bf 0601}, 164 (2006)
  [hep-ph/0601084].
  P.~S.~B.~Dev, P.~Di Bari, B.~Garbrecht, S.~Lavignac, P.~Millington and D.~Teresi,
  [\xlink{1711.02861} [hep-ph]]. 
  \bibitem{Dev:2014laa} 
  P.~S.~Bhupal Dev, P.~Millington, A.~Pilaftsis and D.~Teresi,
  Nucl.\ Phys.\ B {\bf 886}, 569 (2014)
  doi:10.1016/j.nuclphysb.2014.06.020
  [arXiv:1404.1003 [hep-ph]].
  \bibitem{Dev:2017wwc} 
  B.~Dev, M.~Garny, J.~Klaric, P.~Millington and D.~Teresi,
  Int.\ J.\ Mod.\ Phys.\ A {\bf 33}, 1842003 (2018)
  doi:10.1142/S0217751X18420034
  [arXiv:1711.02863 [hep-ph]].
  \bibitem{Buchmuller:2003gz} 
  W.~Buchmuller, P.~Di Bari and M.~Plumacher,
  Nucl.\ Phys.\ B {\bf 665}, 445 (2003)
 [\xlink{hep-ph/0302092} [hep-ph]].

\bibitem{Edsjo:1997bg} 
  J.~Edsjo and P.~Gondolo,
  Phys.\ Rev.\ D {\bf 56}, 1879 (1997)
  [\xlink{hep-ph/9704361}].
\bibitem{Adhikary:2014qba} 
  B.~Adhikary, M.~Chakraborty and A.~Ghosal,
  Phys.\ Rev.\ D {\bf 93}, no. 11, 113001 (2016)
  [\xlink{1407.6173} [hep-ph]]. 
 R.~Samanta, M.~Chakraborty, P.~Roy and A.~Ghosal,
  JCAP {\bf 1703}, no. 03, 025 (2017)
 [\xlink{1610.10081} [hep-ph]].
   \bibitem{Dev:2017wwc} 
  B.~Dev, M.~Garny, J.~Klaric, P.~Millington and D.~Teresi,
  [\xlink{1711.02863}[hep-ph]].
\bibitem{Buchmuller:2004nz} 
  W.~Buchmuller, P.~Di Bari and M.~Plumacher,
  Annals Phys.\  {\bf 315}, 305 (2005)
  [\xlink{hep-ph/0401240}].
\bibitem{Adhikary:2013bma} 
  B.~Adhikary, M.~Chakraborty and A.~Ghosal,
  JHEP {\bf 1310}, 043 (2013)
  Erratum: [JHEP {\bf 1409}, 180 (2014)]
  [\xlink{1307.0988} [hep-ph]].
\bibitem{Asakura:2015ajs} 
  K.~Asakura {\it et al.} [KamLAND-Zen Collaboration],
  Nucl.\ Phys.\ A {\bf 946}, 171 (2016)
  [\xlink{1509.03724} [hep-ph]].

\bibitem{Auger:2012ar} 
  M.~Auger {\it et al.} [EXO-200 Collaboration],
  Phys.\ Rev.\ Lett.\  {\bf 109}, 032505 (2012)
  [\xlink{1205.5608} [hep-ph]].
\bibitem{Majorovits:2015vka} 
  B.~Majorovits [GERDA Collaboration],
  AIP Conf.\ Proc.\  {\bf 1672}, 110003 (2015)
   [\xlink{1506.00415} [hep-ph]].
  \bibitem{Abgrall:2013rze} 
  N.~Abgrall {\it et al.} [Majorana Collaboration],
  Adv.\ High Energy Phys.\  {\bf 2014}, 365432 (2014)
  [\xlink{1308.1633} [hep-ph]]. 
\bibitem{Agostini:2017jim} 
  M.~Agostini, G.~Benato and J.~Detwiler,
  Phys.\ Rev.\ D {\bf 96}, no. 5, 053001 (2017)
 [\xlink{1705.02996} [hep-ph]].
  \bibitem{nufit}\url{http://www.nu-fit.org/?q=node/166}
  \bibitem{Blanchet:2006dq} 
  S.~Blanchet and P.~Di Bari,
  JCAP {\bf 0606}, 023 (2006)
  doi:10.1088/1475-7516/2006/06/023
  [hep-ph/0603107].
  \bibitem{ice2018}
 \url{ http://npc.fnal.gov/wp-content/uploads/2018/09/180830_fermilab2.pdf.}
  \bibitem{Adrian-Martinez:2015ver} 
  S.~Adrian-Martinez {\it et al.} [ANTARES and IceCube Collaborations],
  Astrophys.\ J.\  {\bf 823}, no. 1, 65 (2016)
  doi:10.3847/0004-637X/823/1/65
  [arXiv:1511.02149 [hep-ex]].
  \bibitem{Becker:2007sv} 
  J.~K.~Becker,
  Phys.\ Rept.\  {\bf 458}, 173 (2008)
  doi:10.1016/j.physrep.2007.10.006
  [arXiv:0710.1557 [astro-ph]].
  \bibitem{Ahlers:2015lln} 
  M.~Ahlers and F.~Halzen,
  Rept.\ Prog.\ Phys.\  {\bf 78}, no. 12, 126901 (2015).
  doi:10.1088/0034-4885/78/12/126901
  \bibitem{Hummer:2010vx} 
  S.~Hummer, M.~Ruger, F.~Spanier and W.~Winter,
  Astrophys.\ J.\  {\bf 721}, 630 (2010)
  doi:10.1088/0004-637X/721/1/630
  [arXiv:1002.1310 [astro-ph.HE]].
  \bibitem{Learned:1994wg} 
  J.~G.~Learned and S.~Pakvasa,
  Astropart.\ Phys.\  {\bf 3}, 267 (1995)
  doi:10.1016/0927-6505(94)00043-3
  [hep-ph/9405296, hep-ph/9408296].
  \bibitem{Pakvasa:2007dc} 
  S.~Pakvasa, W.~Rodejohann and T.~J.~Weiler,
  JHEP {\bf 0802}, 005 (2008)
  doi:10.1088/1126-6708/2008/02/005
  [arXiv:0711.4517 [hep-ph]].
  \bibitem{Rodejohann:2006qq} 
  W.~Rodejohann,
  JCAP {\bf 0701}, 029 (2007)
  doi:10.1088/1475-7516/2007/01/029
  [hep-ph/0612047].
  \bibitem{Xing:2008fg} 
  Z.~z.~Xing and S.~Zhou,
  Phys.\ Lett.\ B {\bf 666}, 166 (2008)
  doi:10.1016/j.physletb.2008.07.011
  [arXiv:0804.3512 [hep-ph]].
  \end{thebibliography}
\end{document}